\documentclass[a4paper]{amsart}
\newtheorem{theorem}{Theorem}[section]
\newtheorem{corollary}[theorem]{Corollary}
\newtheorem{lemma}[theorem]{Lemma}

\newtheorem{remark}[theorem]{Remark}
\newtheorem{remarks}[theorem]{Remarks}

\def\dist{{\rm dist\,}}

\def\OO{{\mathcal O}}
\def\FF{{\mathcal{F}}}

\title{Effective Hamiltonians for atoms in very strong magnetic
fields}

\author{Raymond Brummelhuis$^1 $ and Pierre Duclos$^2 $ }

\address{$^1 $ Birkbeck College - University of London, School of
Economics, Mathematics and Statistics, Malet Street WC1E 7HX
London, United Kingdom, e-mail:
r.brummelhuis@statistics.bbk.ac.uk}

\address{$^2 $  Centre de Physique Th\'eorique UMR 6207 - Unit\'e Mixte de Recherche du
CNRS et des Universit\'es Aix-Marseille I, Aix-Marseille II et de
l' Universit\'e du Sud Toulon-Var - Laboratoire affili\'e \`a la
FRUMAM, Luminy Case 907, F-13288 Marseille Cedex 9 France, email:
duclos@univ-tln.fr}

\begin{document}
%\Large{

\maketitle
\begin{abstract} We propose three effective Hamiltonians which
approximate atoms in very strong homogeneous magnetic fields $B$
modelled by the Pauli Hamiltonian, with fixed total angular
momentum with respect to magnetic field axis. All three
Hamiltonians describe $N$ electrons and a fixed nucleus where the
Coulomb interaction has been replaced by $B$-dependent
one-dimensional effective (vector valued) potentials but without
magnetic field. Two of them are solvable in at least the one
electron case. We briefly sketch how these Hamiltonians can be
used to analyse the bottom of the spectrum of such atoms.   
\end{abstract}

%%%%%%%%%%%%%%%%%%%%%%%%%%%%%%%
\section{{\bf Introduction } }%
%%%%%%%%%%%%%%%%%%%%%%%%%%%%%%%

The Pauli Hamiltonian of a non-relativistic atom with an
infinitely heavy nucleus and electrons with spin in a constant
magnetic field $\mathbb{B } $ of strength $B $ is given by:
\begin{equation} \label{HB}
H^B(Z,N) =\sum_{j=1}^N\left(\frac{1}{2}({1\over
i}\nabla_j-\frac{1}{2}\mathbb{B} \wedge r_j)^2
+{\vec\sigma}_j\cdot\mathbb{B }-\frac{Z }{|r_j |}\right) + \sum_{1
\leq j < k \leq N } \frac{1 }{|r_j - r_k |},
\end{equation}
where $r_j = (x_j , y_j , z_j ) \in \mathbb{R }^3 $ are the
coordinates of the $j $-th electron, $\vec\sigma _j$ is its spin,
and $\nabla _j $ is the gradient with respect to $r_j $. Note that
we have made the choice of $\frac{1 }{2 } \mathbb{B } \wedge r $
for the vector potential of $\mathbb{B } $, and that we are
working in atomic units.
\medskip

We fix the direction of $\mathbb{B } $ to be the $z $-direction:
$\mathbb{B } = B (0, 0, 1 ) $ with $B\ge0$ without loss of
generality. Recall that the $z $-component of $\vec \sigma _j $ is
given, in the Pauli representation, by
$$I \otimes \cdots \sigma _{z_j} \cdots \otimes I,
\quad
\sigma_{z_j}=
\frac{1 }{2 } \left(\begin{array}{cc}-1 & 0 \\
                                      0 & 1
                    \end{array}
               \right),
$$
acting on the $N $-fold tensor product $\mathbb{C }^2 \otimes
\cdots \otimes \mathbb{C }^2 $. It is known (see
\cite{KaKu}) that $H^B:=H^B(Z,N)$ defines an essentially
self-adjoint operator on
\begin{equation} \label{Boltz}
\mathcal{H} = \bigotimes _{j = 1 } ^N L^2 (\mathbb{R }^3 )\otimes
\mathbb{C }^2 ,
\end{equation}
the Hilbert space of distinguishable electrons or "boltzons"
(meaning particles satisfying the Boltzman statistics: we thank
Beth Ruskai for introducing us to this expression). Physical atoms
are of course modeled by $H^B$ restricted to the fermionic
subspace
\begin{equation} \label{fermion}
\mathcal{H }_{\rm{f } } = \bigwedge _{j = 1 } ^N L^2 (\mathbb{R
}^3 ) \otimes \mathbb{C }^2 ,
\end{equation}
of totally anti-symmetric wave-functions in $\mathcal{H } $, where
$\wedge $ stands for exterior product. A useful alternative
description of $\mathcal{H } $, used in atomic physics, is given
by the unitary map
$$
\mathcal{H } \to L^2 \left( (\mathbb{R } ^3 \times \{ \pm1\} )^N
\right) , \ \psi \to U(\psi ) (r_1, s_1; \cdots ; r_N, s_N ) ,
$$
the isometry being defined by taking the components of $\psi (r_1
, \cdots , r_N ) $ with respect to the natural basis of $\mathbb{C
}^2 \otimes\cdots \otimes \mathbb{C }^2 $ consisting of the
products $\otimes _{j } e_{s_{j } } $ of the normalized
eigenvectors $e_{\pm} $ of $\sigma _{z_j} $, $e_{-} $
corresponding to "spin down" and $e_+  $ to "spin up". In this new
representation, $\sigma _{z_j } $ acts as the multiplication
operator by $s_j / 2 $, and the fermionic subspace $\mathcal{H
}_{\rm{f } }$ of (\ref{Boltz}) is simply obtained by
anti-symmetrizing with respect to the $4 $-tuples of variables
$(r_1, s_1), \cdots , (r_N , s_N ) $.
\medskip

The main results of this paper can be summarized in the following
five theorems below. It is worthwhile to observe that $H^B$
commutes with each individual spin operator $\sigma_{z_j} $ and
therefore decomposes in a direct sum which is unitarily equivalent
to
$$
\bigoplus_{s_{z_j}\in\{\pm
1\}}H^{B,\mathbb{S}_z=-{NB\over2}}+\sum_{j=1}^N(1+s_{z_j})
{B\over2} ,
$$
where $\mathbb{S}_z:=\sum_{j=1}^N\sigma_{z_j}$ denotes the $z$
component of the total spin operator. So from now on we will
consider only $H^{B,\mathbb{S}_z=-{NB\over2}}$ and denote it again
by $H^B$. Notice that this operator simply acts in
$\otimes_{j=1}^NL^2(\mathbb{R}^{3}) . $ Since the Hamiltonian $H^B
$ also commutes with the total angular momentum operator in the
field direction, which we will call $\mathbb{L }_z $, we can fix a
value $\mathbb{M }\geq 0 $ of the latter. Our results will imply
that the bottom of the spectrum of $H^B $ will necessarily occur
for a non-negative value of $\mathbb{M } $, and we will therefore
restrict ourselves to $\mathbb{M } \geq 0 $. Let $H^{B, \mathbb{M
} } $ be the restriction of $H^B $ to the $\mathbb{M } $-th
angular momentum channel (in the field direction), and let $\Pi
^{B, \mathbb{M } } _{\rm eff} $ be the orthogonal projection onto
the lowest Landau states with $z $-angular momentum $\mathbb{M } $
(cf. (\ref{Pi-eff}) for the precise definition). We define the
{\sl effective} Hamiltonian $h_{\rm eff } = h_{\rm eff } ^{B ,
\mathbb{M } } $\ by
\begin{equation} \label{HsB}
h_{\rm eff } ^{B , \mathbb{M } } = \Pi _{\rm eff } ^{B ,
\mathbb{M } } \; H^{B, \mathbb{M } } \; \Pi _{\rm eff } ^{B ,
\mathbb{M } } ,
\end{equation}
and let
$$
\Pi _{\perp } ^{B, \mathbb{M } } = I - \Pi _{\rm eff } ^{B,
\mathbb{M } }
$$
be the projection onto the orthogonal complement (always
restricting ourselves to the $\mathbb{M } $-th $z $-angular
momentum channel). The operator $h_{\rm eff} ^{B , \mathbb{M } } $
is the first, and most encompassing, of three "effective
Hamiltonians" we will consider in this paper. The two other ones,
called $h_C ^{B , \mathbb{M } } $ and $h_{\delta } ^B $, will be
defined below. It will be convenient to complete $h_{\rm
eff}^{B,\mathbb{M}}$ as follows
\begin{equation}\label{GrandHeff}
H_{\rm eff}^{B,\mathbb{M}}:=h_{\rm eff}^{B,\mathbb{M}}\oplus
H_\perp^{B,\mathbb{M}},\quad{\rm with}\quad
H_\perp^{B,\mathbb{M}}:=\Pi_\perp^{B,\mathbb{M}}
H^{B,\mathbb{M}}\Pi_\perp^{B,\mathbb{M}}.
\end{equation}
For any self-adjoint operator $A$, we let $\sigma (A) $ denote the
spectrum of $A $ and $\rho (A) $ its resolvent set. Our first main
result is:
\begin{theorem} \label{FMT} Let
$\alpha = \alpha (B) $ be the unique positive solution of the
equation
\begin{equation} \label{FMT:alpha}
\alpha + \log \alpha = \frac{1 }{2 } \log B ,
\end{equation}
and let $d_{\rm eff } (\xi ) = \mbox{\dist} \left( \xi , \sigma (
h_{\rm eff } ^{B, \mathbb{M } } ) \right) $.  There exist positive
constants $B_{\rm eff } $, $c_{\rm eff } $ and $C_{\rm eff } $,
which only depend on $Z, N $ and $\mathbb{M } $, such that for all
$B \geq B_{\rm eff } $, and all real $\xi$ satisfying
$$
c_{\rm eff } \frac{\alpha }{\sqrt{B } } \leq d_{\rm eff }(\xi )
\leq \frac{1 }{2 } \alpha ^2 .
$$
we have that $\xi \in \rho (H^{B, \mathbb{M } } ) $, and
\begin{equation} \label{FMTa}
\| \; (H ^{B, \mathbb{M } }  - \xi )^{-1 } - (H_{\rm eff }
^{B,\mathbb{M } } - \xi )^{-1 } | \leq  C_{\rm eff} \frac{\alpha
(B )^2 }{d_{\rm eff } (\xi )^2 \; \sqrt{B } } .
\end{equation}
\end{theorem}

\begin{remarks}
\rm{(i) We shall see in Theorem~\ref{ME} that
$\sigma(H_\perp^{B,\mathbb{M}})\subset\mathbb{R}_+$ for $B\ge
B_{(\ref{constant-B})}$, where the latter is defined by formula
(\ref{constant-B})\footnote{see our convention on constants at the
end of this introduction}. Since $B_{\rm eff}\ge
B_{(\ref{constant-B})}$ by (\ref{FMT:B-eff}), and since one can
see from (\ref{HsB1}) and (\ref{defVeffietjk}) below that
$\mathbb{R}_+\subset\sigma(h_{\rm eff}^{B,\mathbb{M}})$, it
follows that $\sigma(h_{\rm eff}^{B,\mathbb{M}})=\sigma(H_{\rm
eff}^{B,\mathbb{M}})$ when $B\ge B_{\rm eff}$.

\medskip

\noindent (ii) The equation for $\alpha(B) $ is equivalent to
$\alpha e^{\alpha } = \sqrt{B } $, and can therefore be written as
$$
\alpha (B) = W (\sqrt{B} ) ,
$$
where $W $ is the principal branch of the Lambert W function, see e.g. \cite{CoGoHaJeKn}.
Using known properties of the Lambert W-function, or by elementary
arguments, one shows that
\begin{equation}\label{AlphAsymptotics}
\alpha (B) = \frac{1 }{2 } \log B - \log ^{(2) } B + \log 2 +
O\left(\frac{\log^{(2)} B }{\log B } \right) , \ \ B \to \infty ,
\end{equation}
where $\log ^{(2)} (x) := \log (\log x ) $, for $x > 1 $. In
particular, $\alpha (B) \simeq \log (\sqrt{B } ) $, as $B \to
\infty . $
\medskip

\noindent (iii) Our proof yields explicit constants $B_{\rm eff }
$, $c_{\rm eff } $ and $C_{\rm eff } $, for given $N, \mathbb{M }
$ and $Z $. This is also true for the constants in Theorems
\ref{SMT} and \ref{CorrSMT} below.
\medskip

\noindent (iv) The upper bound $\alpha ^2 / 2 $ on $d_{\rm eff }
(\xi ) $ is only there to allow a simple expression for the upper
bound in (\ref{FMTa}), and is by no means essential. The same
remark applies to Theorems \ref{SMT},  \ref{CorrSMT} and
\ref{FermionizedMThms} below. }
\end{remarks}

Some applications of Theorem \ref{FMT}, as well as of Theorems
\ref{SMT}, \ref{CorrSMT}, \ref{FermionizedMThms} and \ref{ThMT}
below, to the study of spectral properties of $H^{B,\mathbb{M}}$
are given in the concluding remarks section.

\medskip
The operator $h_{\rm eff} ^{B, \mathbb{M } } $ has the structure
of a multi-particle Schr\"odinger operator on the real line,
$\mathbb{R } $:
\begin{equation} \label{HsB1}
h _{\rm eff} ^{B, \mathbb{M } } = -\frac{1 }{2 } \Delta - Z \sum
_j V_j ^{B, \mathbb{M } } (z_j ) + \sum _{j < k } V_{jk } ^{B,
\mathbb{M } } (z_j - z_k ) ,
\end{equation}
with operator-valued potentials acting point-wise on a certain
finite-dimensional Hilbert-space  $F_{\mathbb{M } }^B $, defined
in (\ref{VBM}) below.  Essentially, $F_{\mathbb{M } }^B $ is the
vector space spanned by the lowest Landau states with angular
momentum $\mathbb{L }_z = \mathbb{M } $ . As we will see,
$\mbox{Ran } \Pi ^{B , \mathbb{M } } _{\rm{eff } } $ is
canonically isomorphic to the space $L^2 (\mathbb{R }^N ,
F_{\mathbb{M } }^B ) $ of vector-valued $L^2 $-functions , and the
potentials in (\ref{HsB1}) are simply obtained by projecting the
respective Coulomb terms in (\ref{HB}) along $\Pi _{\rm eff } ^{B,
\mathbb{M } } $:
\begin{equation} \label{defVeffietjk}
V_j ^{B, \mathbb{M } } (z_j )  :=\Pi_{\rm eff }^{B,
\mathbb{M } } {1\over|r_j|} \Pi_{\rm eff } ^{B, \mathbb{M } }
\qquad V_{jk} ^{B, \mathbb{M } }  (z_j - z_k )
:=\Pi_{\rm eff } ^{B, \mathbb{M } } {1\over|r_j-r_k|} \Pi_{\rm eff
} ^{B, \mathbb{M } }.
\end{equation}
We will show that the potentials (\ref{defVeffietjk})
can be approximated by certain simpler ones, which will give rise
to our two other effective Hamiltonians. Define the tempered
distribution $q^B $ on $\mathbb{R } $ by:
\begin{equation} \label{qBintro}
q^B (z) = \log B \; \delta (z) + \mbox{Pf }\left( \frac{1 }{|z| }
\right) ,
\end{equation}
where
$$
\mbox{Pf} \left( \frac{1 }{|z | } \right) := {\frac{{\rm d} }{{\rm
d } x } } \left( \mbox{sgn} (z ) \log |z | \right) ,
$$
(with distributional derivative)  is the finite part (in the sense
of Hadamard) of the singular function $1/|z | $; $\mbox{Pf} (|z
|^{-1 } ) $ should be interpreted as a regularization of the
(3-dimensional) Coulomb potential restricted to the line. Also
introduce (constant) finite dimensional operators $C_j ^{n, B,
\mathbb{M } } $ and $C_{jk } ^{e, B, \mathbb{M } } $, acting on
the vector space $F_{\mathbb{M } }^B $ introduced above, defined
by
\begin{equation} \label{CjB}
C_j ^{n, B, \mathbb{M } } := - \Pi _{\rm eff } ^{B, \mathbb{M } }
\log \left( \frac{B }{4 } (x_j ^2 + y_j ^2 ) \right) \Pi _{\rm eff
} ^{B, \mathbb{M } } ,
\end{equation}
and
\begin{equation} \label{CjkB}
C_{jk } ^{e, B, \mathbb{M } } := - \Pi _{\rm eff } ^{B, \mathbb{M
} } \log \left( \frac{B }{4 } \left((x_j - x_k )^2 + (y_j - y_k )
^2 \right) \right) \Pi _{\rm eff } ^{B, \mathbb{M } } .
\end{equation}
The superscripts "$n $" and "$e $" stand for "nucleus" and
"electron", respectively, as a reminder that (\ref{CjB}) is a
vestige of the interaction between the $j $-th electron and the
nucleus, while (\ref{CjkB}) originates in the electron-electron
interaction between electrons $j $ and $k $.  Finally, define
an operator $h_C ^{B , \mathbb{M } } $ on $L^2 (\mathbb{R } ,
F_{\mathbb{M } }^B ) $ (the `C' standing for `Coulomb') by
\begin{eqnarray} \label{h0Bintro}
h_C ^{B , \mathbb{M } } &=& -\frac{1 }{2 } \Delta - Z \sum _j
\left( q^B
(z_j ) + C_j ^{n, B , \mathbb{M } } \delta (z_j ) \right) \\
&\ & + \sum _{j < k } \left( q^B (z_j - z_k ) + C_{jk } ^{e, B ,
\mathbb{M } } \delta (z_j - z_k ) \right) . \nonumber
\end{eqnarray}
 As we will see in section 4, the right hand side of
(\ref{h0Bintro}) defines a self-adjoint operator $h_C ^{B ,
\mathbb{M } } $ on $L^2 (\mathbb{R }^N ) $, despite the
distributional potentials. This will be a consequence of the
Kato-Lax-Lions-Milgram-Nelson Theorem. The form domain of $h_C ^{B
, \mathbb{M } } $ is simply the vector-valued first Sobolev space
$H^1 (\mathbb{R }^N; F_{\mathbb{M } } ) $, while its operator
domain will be characterized in appendix A.  As in
(\ref{GrandHeff}) we introduce
$$
H_C^{B,\mathbb{M}}:=h_C^{B,\mathbb{M}}\oplus
H_\perp^{B,\mathbb{M}} .
$$
Our second main theorem then is the following:

\begin{theorem} \label{SMT} Let
$\alpha = \alpha (B ) $ be as in Theorem \ref{FMT}, and put $d_C
(\xi ) := \mbox{\dist} \left( \xi , \sigma ( h_C ^{B, \mathbb{M }
} ) \right) . $ There exists  positive  constants
$B_C$, $c_C$ and $C_C$ which depend only on $Z, N $ and $\mathbb{M
} $, such that for all $B \geq B_C $ and all real $\xi $
satisfying
$$
c_C \frac{\alpha^{3/2} }{B^{1/4}  } \leq  d_C (\xi ) \leq
\frac{1 }{4 } \alpha ^2 .
$$
we have that $\xi \in \rho (H^{B, \mathbb{M } } ) $, and
\begin{equation} \label{SMT:DiffRes}
 \| (H^{B, \mathbb{M } } - \xi )^{-1 }  - (H_C ^{B, \mathbb{M
} } - \xi )^{-1 } \| \leq  \frac{C_C \alpha^{3\over2}}{ B^{1/4 }
d_C (\xi )^2 }.
\end{equation}
\end{theorem}

\begin{remark} \label{interpr:v_C} \rm{In top order in $B $, all that remains
of the electrostatic potentials in $H^{B , \mathbb{M } } $ are the
extremely short-range $\delta $-potentials. In next order, the
long range character of the original Coulomb potentials reasserts
itself in two ways: in the magnetic field direction, through the
$\mbox{Pf } (| \cdot | ^{-1 } ) $-terms in $h_C ^{B , \mathbb{M }
} $, and in the transversal directions, through the $C_j ^{n, B,
\mathbb{M } }$- and $C_{jk } ^{e, B, \mathbb{M } } $- terms. The
latter are in fact  simply the quantum mechanical mean, with
respect to the projection onto the lowest Landau band states of
total angular momentum $\mathbb{M } $ (in the field direction), of
a  2-dimensional logarithmic potential, minus a $B $-dependent
constant. This  logarithmic potential is the natural electrostatic
potential for the plane. Physically, this can be understood as
follows: under the influence of the strong magnetic field the
electrons will spiral closely around the field lines, along
circles of radius $O(B^{-1/2 } ) $ in the plane transversal to the
field, while occupying an interval of size $O((\log B )^{-1 } ) $
in the field direction itself, as a consequence of the nuclear
attraction. For big $B $, and at different locations in the $(x, y
) $-plane, they will see each other and the nucleus as so many
infinitely long charged wires, and, as is known from classical
electrostatics, such wires interact via a logarithmic potential. }
\end{remark}

A simpler effective Hamiltonian, our third and last one, and
historically the first to be proposed (cf. \cite{LSY},
\cite{BaSoY}, \cite{BD}), is roughly speaking obtained by only
keeping the leading term in the potential of $h_C ^{B, \mathbb{M }
} $. More precisely, we put:
\begin{equation} \label{Hdelta1}
h_{\delta } ^{B} = -\frac{1 }{2 } \Delta _z + 2\alpha(B)
\; v_{\delta } ,
\end{equation}
where
\begin{equation} \label{Hdelta2}
v_{\delta } (z) = - Z \sum _{j = 1 } ^N \delta (z_j ) + \sum _{j <
k } \delta (z_j - z_k ) .
\end{equation}
Looking back at (\ref{qBintro})  it would seem natural to
take as potential  $\log B v_\delta $, but it turns out that
$2\alpha(B) v_\delta$ leads to smaller error estimates; notice
that in view of (\ref{AlphAsymptotics}), $2\alpha v_\delta$ is
also a $O(\log B)$ part of  the potential in $h_C^B . $
Furthermore, with this choice the coupling constant $2\alpha (B) $
is positive for all $B>0$ which is not the case for $\log B $.
Contrary to our previous two effective Hamiltonians, $h_{\delta }
^B $ does not explicitly depend on $\mathbb{M } $ anymore, but it
will operate on an $\mathbb{M } $ and $B $-dependent Hilbert
space, namely $L^2 (\mathbb{R }^N , F_{\mathbb{M } }^B ) \simeq
L^2 (\mathbb{R } ^N) \otimes F_{\mathbb{M } }^B $ (which in fact
are canonically isomorphic for different $B $). Considering
(\ref{Hdelta1}) as acting on scalar $L^2 (\mathbb{R } ^N ) $, we
define the $\delta $-model as being the operator
\begin{equation}\label{hdeltaBM}
 h_{\delta } ^{B,\mathbb{M}} :=
h_{\delta } ^B \otimes I_{F_{\mathbb{M } }^B } ,\quad\mbox{
$I_{F_{\mathbb{M } }^B } $ being the identity operator.}
\end{equation}
We will often simply write $h_{\delta } ^B $, except when we want
to stress the vector-valued nature of the $L^2 $-functions in the
domain. Again as in (\ref{GrandHeff}) we introduce
$$
H_\delta^{B,\mathbb{M}}:=h_\delta^{B,\mathbb{M}}\oplus
H_\perp^{B,\mathbb{M}}.
$$
Our third approximation theorem is:

\begin{theorem} \label{CorrSMT} Let $\alpha:=\alpha(B)$ be as in Theorem
\ref{FMT}, and put $d_{\delta } (\xi ) := \mbox{\dist} \left(\xi ,
\sigma (h_{\delta } ^{B , \mathbb{M } } ) \right) $. There exist
positive constants $B_\delta$, $c_\delta$ and $C_\delta$,
depending on $N$, $Z$ and $\mathbb{M } $, such that for all $B\ge
B_\delta $ and real $\xi$ satisfying
\begin{equation}\label{IsolationConditionHHdelta}
c_\delta \alpha \le d_{\delta } (\xi ) \le {1\over4}\alpha^2 ,
\end{equation}
we have that $\xi\in\rho(H^{B,\mathbb{M }})$, and
\begin{equation} \label{4.11}
\| (H^{B, \mathbb{M } } - \xi )^{-1 }  - (H_{\delta } ^{B ,
\mathbb{M } } - \xi )^{-1 } \| \leq \frac{C_\delta\, \alpha}{
d_{\delta } (\xi )^2 } .
\end{equation}
\end{theorem}

\noindent See \cite{BD} for  weaker versions of this theorem. We
also mention \cite{BaSoY}, which established\footnote{
\cite{BaSoY}, following \cite{LSY}, first did a re-scaling of
$H^B$'s ground state energy which allowed them to compare with
$h_{\delta } ^B $ for a {\sl fixed} $B $ (e.g. fixing $2 \alpha
(B) = 1 $). Since this homogeneity property is not valid anymore
for our other two effective hamiltonians, we prefer not to do this
here (contrary to our earlier papers \cite{BD}), in order to have
a coherent presentation.} the convergence of the ground state
energy of fermionic $H ^B $ (see below) to that of bosonic
(scalar) $h_{\delta }^B $ on $L^2 (\mathbb{R }^N ) $, using
variational arguments: these authors did not fix $\mathbb{M } $,
but they only proved convergence of the ground state energy, while
we can conclude much more from the norm resolvent convergence to
the effective hamiltonians; see \S9 for a list of applications of
the results of the present paper.
 Earlier,
\cite{LSY} had shown that the ground state of the Hartree
mean-field model associated to (\ref{Hdelta1}) approximates the
quantum mechanical ground state energy in the so-called
hyper-strong limit $Z, B/Z^3 \to \infty $, assuming $N/Z $
uniformly bounded. The idea that a model such as the $\delta
$-model could be relevant in  the context of strong magnetic
fields  is not new in the physics literature, see e.g. \cite{Spr}.
\medskip

We next turn to the effects of particle symmetry. Electrons in
physical atoms are fermions, and we now  consider the
analogues of Theorems \ref{FMT}, \ref{SMT} and \ref{CorrSMT} for
$H^B $  restricted to the fermionic subspace $\mathcal{H }_{\rm f
} = P^{AS } (\mathcal{H } ) $, where $P^{AS } $ is
 the orthogonal projection onto the subspace of
anti-symmetric symmetric wave-functions  defined by:
\begin{eqnarray*} \nonumber
P^{\rm AS } \psi (r_1 , s_1 ; \cdots ; r_N , s_N ) := \frac{1 }{N
! } \sum _{\sigma \in S_N } (-1 )^{\sigma } \psi \left( r_{\sigma
(1 ) } , s_{\sigma (1 ) } ; \cdots ; r_{\sigma (N) } , s_{\sigma
(N ) } \right) ,
\end{eqnarray*}
 with  $(-1 )^{\sigma } := (-1 )^{\mbox{sgn } (\sigma )
} $. The projection $P^{\rm AS } $ commutes with $H^B $,
$\mathbb{L }_z $, $ \mathbb{S }_z $ and with the $N $-particle
Landau Hamiltonian $H_0 ^B $ defined in (\ref{LandauHam}) below,
and therefore also with $\Pi _{\rm eff } ^{B, \mathbb{M } } $ (see
section 2). Recalling that we have fixed our spins to $\mathbb{S
}_z = -N B / 2 $, $P^{AS } $ for us will only act in the `spatial'
variables $(r_1 , \cdots , r_N ) $.  Let $H_{\rm f }
^{B,\mathbb{M}} := P^{\rm AS } H^{B,\mathbb{M}} P^{\rm AS } =
H^{B,\mathbb{M}} P^{\rm AS } $, the fermionic Pauli operator with
$z $-angular momentum $\mathbb{M } $. Similarly, introduce
`fermionized' versions of the other operators: $\Pi _{{\rm eff } ,
{\rm f } } ^{B , \mathbb{M } } := P^{\rm AS } \Pi _{\rm eff } ^{B
, \mathbb{M } } $, $\Pi _{\perp, {\rm f } } ^{B , \mathbb{M } } :=
P^{\rm AS } \Pi _{\perp } ^{B , \mathbb{M } } $, $H_{{\rm eff } ,
{\rm f } } ^{B , \mathbb{M } } := P^{AS } H_{{\rm eff } } ^{B ,
\mathbb{M } } =P^{AS } h_{{\rm eff } } ^{B , \mathbb{M } }\oplus
H_{\perp\rm,f}^{B , \mathbb{M } }=: h_{{\rm eff,f } } ^{B ,
\mathbb{M } }\oplus H_{\perp\rm,f}^{B , \mathbb{M } } $, $h_{C ,
{\rm f } } ^{B ,\mathbb{M } } = P^{AS } h_C ^{B, \mathbb{M } }$,
$h_{\delta , {\rm f } } ^B := P^{AS } h_{\delta } ^B$, $H_{C ,
{\rm f } } ^{B , \mathbb{M } } = P^{AS } H_C ^{B, \mathbb{M } }$
and finally $H_{\delta , {\rm f } } ^{B,\mathbb{M}} := P^{AS }
H_{\delta } ^{B,\mathbb{M}} . $ A careful examination of the
proofs of Theorems \ref{FMT}, \ref{SMT} and \ref{CorrSMT} will
show:

\begin{theorem} \label{FermionizedMThms} Theorems \ref{FMT},
\ref{SMT} and \ref{CorrSMT} also hold true for the fermionized
operators. For example, if  $\xi $ satisfies the conditions of
theorem \ref{FMT} with $d_{\rm eff } (\xi ) $ replaced by $d_{\rm
eff, f } (\xi ) := \mbox{\dist} \left( \xi , \sigma ( h_{\rm eff ,
f } ^{B, \mathbb{M } } ) \right) $, then
\begin{equation} \label{fer1}
\| \left( H_{\rm f } ^{B, \mathbb{M } } - \xi \right) ^{-1 }
-\left( H_{\rm eff , f } ^{B , \mathbb{M } }  - \xi \right) ^{-1 }
\| \leq C_{\rm eff  } \frac{\alpha (B) ^2 }{d_{\rm eff , f }(\xi
)^{2 }\sqrt{B }  } ,
\end{equation}
with the same constants as before.  Similarly for
$H_{C, \rm{f } } ^{B , \mathbb{M } } $, $H_{\delta , {\rm f } }
^{B, \mathbb{M } } . $
\end{theorem}

\begin{remark} \rm{ Theorem \ref{FermionizedMThms} is not simply obtained by
"sandwiching" Theorems \ref{FMT}, \ref{SMT} and \ref{CorrSMT}
between $P^{AS } $, since the statements thus obtained would not
involve the distances to the spectra of the fermionized operators.
Also, we established the fermionic versions with the same
constants as for the boltzonic ones, but it is conceivable that
one could have smaller constants in the fermionic case.  }
\end{remark}

The operators $h_{{\rm eff , f } } ^{B, \mathbb{M } } $ and $h_{C,
{\rm f } } ^{B, \mathbb{M } } $ are easily described:
\begin{equation} \label{FHsB}
h_{{\rm eff , f } } ^{B, \mathbb{M } } = -\frac{1 }{2 } \Delta - Z
\sum _{j = 1 } ^N V^{B, \mathbb{M } } _{{\rm av }:1 } (z_j ) +
\sum _{j < k } V^{B, \mathbb{M } } _{{\rm av}:2 } (z_j - z_k ) ,
\end{equation}
where, using the notation of (\ref{defVeffietjk}),
$$
V^{B, \mathbb{M } } _{{\rm av }:1 } (z) := \frac{1 }{N } \sum _j
V_{j }^{B, \mathbb{M } } (z) ,
$$
and
$$
V^{B, \mathbb{M } } _{{\rm av }:2 } (z) := \left( \begin{array}{cc} N \\
2
\end{array} \right) ^{-1 } \sum _{j < k } V_{jk } ^{B , \mathbb{M
} } (z) ,
$$
the average one-, respectively two-particle potentials. Similarly,
$h_{C, {\rm f } } ^{B , \mathbb{M } } $ equals
\begin{eqnarray} \label{Fh0}
h_{C , {\rm f } } ^{B, \mathbb{M } } &= & -\frac{1 }{2 } \Delta -
Z \sum _{j = 1 } ^N \left( \; q^B (z_j ) + C^{n, B, \mathbb{M } }
_{{\rm
av }:1 } \delta (z_j ) \; \right) \\
&\ & + \sum _{j < k } \left( \; q^B (z_j - z_k )+ C^{e, B,
\mathbb{M } } _{{\rm av }:2 } \delta (z_j - z_k ) \; \right) ,
\nonumber
\end{eqnarray}
with
$$
C^{n, B, \mathbb{M } } _{{\rm av }:1 } := \frac{1 }{N } \sum _{j =
1 } ^N C_j ^{n, B, \mathbb{M } } ,
$$
and
$$
C^{e, B, \mathbb{M } }_{{\rm av }:2 } := \left(\begin{array}{cc} N
\\ 2
\end{array} \right) ^{-1 }
\sum _{j < k } C _{jk } ^{e, B, \mathbb{M } } ,
$$
while $h_{\delta , {\rm f } } ^{B, \mathbb{M } } $ is given by the
same expression as $h_\delta^B $, except of course that its domain
changes.
\medskip

To complete the picture, we  finish with an explicit description
of $\mbox{Ran} (\Pi _{\rm eff , f } ^{B , \mathbb{M } } )$, which
is equal to $P^{\rm AS } \left( \mbox{Ran} \; \Pi _{\rm eff } ^{B
, \mathbb{M } } \right) = P^{AS } \left( L^2 (\mathbb{R }^N )
\otimes F_{\mathbb{M } }^B \right) $. Fix $\mathbb{M } \geq 0 $,
and let
$$
\Sigma (\mathbb{M } ) = \{ m \in \mathbb{N }^N : |m | = \mathbb{M
} \},
$$
the set of partitions of $\mathbb{M } $, where, as usual, $|m | =
m_1 + \cdots + m_N $ if $m = (m_1 , \cdots , m_N ) $. The vector
space $F_{\mathbb{M } }^B $ is spanned by the lowest Landau
(generalized) eigenstates indexed by $m \in \Sigma (\mathbb{M } )
$, cf. section 2 below. The symmetric group $S_N $ acts on $\Sigma
(\mathbb{M } ) $ in the natural way, by permuting the indices of
an element $m = (m_1 , \cdots , m_N ) $ of $\Sigma (\mathbb{M } )
. $ Under this action, $\Sigma (\mathbb{M } ) $ will decompose as
a finite union of disjoint orbits:
$$
\Sigma (\mathbb{M } ) = \bigcup _{\overline{m } \in \mathcal{M } }
S_N \cdot \overline{m } ,
$$
$\mathcal{M } $ being a set of representatives of the orbits. If
$G_{\overline{m } } $ denotes the stabilizer of
$\overline{m } \in \mathcal{M } $, then we write $\mathcal{M } $
 as a disjoint union  $\mathcal{M } = \mathcal{M }_1
\cup \mathcal{M }_2 $,  with  $\mathcal{M }_1 $ the
subset of those $\overline{m } \in \mathcal{M } $ such that
 $G_{\overline{m } } = \{ e \} $, and $\mathcal{M }_2 =
\mathcal{M } \setminus \mathcal{M }_1 $, its complement.  In other
words, $\overline{m } \in \mathcal{M }_1 $ iff no two components
of $\overline{m} $ are the same, and $\overline{m } \in \mathcal{M
}_2 $ iff at least two of its  components are
identical. Let $L^2_{\rm AS } (\mathbb{R }^N) $ be the space of
anti-symmetrical wave functions in $L^2 (\mathbb{R }^N ) $. We
then will prove, in section 8, that:

\begin{theorem} \label{ThMT} There is a natural unitary isomorphism
\begin{equation} \nonumber %\label{ThMT1}
U_{\mathbb{M } } ^B : \mbox{\rm Ran} (\Pi _{{\rm eff , f } } ^{B,
\mathbb{M } } ) \to \sum ^{\oplus } _{\overline{m } \in \mathcal{M
}_1 } L^2 (\mathbb{R }^N ) \oplus \sum ^{\oplus }_{\overline{m }
\in \mathcal{M }_2 } L^2 _{AS } (\mathbb{R }^N ) ,
\end{equation}
and
$$
U_{\mathbb{M } }  ^B \; h_{\delta , {\rm f } } ^{B, \mathbb{M } }
\; U_{\mathbb{M } } ^{ B * } = \sum _{\overline{m} \in \mathcal{M
}_1 } ^{\oplus } {h_{\delta } ^B}{|_{L^2(\mathbb{R}^N)}} \oplus
\sum _{\overline{m} \in \mathcal{M }_2 } ^{\oplus } h_{\delta }
^B{|_{L^2_{\rm AS}(\mathbb{R}^N)}} .
$$
\end{theorem}

\begin{remark} \rm{The operators $U_{\mathbb{M } }  ^B \; h_{C , {\rm f } } ^{B, \mathbb{M } }
\; U_{\mathbb{M } } ^{B * } $ and $U_{\mathbb{M } }  ^B \; h_{\rm
eff, f }^{B, \mathbb{M } } \; U_{\mathbb{M } } ^{B * } $ can mix
different components of $\mbox{Ran } (U_{\mathbb{M } } ^B ) $, as
we will see at the end of section 8, and will therefore have a
more complicated structure. }
\end{remark}

The paper is organized as follows. Section 2 contains the
precise definition of our effective projector $\Pi ^{B, \mathbb{M
} } _{\rm eff } . $ In section 3 we establish, with the help of
the Feschbach decomposition, a first approximation theorem,
comparing $H^{B, \mathbb{M } } $'s resolvent at $\xi $ with that
of $H_{\rm eff } ^{B, \mathbb{M } } + \mathcal{W }^{B, \mathbb{M }
 }(\xi ) $, where the last term is an auxiliary `potential'
which itself depends on the spectral parameter $\xi $. Section 4
analyzes the large-$B $ behavior of the potential of $H^{B,
\mathbb{M } } _{\rm eff } $, as well as that of $\mathcal{W }^{B ,
\mathbb{M } } . $ Sections 5, 6 and 7 are devoted to the proofs
of, respectively, theorems \ref{FMT}, \ref{SMT} and \ref{CorrSMT}.
In section 8 we prove theorems \ref{FermionizedMThms} and
\ref{ThMT}. Section 9, finally, concludes with some applications
to the spectral theory of $H^{B, \mathbb{M } } $, and some general
observations.
\medskip

\noindent {\bf Convention on constants.} In the course of this
work, we have had to introduce a large number of constants. To
keep track of them, we will use the convention that whenever the
subscript of a constant is a number, the number refers to the
formula where the constant in question was first introduced. That
is, $C_{(x) } $ := constant defined in formula $(x) . $
\medskip

\noindent{\bf Acknowledgements.} We first of all would like to
thank Beth Ruskai for having introduced us to the fascinating
world of atoms in strong magnetic fields. We would also like to
thank the organizers of the Clausthal PDE Conference, Summer 1999,
and of the Critical Stability Workshop at les Houches, October
2001, for having given us the opportunity to report on preliminary
written versions of this work, \cite{BD}.

\section{{\bf Non-interacting electrons and the Lowest Landau Band}
}

We begin by reviewing the spectral decomposition of
\begin{equation} \label{LandauHam}
H_0 ^B := H_0 ^B (N) := \sum _{j = 1 } ^N \frac{1 }{2 }\left(
(\frac{1 }{i } \nabla _{r_j } - \frac{1 }{2 } \mathbb{B } \wedge
r_j )^2  -   NB \right) ,
\end{equation}
the "free" Hamiltonian of $N $ independent electrons interacting
only with the field $\mathbb{B }$. Recall that we have fixed all
 electron spins  in their $s_{z_j}=-1 $-state. The operator $ H_0 ^B $ is just
a direct sum of $N $ one-particle operators ${1\over2}\left(
(i^{-1 }\nabla _{r } - {1 \over 2 } \mathbb{B } \wedge r )^2
-B\right) $, whose spectral decomposition is explicitly known:
$$
\bigoplus _{m \in \mathbb{Z }, n \in \mathbb{N }} \left( {1\over2}
p_z ^2 + {B\over2} ( 2n+|m | - m   ) \right) \Pi ^B _{m, n } .
$$
Here, $p_z $ is the momentum in the field direction, and $\Pi ^B
_{m, n } $ is the projection, in the $x, y $-variables, onto the
normalized eigenfunctions $\chi ^B _{m, n} = \chi ^B _{m, n} (x, y
) \in L^2 (\mathbb{R } ^2 ) $ of the operator
$$
-\frac{1 }{2 } \Delta _{x, y } + \frac{B^2 }{8 } (x^2 + y^2 ) -
\frac{B  }{2 } ,
$$
restricted to the $m $-eigenspace of $L_z = xp_y - yp_x $, the
angular momentum in the field-direction. These eigenfunctions are
explicitly known in terms of Laguerre functions, see e.g.
\cite{FW}, but for our purposes we will only need those
with $n = 0 $, $m \geq 0 $. These have a particularly simple
expression: if $(\rho, \varphi ) $ are polar coordinates in the
$x, y $-plane, then
\begin{equation} \label{chi-m}
\chi _m ^B := \chi ^B _{m, 0} : (x,y)\to \left( \frac{B ^{m + 1 }
}{2 \pi 2^m m !  } \right)^{1/2}  \rho^{m} e^{ + i m \varphi }
e^{-B \rho^2 /4}  .
\end{equation}
\medskip

The spectral decomposition of $H_0 ^B $ is simply the sum of the
one-particle decompositions, and the projections onto its
eigenstates will be indexed by $N $-tuples $m = (m_1 , \cdots ,
m_N ) \in \mathbb{Z }^N$, $n = (n_1 , \cdots n_N ) \in \mathbb{N
}^N $. If we let
$$
\Pi ^B _{m, n} := \Pi ^B _{m_1, n_1  } \otimes \cdots \otimes \Pi
_{m_N, n_N  } ,
$$
then
\begin{equation} \label{free}
H_0 ^B = \bigoplus _{m, n} \left( \sum_{j = 1 } ^N \frac{1 }{2 }
p_{z_j} ^2 + {B\over2} (2n_j + \vert m_j \vert - m_j ) \right) \Pi
^B _{m, n} .
\end{equation}
The Lowest Landau Band of $H^B _0 $ is defined as
\begin{equation} \label{LLB}
\mathcal{L }_0 ^B = \bigoplus _{m \in \mathbb{Z }_+ ^N } \mbox{Ran
} \Pi _{m, {\bf 0 }} ^B  ,
\end{equation}
where ${\bf 0}:= (  0, \cdots, 0  )  $, and
if we put
\begin{equation} \label{LLBF}
X^B _m (x, y) := \prod_{j = 1 } ^N \chi ^B _{m_j } (x_j , y_j ) ,
\quad \ m_1, \cdots , m_N \geq 0 ,
\end{equation}
then $\mathcal{L }_0 $ will be spanned by the tensor products $X^B
_m \otimes u $, with $u = u(z) \in L^2 (\mathbb{R }^N ) . $
We will call the $X_m ^B $  the lowest Landau band states
(these are not eigenvectors of $H_0 ^B $, but $X_m ^B
\otimes 1 $ would be generalized eigenvectors with eigenvalue
0). The operator $H^B _0 $ restricted to $\mathcal{L }_0 ^B $
simply is the free Laplacian in the field direction,
$$
\displaystyle{ {1\over2}\sum _j p_{z_j} ^2  = - \frac{1 }{2
} \Delta _z },
$$
where $z = (z_1 , \cdots , z_N ) $.
\medskip

We next reduce the Hamiltonians $H^B $ and $H^B _0 $ to their
angular momentum sectors with respect to the field direction. The
total orbital angular momentum in the direction of $\mathbb{B } =
(0 , 0, B ) $,
\begin{equation} \nonumber
\mathbb{L }_z = \sum _j (x_j p_{y_j } - y_j p_{x_j } ), \ \ (p_x,
p_y, p_z ) = \frac{1 }{i } \nabla _{r} ,
\end{equation}
commutes with $H^B $ and $H_0 ^B $, and is therefore a constant of
motion for both Hamiltonians. If $P^{\mathbb{M } } $ is the
orthogonal projection onto the $\mathbb{M } $-th eigenspace of
$\mathbb{L }_z $, then we let
\begin{equation} \label{HMB:s-1/2}
H^{B, \mathbb{M } } := H^B P^{\mathbb{M } } ,\qquad H_0^{B, \mathbb{M } } := H_0^B P^{\mathbb{M } }
\end{equation}
acting on $\mathcal{H } := L^2 (\mathbb{R }^{3N } )\otimes
\mathbb{C}^{2N} $. Since we are primarily interested in the
spectral behavior of $H^B $ near the bottom of its spectrum, we
will restrict $\mathbb{M } $ to $\mathbb{Z } _+ $, for $
\mathcal{L }_0 ^B \cap \mbox{Ran } P^{\mathbb{M } } \neq \{ 0 \}
\Leftrightarrow \mathbb{M } \geq 0 . $ Indeed, notice that since
$H^{B,-\mathbb{M}}$ is unitarily equivalent to $H^{B,\mathbb{M}}+
\mathbb{M}B $, one has $ \inf \sigma (H^B \vert _{\mathbb{M } \geq
0 } ) < \inf \sigma (H^B \vert _{\mathbb{M } < 0 } ) $ as soon as
$B>0$.

We next let
\begin{equation} \label{SigmaM}
\Sigma (\mathbb{M } ) = \{ m = (m_1 , \cdots m_N ) \in \mathbb{Z }
^N: m_j \geq 0,\ m_1 + \cdots + m_N = \mathbb{M } \} ,
\end{equation}
the set of partitions of $\mathbb{M } $, and define the {\sl
effective projection} $\Pi _{\rm eff } ^{B , \mathbb{M } } $ by
\begin{equation} \label{Pi-eff}
\Pi _{\rm eff } ^{B, \mathbb{M } } := \sum _{m \in \Sigma
(\mathbb{M } ) } \Pi ^B _{m, {\bf 0 } } .
\end{equation}
This is simply the orthogonal projection onto $\mathcal{L } _0
\cap (\mathbb{L }_z = \mathbb{M } \} $. We also let $\Pi _{\perp }
^{B , \mathbb{M } } $ be the orthogonal projection onto the
orthogonal complement of $\mbox{Ran} (\Pi _{\rm eff } ^{B ,
\mathbb{M } } ) $ in $\mbox{Ran} (P^{\mathbb{M } } ) $.
Observe  that
$$
\Pi_{\perp } ^{B, \mathbb{M } } =\bigoplus_{
\begin{array}{cc}
m_1+\ldots+m_N=\mathbb{M } , \\
\sum_j 2n_j +|m_j|-m_j \geq 2
\end{array} }
\Pi_{m, n } .
$$
If we let $F_{\mathbb{M } }^B$ be the finite dimensional vector
space spanned by the lowest Landau states with total angular
momentum $\mathbb{M } $,
\begin{equation} \label{VBM}
F^B _{\mathbb{M } } := \mbox{Span} \{ X^B _m : m \in \Sigma
(\mathbb{M } ) \} ,
\end{equation}
then we can identify the range of $\Pi _{\rm eff }^{B, \mathbb{M }
} $ with the space $L^2 (\mathbb{R }^N , F^B _{\mathbb{M } } ) $
of $ F^B _{\mathbb{M } } $-valued $L^2 $-functions, as we will do
without further comment.
\medskip

To lighten the notations, we will often suppress one or both
upper-indices $B $ or $\mathbb{M } $,  unless where this
would cause confusion.  This will always be clearly
indicated, usually at the beginning of a section.

\section{{\bf Estimates for Feschbach decompositions } }

We fix a non-negative integer $\mathbb{M } \geq 0 $. In this
section we will drop all upper-indices $B $, $\mathbb{M }$, and
simply write $H $ for $H^{B , \mathbb{M } } $, $H_0 $ for $H_0^{B , \mathbb{M } } $ and $\Pi _{\rm eff }
$ respectively $\Pi _{\perp } $ for $\Pi _{\rm eff} ^{B, \mathbb{M
} } $ and $\Pi _{\perp } ^{B, \mathbb{M } } $. We write our atomic
Hamiltonian $H^{B , \mathbb{M } }$  as
\begin{equation} \nonumber
H = H_0 + \mathcal{V } ,
\end{equation}
where
\begin{equation} \label{Coulomb}
\mathcal{V } := - \sum _j \frac{Z }{|r_j | } + \sum _{j < k }
\frac{1 }{|r_j - r_k | } ,
\end{equation}
the electrostatic potential, and introduce the operators
$$
\mathcal{V }_{\rm eff } := \Pi _{\rm eff } \mathcal{V } \Pi _{\rm
eff }, \ \mathcal{V }_{\perp } := \Pi _{\perp } \mathcal{V } \Pi
_{\perp } , \ \mathcal{V }_{\perp  , {\rm eff } } := \Pi _{\perp }
\mathcal{V } \Pi _{\rm eff } ,
$$
and its adjoint, $\mathcal{V }_{{\rm eff } , \perp } = \Pi _{\rm
eff } \mathcal{V } \Pi _{\perp } $. These are to be considered as
operators on $\mbox{Ran } \Pi _{\rm eff } $, $\mbox{Ran } \Pi
_{\perp } $, and between these two Hilbert spaces, respectively.
We furthermore put
$$
 h_{\rm eff} := \Pi _{\rm eff } H \Pi _{\rm eff }  \ ,
H_{\perp } := \Pi _{\perp } H \Pi _{\perp } , \ H_{\perp  , {\rm
eff } } := \Pi _{\perp } H \Pi _{\rm eff } = H_{ {\rm eff } ,
\perp  } ^* ,
$$
and for $\xi \in \mathbb{C } $ introduce the resolvents
(wherever defined)
$$
R := R(\xi ) := \left( H_{\perp } - \xi \right)^{-1 } , \ R_{\rm
eff } ^{\mathcal{W } } := R_{\rm eff } ^{\mathcal{W } } (\xi ) :=
\left( h_{\rm eff} + \mathcal{W } (\xi) - \xi \right)^{-1 } ,
$$
where
$$
\mathcal{W } :=\mathcal{W } (\xi) = - \mathcal{V } _{{\rm eff },
\perp } R (\xi) \mathcal{V }_{\perp , {\rm eff } } .
$$
Strictly speaking $R_{\rm eff}^{\mathcal{W }}$ is not a resolvent
since  the potential $\mathcal{W } (\xi ) $ depends on the
spectral parameter $\xi $. The  operators  $R $
and $R_{\rm eff } ^{\mathcal{W } } $  act  on,
respectively, the ranges of $\Pi _{\perp } $ and of $\Pi _{\rm eff
} $. Finally, let
$$
T := H_0 P^{\mathbb{M } } , \ T_{\rm eff } := \Pi _{\rm eff } T
\Pi _{\rm eff } , \ T_{\perp } := \Pi _{\perp } T \Pi _{\perp } ;
$$
$T $ commutes with $\Pi _{\rm eff } $ and $\Pi _{\perp } $,
and $\Pi _{\perp } T \Pi _{\rm eff } = 0$. Note that
$$
T_{\perp } \geq  B \; \Pi _{\perp } ,
$$
on the range of $P^{\mathbb{M } } . $
\medskip

Using matrix notation associated to the decomposition
$P^{\mathbb{M } } H = \mbox{Ran } \Pi _{\rm eff } \oplus \mbox{Ran
} \Pi _{\perp } $, we decompose $H $ as
\begin{eqnarray}
H &=&
\left(
\begin{array}{cc}
 h_{\rm eff} & H_{ {\rm eff } , \perp  } \\
H_{\perp  , {\rm eff } } & H_{\perp }
\end{array}
\right)
= \left(
\begin{array}{cc}
T_{\rm eff }  + \mathcal{V } _{\rm eff } & \mathcal{V } _{ {\rm eff } , \perp  } \\
\mathcal{V }_ {\perp , {\rm eff } } & T_{\perp } ^B + \mathcal{V }
_{\perp } )
\end{array}
\right) .
\end{eqnarray}
By the classical Feschbach formula, we then have
\begin{equation} \label{Feschbach}
(H - \xi )^{-1 } = \left(
\begin{array}{cc}
R_{\rm eff } ^{\mathcal{W } } & - R_{\rm eff } ^{\mathcal{W } } \mathcal{V }_{ {\rm eff } , \perp  } R \\
- R \mathcal{V }_{\perp  , {\rm eff } }R_{\rm eff } ^{\mathcal{W }
} & R + R \mathcal{V } _{\perp  , {\rm eff } } R_{\rm eff }
^{\mathcal{W } } \mathcal{V } _{ {\rm eff } , \perp  } R
\end{array}
\right) ,
\end{equation}
for  those  $\xi \in \mathbb{C } $ for which the right
hand side makes sense. The following theorem is the main result of
this section: recall that $\rho (A ) $ denotes the resolvent set
of an operator $A $, and $\sigma (A ) $ its spectrum.

\begin{theorem} \label{ME} Let
\begin{equation} \label{constant-B}
B_{(\ref{constant-B}) } := 16 Z^2 N (\mathbb{M} + N + 2 ) ,
\end{equation}
and
\begin{equation} \label{constant-C}
C_{(\ref{constant-C}) }  :=  c_0  + \frac{c_0 ^2
}{\sqrt{B_{(\ref{constant-B}) } } } ,\qquad c_0 ^2 = (32 Z^2
N + 8 N (N - 1 )^2 )(\mathbb{M }+ N + 2 ).
\end{equation}
If $\xi \leq 0 $ and if the field strength $B \geq B_{(\ref{constant-B}) } $, then $\xi
\in \rho (H_{\perp } ) $. If, moreover, $\xi \notin \sigma (H_{\rm
eff} + \mathcal{W } ) $, then $\xi \in \rho (H ) $, and
\begin{equation} \label{MainEstimate}
\| \ (H - \xi )^{-1 } - (  h  _{\rm eff} + \mathcal{W }
- \xi )^{-1 } \oplus  R(\xi)  \| \leq \frac{
C_{(\ref{constant-C}) } }{d_{\rm eff }^\mathcal{W } (\xi ) \sqrt{B
} }
 .
\end{equation}
where $d_{\rm eff }^\mathcal{W } (\xi ) $ is the distance of $\xi
$ to $\sigma ( h_{\rm eff} + \mathcal{W } ) $.

\end{theorem}

\noindent {\it Proof of Theorem \ref{ME}.} The proof consists of
estimating the relevant matrix elements in the Feschbach formula.
This will be done in several steps. Let
$$
R_0 = (T_{\perp } - \xi )^{-1 } ,
$$
the resolvent of $T_\perp$ on $\mbox{Ran } \Pi _{\perp } . $
\medskip

\noindent {\bf Bound on $R_0 $.} Since $T_{\perp } \geq B $ on
$\mbox{Ran } \Pi _{\perp } $ and since $\xi \leq 0 $, it
immediately follows that $\| R_0 \| \leq B^{-1 } . $
Write $\mathcal{V } $ as
\begin{equation}
\mathcal{V } = \mathcal{V }_n + \mathcal{V }_e , \quad{\rm
where}\quad \mathcal{V }_n  := - \sum _j \frac{Z }{|r_j | } ,
\quad{\rm and}\quad \mathcal{V }_e  := \sum _{j < k } \frac{1
}{|r_j - r_k | } ,
\end{equation}
the sum of electron-nucleus and the electron-electron interactions.
\medskip

\noindent {\bf A remark on notation:} we will often leave the
projection $\Pi _{\perp } $ understood when multiplying operators
on the left and/or right by $R_0 $ or $R $, and for example simply
write $R_0 ^{1/2 } \mathcal{V }_n R_0 ^{1/2 } $  instead of the
more  explicit  $R_0 ^{1/2 } \Pi _{\perp } \mathcal{V }_n \Pi
_{\perp } R_0 ^{1/2 } . $
\medskip

\noindent {\bf Bound on $R_0 ^{1/2 } \mathcal{V }_n R_0 ^{1/2 }
$.} First, since $R_0 \leq B^{-1 } $ on the range of $\Pi _{\perp
} $,
$$
0 \leq (\sqrt{R_0 } \mathcal{V }_n \sqrt{R_0 } ) ^2 \leq
B^{-1 }\sqrt{R_0 } \mathcal{V }_n ^2 \sqrt{R_0 } .
$$
By Cauchy-Schwarz,
$$
\sqrt{R_0}\mathcal{V }_n ^2 \sqrt{R_0}\le
Z^2 N \sqrt{R_0} \sum_j|r_j|^{-2} \sqrt{R_0}.
$$
Next, using Hardy's inequality on $\mathbb{R } ^3 $: $|r_j |^{-2 }
\leq - 4 \Delta _j $, and the fact that
$$
H_0 ^B = \sum _j - \frac{1 }{2 } \Delta _j + \frac{1 }{8 }
|\mathbb{B } \wedge r_j |^2  - \frac{B }{2 } (\mathbb{L }_z
+ N ),
$$
we find that
\begin{eqnarray*}
\sqrt{R_0}\left(\sum_j|r_j|^{-2}\right) \sqrt{R_0}
&\le & 8\sqrt{R_0}\sum_j
\left(-{\Delta_j\over2}+{|\mathbb{B } \wedge r_j |^2\over8}\right)\sqrt{R_0} \\
&= & 8\sqrt{R_0}\left(T_{\perp }+{B\over2}(\mathbb{M } +N)\right)\sqrt{R_0} \\
&= & 8 \Pi _{\perp } + 8 \xi R _0 + 4 (\mathbb{M } + N ) B R _0 \\
&\le & \left( 8 + 4 (\mathbb{M } + N ) \right) \Pi _{\perp } ,
\end{eqnarray*}
since $\xi \leq 0 $. It follows from these estimates that
$$
\| R_0 ^{1/2 } \mathcal{V }_n R_0 ^{1/2 } \| \leq 2 Z B^{-1/2 }
\sqrt{N (\mathbb{M } + N + 2 ) } .
$$
We note, as a consequence, that if $b_0 = 4 Z^2 N (\mathbb{M } + N
+ 2 ) $, then $\| R_0 ^{1/2 } \mathcal{V }_n R_0 ^{1/2 } \| \leq
(b_0 B^{-1 } )^{1/2 } < 1 $ if $B > b_0 $. For later reference we
also note the:
\medskip

\noindent {\bf Bound on $R_0 ^{1/2 } \mathcal{V }_n ^2 R_0 ^{1/2 }
$:} the estimates above immediately imply that this positive
operator is bounded from above by $ 4 Z^2 N (\mathbb{M } + N + 2 )
. $
\medskip

\noindent {\bf Existence of and bound on $R $.} Since the
electron-electron repulsion $\mathcal{V }_e \geq 0 $, it follows
that $R \leq R_{\rm{NI} } $, where $R_{\rm{NI} } = (T_{\perp } +
\mathcal{V }_{n, \perp } - \xi )^{-1 } $, the resolvent of an atom
with non-interacting electrons. Using the symmetrized resolvent
formula,
\begin{equation}\label{formulePourRNI}
R_{\rm NI }=\sqrt{R_0} \left( 1 + \sqrt{R_0}
\mathcal{V }_n \sqrt{R_0}\right)^{-1} \sqrt{R_0} .
\end{equation}
we see that $ R_{\rm NI} $ exists and is positive if $B > b_0 $
and $\xi \le 0 $. Hence $ T_{\perp } + \mathcal{V }_{n, \perp } -
\xi \geq 0 $ and therefore also $H_{\perp } - \xi $ and $R $.
Moreover, if $B > 4b_0 = 16 Z^2 N (\mathbb{M } + N + 2 )
=B_{(\ref{constant-B}) }$, then every $\xi\le 0 $ belongs to
$\rho(H_\perp)$ and
\begin{eqnarray*}
0 \leq R &\leq & \frac{\|R_0 \| }{1 - \| R_0 ^{1/2 } \mathcal{V
}_n R_0 ^{1/2 } \| }\leq  B^{-1 } \left( \frac{1 }{1 - (b_0 B^{-1
} )^{1/2 } } \right) \leq  2 B^{-1 }.
\end{eqnarray*}
\medskip

\noindent {\bf Bound on $R_0 ^{1/2 } \mathcal{V }_e ^2 R_0 ^{1/2 }
$:} The following elementary operator inequality is very useful to
estimate the electron-electron interactions.

\begin{lemma} \label{FE-special}
\begin{equation} \label{Hardy'}
\frac{1 }{|r_j - r_k |^2 } \leq 2 \left( -\Delta _j - \Delta _k \right).
\end{equation}
\end{lemma}

\noindent {\it Proof.} The unitary transformation induced by the
following orthogonal transformation of $\mathbb{R }^3 \times
\mathbb{R }^3 $,
\begin{equation} \label{RdD}
s := \frac{r_1 - r_2 }{\sqrt{2 } } ,\quad t := \frac{r_1 + r_2
}{\sqrt{2 } } .
\end{equation}
commutes with the Laplacian, and transforms $|r_j - r_k |^{-2 } $
into $2^{-1 }  |s |^{-2 } $. By Hardy's inequality,
$$
\frac{1 }{2 |s |^2 } \leq - 2 \Delta _s \leq -2 (\Delta _s + \Delta _t ) ,
$$
and transforming back to the $(r_j , r_k ) $-coordinates yields (\ref{Hardy'}). \hfill QED
\medskip

\noindent We can then estimate:
\begin{eqnarray*}
\mathcal{V }_e^2 &=& \left(\sum_{i<j}{1\over|r_i-r_j|}\right)^2 \leq  {N(N-1)\over2}\,\sum_{i<j}{1\over|r_i-r_j|^2}\\
&\leq & N(N-1)\sum_{i<j}(-\Delta_i-\Delta_j)
= N(N-1)^2\sum_i(-\Delta_i)\\
&\leq & 2N(N-1)^2\sum_i(-{\Delta_i\over2}+{{|\mathbb{B } \wedge
r_j |^2 }\over8}) ,
\end{eqnarray*}
and therefore, by similar arguments as before,
\begin{eqnarray*}
\sqrt{R_0}\mathcal{V }_e^2\sqrt{R_0}&\le&
2N(N-1)^2\sqrt{R_0}(T_{\perp }+{B\over2}(\mathbb{M } + N))\sqrt{R_0}\\
%&=&2N(N-1)^2{T_{\perp }+{B\over2}(\mathbb{M } + N)\over T_{\perp }-\xi}\\
&\leq & N (N - 1 )^2 (\mathbb{M } + N + 2 ) ,
\end{eqnarray*}
on $\mbox{Ran } (\Pi _{\perp } ) . $
\medskip

\noindent {\bf Bound on $\mathcal{V } R_0 ^{1/2 } $:} By the
general identity $\| A A^* \| = \| A \|^2 $, we have:
\begin{eqnarray*}
\| \mathcal{V } R_0 ^{1/2 } \|^2 &=& \| R_0 ^{1/2 } \mathcal{V }^2 R_0 ^{1/2 } \| \leq  2 (\| R_0 ^{1/2 } \mathcal{V }_n ^2 R_0 ^{1/2 } \| +
\| R_0 ^{1/2 } \mathcal{V }_e ^2 R_0 ^{1/2 } \| )
\\&\leq&  (8 Z^2 N + 2 N (N - 1 )^2 ) (\mathbb{M } + N + 2 ) .
\end{eqnarray*}
\medskip

\noindent {\bf Bound on $\mathcal{V }_{ {\rm eff } , \perp  }
R^{1/2 } $:} (Remember that we have shown that $R \geq 0 $, so its
square root is well-defined.) We first estimate $\| \mathcal{V } R
\mathcal{V } \| $, as follows. Recalling the non-interacting
resolvent $R_{\rm{NI} } $ introduced above, we have
that:
$$
0 \leq \mathcal{V } R \mathcal{V } \leq  \mathcal{V } R_{\rm{NI} }
\mathcal{V } = \mathcal{V } R_0 ^{1/2 } \left( 1 + R_0 ^{1/2 }
\mathcal{V }_n R_0 ^{1/2 } \right) ^{-1 } R_0 ^{1/2 } \mathcal{V }
.
$$
Hence its norm can be estimated by:
$$
\| \mathcal{V } R \mathcal{V } \| \leq \frac{\| \mathcal{V } R_0
^{1/2 } \|^2 }{1 - \| R_0 ^{1/2 } \mathcal{V }_n R_0 ^{1/2 } \| }
,
$$
from which we obtain an estimate for $\| \mathcal{V } R^{1/2 } \|
$ by taking square roots. Therefore, if $B \geq
B_{(\ref{constant-B}) } = 4 b_0 $ as above,
$$
\| \mathcal{V }_{ {\rm eff } , \perp  } R^{1/2 } \| \leq
\|\mathcal{V } R^{1/2 } \| \leq  \sqrt{ \left( 16 Z^2 N +  4 N (N
- 1 )^2 \right) \left( \mathbb{M } + N + 2 \right) } .
$$
\medskip

We now come to the proof of (\ref{MainEstimate}). By Feschbach's formula, we have
\begin{eqnarray} \nonumber
&\ & \| \ (H - \xi )^{-1 } - R_{\rm eff } ^{\mathcal{W } }\oplus
R(\xi) \| \nonumber =  \left\| \left(
\begin{array}{cc}
0 & - R_{\rm eff } ^{\mathcal{W } }\mathcal{V }_{ {\rm eff } , \perp  } R \\
- R \mathcal{V }_{\perp  , {\rm eff } }R_{\rm eff } ^{\mathcal{W }
}& R \mathcal{V } _{\perp  , {\rm eff } } R_{\rm eff }
^{\mathcal{W } }\mathcal{V } _{ {\rm eff } , \perp  } R
\end{array}
\right) \right\| \nonumber \\
&\leq & \| R_{\rm eff } ^{\mathcal{W } }\mathcal{V }_{ {\rm eff }
, \perp  } R \| + \| R \mathcal{V } _{\perp  , {\rm eff } } R_{\rm
eff } ^{\mathcal{W } }
\mathcal{V } _{ {\rm eff } , \perp  } R \| \nonumber \\
&\leq & \| R_{\rm eff } ^{\mathcal{W } }\| \ \|\mathcal{V }_{ {\rm
eff } , \perp  } R^{1/2 } \| \ \|R^{1/2 } \| + \|R_{\rm eff }
^{\mathcal{W } }\| \ \|\mathcal{V } _{{\rm eff },\perp } R ^{1/2 }
\|^2 \ \|R ^{1/2 } \| ^2 , \label{block-matrix-norm}
\end{eqnarray}
where we have used the following elementary estimate for the norm
of matrices of operators:
$$
\| \left(
\begin{array}{cc}
0 & A \\
A^\star & B
\end{array}
\right) \| \leq \|A \| + \|B \| .
$$
Hence if $B > B_{(\ref{constant-B}) }  $ and if $\xi \notin \sigma
(R_{\rm eff } ^{\mathcal{W } }) $, we obtain
\begin{eqnarray*}
\| \ (H - \xi )^{-1 } - R_{\rm eff }^{\mathcal{W } } \oplus
 R(\xi) \|&
\leq &
\frac{1 }{d_{\rm eff }^\mathcal{W } (\xi ) } \left( \frac{\sqrt{2 } \|
\mathcal{V } R^{1/2 } \| }{\sqrt{B } } +
\frac{2 \| \mathcal{V } R^{1/2 } \|^2   }{B } \right) \\
&\leq &  \frac{1 }{d_{\rm eff }^\mathcal{W } (\xi ) B^{1/2 }}
\left(  c_0 + \frac{ c_0 ^2  }{
\sqrt{B_{(\ref{constant-B}) }  }  } \right),
\end{eqnarray*}
with $ c_0 ^2 = (32 Z^2 N + 8 N (N - 1 )^2 )(\mathbb{M }+ N + 2 )
. $ \hfill QED

\begin{corollary} \label{ME:Fermi} ({\em of the proof}) Theorem
\ref{ME} also holds, if we replace $H $ and $h_{\rm
eff } + \mathcal{W } $ by their fermionized versions $H_{\rm f }
$, $h_{\rm eff , f } + \mathcal{W }_{\rm f } $, and
$d_{\rm eff } ^{\mathcal{W } } (\xi ) $ by the distance of $\xi $
to the spectrum of $h_{\rm eff, f } + \mathcal{W
}_{\rm f}.$
\end{corollary}

\noindent {\it Proof.} Simply write down the Feschbach's formula
(\ref{Feschbach}) for $H_{\rm f } $ with respect to the
decomposition $I = \Pi _{\rm eff , f } + \Pi _{\perp , {\rm f } }
$ of $\mathcal{H }_{\rm f } $, and estimate as in
(\ref{block-matrix-norm}), where all operators will now have a
sub-index `f'. Next use that $P^{AS } $ commutes with everything,
and trivially estimate $\| A_{\rm f } \| = \| P^{AS } A \| \leq \|
A \| $, for $A = R , R^{1/2 } $ and $\mathcal{V }_{{\rm eff } ,
\perp } R^{1\over2 } $, except for $ \| R^{\mathcal{W } } _{\rm
eff, f } \| $, which will be estimated by 1 over the distance of
$\xi$ to the spectrum of $H_{\rm eff, f } + \mathcal{W }_{\rm f} .
$ \hfill QED
\medskip

\noindent The proof shows that in the fermionic case, Theorem
\ref{ME} will at least be true with the same constants as for the
boltzonic case. The optimal constants for fermions might be
smaller, though.

\begin{remark} \rm{In the proof of theorem\ref{ME} we
systematically used Hardy's inequality. Alternatively,  one can
use, at least when $N=1$, the bounds on the matrix elements of the
Coulomb potential with respect to the Landau levels which were
derived in \cite{FW}. }
\end{remark}

\section{{\bf Effective potentials for large fields} }

The operator $h_{\rm eff} + \mathcal{W } = h_{\rm eff} ^{B, \mathbb{M } } +
\mathcal{W } ^{B, \mathbb{M } } $ of Theorem \ref{ME} acts on $
\mbox{Ran} (\Pi _{\rm eff } ^{B, \mathbb{M } } ) = \Pi _{\rm eff } ^{B ,
\mathbb{M } } (\mathcal{H } ) $, a Hilbert space which depends on both $B
$ and $\mathbb{M } $, and which is canonically isomorphic to the
space of $F_{\mathbb{M } }^B $-valued $L^2 $-functions on
$\mathbb{R } ^N $,
\begin{equation}
\mbox{Ran } (\Pi _{\rm eff } ^{B, \mathbb{M } }) = L^2 ( \mathbb{R
} ^N , F_{\mathbb{M } }^B ) .
\end{equation}
(Recall that $F_{\mathbb{M } }^B = \mbox{Span } \{ X_m ^B : m \in
\Sigma (\mathbb{M } ) \} $.) We will mostly suppress the
$\mathbb{M } $-dependence from our notations, $\mathbb{M } $ being
fixed in our analysis. The potential term of $h_{\rm eff}
^B $,
\begin{equation}
\mathcal{V }_{\rm eff } ^B = \Pi _{\rm eff } ^B \mathcal{V } \Pi
_{\rm eff } ^B ,
\end{equation}
can be interpreted as an operator valued function of $z = (z_1 ,
\cdots , z_N ) \in \mathbb{R }^N $, with values in the space of
linear operators on $F_{\mathbb{M } }^B $ and acting in the
natural way on $L^2 ( \mathbb{R }_z ^N ; F_{\mathbb{M } }^B ) . $
To get rid of the $B $-dependence of our Hilbert spaces we do a
unitary re-scaling. Let us pose $x = (x_1 , \cdots , x_N ) \in
\mathbb{R }^N $ and similarly for $y $ and $z $. Define a unitary
operator $U_{xy} ^B $ on $\mathcal{H } $ by:
\begin{equation} \label{def-U}
U_{xy} ^B \psi (x, y, z ) = B^{N/2 } \psi (\sqrt{B } x, \sqrt{B }
y , z ) .
\end{equation}
Since $X^B _m (x, y ) = B^{N / 2 } X_m ^1 (\sqrt{B } x , \sqrt{B }
y ) $, it follows that
$$
U_{xy} ^{B * } \Pi _{\rm eff } ^B U_{xy} ^B = \Pi _{\rm eff } ^1
.
$$
Let us write $\mathcal{V } _{\rm eff } ^B $ in multi-particle
form:
\begin{equation}
\mathcal{V }_{\rm eff } ^B = - \sum _j Z V_j ^B
 + \sum _{j < k } V_{jk } ^B ,
\end{equation}
with $V_j^B$ and $V_{jk}^B$ defined by (\ref{defVeffietjk}). Then
 $U_{xy} ^{B * } V_j ^B U_{xy} ^B = \Pi _{\rm eff } ^1
U_{xy} ^{B * } | r_j |^{-1 } U_{xy} ^B \Pi _{\rm eff } ^1 =
\sqrt{B } V_j ^1 (\sqrt{B } z_j ) $, with
\begin{equation} \label{def:V1j}
V_j ^1 (z) =   \Pi _{\rm eff } ^1 \frac{1 }{\sqrt{x_j ^2 + y_j ^2
+ z ^2 } } \Pi _{\rm eff } ^1 ,
\end{equation}
and likewise for $V_{jk } ^B $: $U_{xy} ^{B * } V_{jk } ^B U_{xy}
^B (z) = \sqrt{B } V_{jk } ^1 (\sqrt{B } (z_j - z_k ) ) $, with
\begin{equation} \label{def:V1jk}
V_{jk }^1 (z) = \Pi _{\rm eff } ^1 \frac{1 }{\sqrt{(x_j - x_k )^2
+ (y_j - y_k )^2 + z^2 } } \Pi _{\rm eff } ^1 .
\end{equation}
The operator
\begin{equation} \label{Hs-hat}
\widehat{ h  }_{\rm eff } ^B :=  U_{xy} ^{B * } H^B
_{\rm eff } U_{xy} ^B ,
\end{equation}
will now act on the fixed, $B $-independent, Hilbert space, $L^2
(\mathbb{R } ^N , F_{\mathbb{M } } ^1 ) $, and
\begin{eqnarray} \nonumber
\widehat{h }_{\rm eff } ^B &=& - \frac{1 }{2 } \Delta
_z - \sum _j Z \sqrt{B } V_j ^1 (\sqrt{B } z_j ) + \sum _{j < k }
\sqrt{B } V_{jk
} ^1 (\sqrt{B } (z_j - z_k ) ) \nonumber \\
&=& - \frac{1 }{2 } \Delta _z + \sqrt{B } \mathcal{V }_{\rm eff }
^1 (\sqrt{B } z ) . \nonumber
\end{eqnarray}
\medskip

The next step will be to examine the asymptotic behavior of
$\sqrt{B } \mathcal{V }_{\rm eff } ^1 (\sqrt{B } z ) $ as $B \to
\infty $. The main idea is contained in lemma \ref{Asymp-Lemma}
below. We introduce the free Laplacian on $\mathbb{R }^N $,
\begin{equation}
h_{00 } = - \frac{1 }{2 } \Delta _z ,
\end{equation}
and its resolvent:
\begin{equation}
R_{00 } (-\alpha ^2 ) = (h_{00 } + \alpha ^2 )^{-1 } .
\end{equation}
We will need this resolvent both in dimension $N $ and dimension
1. To distinguish between these two cases we will, in the
1-dimensional case, systematically use $\beta ^2 $ as spectral
parameter instead of $\alpha ^2 $, reserving the latter for the
multidimensional case.

If $u $ is a function or tempered distribution on $\mathbb{R }^N
$, with values in some auxiliary Hilbert space $F $, then
$\displaystyle{\| R_{00 } ( - \alpha ^2 ) ^{s/2 } u \|_{L^2
(\mathbb{R }^N ; F ) } }  $ is a norm on the $s $-th Sobolev space
$H^s (\mathbb{R }^N ; F ) $. A linear operator $A $ sends $H^s
(\mathbb{R }^N ; F ) $ continuously into $H^{-s }(\mathbb{R }^N ;
F ) $ iff the $L^2 $-operator norm $\| R_{00 } (- \alpha ^2 )
^{s/2 } A R_{00 } (- \alpha ^2 ) ^{s/2 } \| $ is finite. The case
of interest for us will be $s = 1 $. We will also need the Fourier
transform $\mathcal{F } $, but only in dimension 1, for which we
normalize as follows:
$$
\mathcal{F } (u) (\zeta )  = \int _{\mathbb{R }
} u(z ) e^{-iz \zeta } dz .
$$
There will consequently be a factor of $(2 \pi )^{-1 } $ in
the inversion formula.
\medskip

Recall that
$$
\mbox{Pf} \left( \frac{1 }{|x| } \right) = \frac{{\rm d } }{{\rm d
} x } \left( \mbox{sgn} (x) \log |x | \right) ,
$$
with the derivative in distribution sense. Let $F $ be a finite
dimensional complex Hilbert space, and $L(F ) $ the space of
linear operators on $F$.

\begin{lemma} \label{Asymp-Lemma} Let $\mathfrak{v} $ be an $L(F) $-valued
tempered distribution on $\mathbb{R } $, such that its Fourier
transform can be identified with a locally integrable function
$\FF \mathfrak{v}  =  \FF \mathfrak{v} (\zeta ) $. Assume also:
\medskip

\noindent (i) There exist $C_0 , C_1 \in L(F) $ and $a > 1/2 $,
such that:
\begin{equation} \label{AsympF-Hat}
\FF \mathfrak{v} (\zeta ) = - C_0 \log |\zeta | + C_1 + O(|\zeta
|^a ) , \ \ \zeta \to 0 .
\end{equation}
\medskip

\noindent (ii) If
$$
\mathfrak{e}(\zeta ) := \FF \mathfrak{v}  (\zeta ) + C_0 \log
|\zeta | - C_1 ,
$$
denotes the error in the approximation (\ref{AsympF-Hat}), then
\begin{equation} \label{ErrorAsymp}
C_{\mathfrak{v} } ^2 := \int _{\mathbb{R } } \frac{\| \mathfrak{e}
(\zeta ) \|^2 }{|\zeta |^2 } d\zeta < \infty .
\end{equation}

\noindent For each $\lambda>0$ let
\begin{equation} \label{AsympMC}
\mathfrak{v} _{\infty , \lambda } := C_0 \log \lambda \cdot \delta
+ \frac{1 }{2 } C_0 \cdot \mbox{{\rm Pf} } \left( \frac{1 }{|x | }
\right) + (\gamma C_0 + C_1 ) \cdot \delta ,
\end{equation}
where $\delta $ is Dirac's delta-distribution in 0, and $\gamma =
\Gamma '(1 ) $ is the Euler constant. If $R_{00 } (-\beta ^2 ) $
denotes the free resolvent in dimension 1 and $\beta
> 0 $, then
\begin{equation} \label{Asymp-Est}
\| R_{00 } ( - \beta ^2 ) ^{1/2 } \left( \lambda \mathfrak{v}
(\lambda \cdot ) - \mathfrak{v} _{\infty , \lambda } \right) R_{00
} ( - \beta ^2 ) ^{1/2 } \| \leq \frac{2^{1/4 } C_{\mathfrak{v} }
}{ \sqrt{\beta \lambda \pi } } .
\end{equation}
\end{lemma}

\noindent {\bf Remark.} Observe that the integral
(\ref{ErrorAsymp}) converges in $0 $, since we assumed that $a > 1
/ 2 $ in (\ref{AsympF-Hat}).
\medskip

\noindent {\it Proof.} It is known that
\begin{equation} \label{bd1}
\mathcal{F }^{-1 } (\log |\zeta | ) = -\frac{1 }{2 } {\rm Pf}
\left( \frac{1 }{|x| } \right) - \gamma \delta _0 ,
\end{equation}
where $\gamma $ is Euler's constant: cf. e.g.
\cite{Schw}\footnote{(\ref{bd1}) can also easily be shown
directly, using the observation that $\mbox{Pf } (1 / |x| ) $ and
$ - 2 \mathcal{F }^{-1 } (\log|\zeta|) $ are both solutions of the
distributional equation $x \Lambda = - \mbox{sgn } x $ and
therefore only differ by a multiple of $\delta $, which can then
be computed to be $- 2 \gamma . $}. Therefore
\begin{eqnarray} \nonumber
\mathfrak{v} _{\infty , \lambda } &=& C_0 \left( \log \lambda
\right) \; \delta - C_0 \mathcal{F
} ^{-1 } (\log |\zeta | ) + C_1 \delta \\
&=& \mathcal{F }^{-1 } \left( -C_0 \log (|\zeta | / \lambda ) +
C_1 \right) , \nonumber
\end{eqnarray}
and
$$
\mathcal{F } \left( R_{00 } ( - \beta ^2 ) ^{1/2 } \left( \lambda
\mathfrak{v} (\lambda \cdot ) - \mathfrak{v} _{\infty , \lambda }
\right) R_{00 } ( - \beta ^2 ) ^{1/2 } \right) \mathcal{F }^{-1 }
$$
is an integral operator with kernel:
\begin{equation} \label{AL1}
\frac{1 }{2\pi } \frac{1 }{(\zeta ^2 / 2 + \beta ^2 )^{1/2 } } \;
\mathfrak{e} \left( \frac{\zeta - \zeta ' }{\lambda } \right)
\frac{1 }{({\zeta '}^2 /2 + \beta ^2 )^{1/2 } } ,
\end{equation}
since multiplication by a distribution $a(x ) $ becomes an
integral operator with kernel $(2 \pi )^{-1 } \FF a  (\zeta -
\zeta ' ) $ after conjugation by $\mathcal{F } $. Since
conjugation by the Fourier transform does not change the operator
norm, it follows that the norm in (\ref{Asymp-Est}) can be bounded
by the Hilbert-Schmidt norm of (\ref{AL1}), whose square equals:
\begin{eqnarray*}
&\ & \frac{1 }{\pi ^2 } \int _{\mathbb{R } } \left( \int
_{\mathbb{R } } \frac{1 }{( (\zeta - \eta )^2 + 2 \beta ^2 )
(\zeta ^2 + 2 \beta ^2 ) } d\zeta \right)
| \mathfrak{e} \left( \frac{\eta }{\lambda } \right) |^2 d\eta \\
&= & \frac{\sqrt{2 } }{\pi \beta } \int _{\mathbb{R } }
\frac{ | \mathfrak{e} (\eta / \lambda ) |^2 }{\eta ^2 + 8\beta ^2 } d\eta
= \frac{\sqrt{2 } }{\pi \beta \lambda }
\int _{\mathbb{R } } \frac{|\mathfrak{e} (\eta ) |^2 }{\eta ^2 + (\beta ^2 / 8\lambda ^2 ) } d\eta \nonumber
\leq  \frac{C_{\mathfrak{v} } ^2 \sqrt{2 } }{\pi \beta \lambda }.
\end{eqnarray*}
Here we have used the elementary
integral identity:
\begin{equation} \label{AL2}
\int _{\mathbb{R } } \frac{1 }{( a \zeta ^2 + b ) (a (\zeta - \eta
)^2 + b ) } d \zeta = \frac{2 \pi }{\sqrt{a b } } \frac{1 }{a \eta
^2 + 4 b } \; ,
\end{equation}
where $a , b > 0 $. This finishes the proof of lemma
\ref{Asymp-Lemma}. \hfill QED.
\medskip

We will apply the previous lemma to our potentials
(\ref{defVeffietjk}), but before doing so we first state and prove
a weaker variant, which will be used to prove Theorem \ref{FMT}.
Let us introduce the (numerical) constant:
\begin{equation} \label{AsympPot1a:c}
C_{(\ref{AsympPot1a:c}) } := \left( \frac{1 }{4 \pi } \int
_{\mathbb{R } } \frac{(|\log |\eta | \; | + 2 )^2 }{\eta ^2 + 4 }
\; d\eta \right) ^{1/2 } .
\end{equation}
Numerical evaluation of the integral (using either Mathematica or
Maple 8) gives $C_{(\ref{AsympPot1a:c}) }^2 \simeq 1.53 . $

\begin{lemma} \label{AsympPot1a} Let $\mathfrak{v} = \mathfrak{v}(z) $ be
an $L(F)$-valued tempered distribution on $\mathbb{R } $
such that
\begin{equation} \label{HypAsympPot1a}
C_{\mathfrak{v} } := \sup _{\zeta \in \mathbb{R } }  \| \; \left(
|\log |\zeta | \; | + 1 \right) ^{-1 } \; \FF {\mathfrak{v} }
(\zeta ) \; \| < \infty .
\end{equation}
Then for all $\lambda \geq e $ and all $\varepsilon > 0 $,
\begin{equation} \label{Asymp1a-Est}
\| \ (-\varepsilon \Delta +  \varepsilon ^{-1 } )^{-1/2 }
\left( \frac{\lambda }{\log \lambda } \mathfrak{v} (\lambda z )
\right) (-\varepsilon \Delta +  \varepsilon ^{-1 } )^{-1/2 } \ \|\leq
C_{(\ref{AsympPot1a:c}) } C_{\mathfrak{v } }
\left( |\log \varepsilon | + 2 \right) .
\end{equation}
\end{lemma}

\noindent {\it Proof.} Conjugating as before by the Fourier
transform, and estimating the operator norm by the Hilbert-Schmidt
one, we find that the left hand side of (\ref{Asymp1a-Est}) is
bounded, by the square root of
$$
(2\pi )^{-2 } \int _{\mathbb{R } } \left( \int _{\mathbb{R } }
\frac{1}{(\varepsilon \zeta ^2 + \varepsilon ^{-1 }) (\varepsilon
(\zeta - \eta )^2 + \varepsilon ^{-1 })} d\zeta \right) \frac{1
}{(\log \lambda )^2 } \; \| \; {\mathfrak{v} } (\frac{\eta
}{\lambda } ) \; \|^2 \ d\eta .
$$
By (\ref{HypAsympPot1a}) we can bound
\begin{eqnarray*}
\frac{1 }{(\log \lambda )^2 } \; \| \; {\mathfrak{v } } (\eta
/ \lambda ) \; \| ^2 &\leq &   C_{\mathfrak{v}}^2 \left(
\frac{|\; \log |\eta | \; | }{\log \lambda } + 1 +
\frac{1 }{\log \lambda } \right)^2 \\
&\leq & C_{\mathfrak{v }}^2 \left( | \log |\eta | \; | + 2
\right)^2 ,
\end{eqnarray*}
since we suppose that $\lambda \geq e $. Hence, using (\ref{AL2})
again, we find that our norm is bounded by the square root of
\begin{eqnarray*}
&\ & \frac{ C_{\mathfrak{v } } ^2 }{2\pi } \int _{\mathbb{R } }
\frac{ \left( \; | \; \log |\eta | \; | + 2 \; \right)^2 }
{\varepsilon \eta ^2 + 4 \varepsilon ^{-1 } } \ d\eta  \leq
 \frac{ C_{\mathfrak{v } } ^2 }{2\pi } \int _{\mathbb{R } }
\frac{\left( | \; \log |\eta | \; | + | \log \varepsilon | + 2 \right)^2 }{\eta ^2 + 4 }
d\eta \\ &\leq & \frac{ C_{\mathfrak{v} } ^2 }{\pi } (\log
\varepsilon )^2
\int _{\mathbb{R } } \frac{d\eta }{\eta ^2 + 4 } \ + \ \frac{
C_{\mathfrak{v } } ^2 }{ \pi } \int _{\mathbb{R } } \frac{ (|\log
|\eta | \; | + 2 )^2 }{\eta ^2 + 4 } d\eta  =
C_{\mathfrak{v } } ^2 \left( \frac{(\log \varepsilon )^2 }{2 } + 4
C_{(\ref{AsympPot1a:c}) } ^2 \right) ,
\end{eqnarray*}
by (\ref{AsympPot1a:c}). Since $C_{(\ref{AsympPot1a:c}) } ^2 \geq
1/2 $, we see that (\ref{Asymp1a-Est}) will be bounded by
$C_{(\ref{AsympPot1a:c}) } C_{\mathfrak{v } } \left( |\log
\varepsilon | + 2 \right) $, as claimed. \hfill QED
\medskip

The next step will be to apply lemma \ref{Asymp-Lemma} to the
potentials $V_j ^1 $ and $V_{jk } ^1 $, with $\lambda = \sqrt{B }
$. We introduce the $B $-dependent tempered distribution $q = q^B
$, and linear operators $C_j ^n , C_{jk } ^e \in L(F^1_{\mathbb{M
} } ) $ by:
\begin{equation} \label{qB }
q^B (z) = \log B \ \delta (z) + \mbox{Pf } \left( \frac{1 }{|z| }
\right) ,
\end{equation}
\begin{equation} \label{Cn}
C_j ^{n } :=C_j ^{n ,\mathbb{M}}:= - \Pi _{\rm eff } ^1 \log \left( \tfrac{1 }{4 } \left(
x_j ^2 + y_j ^2 \right) \right) \Pi _{\rm eff } ^1 ,
\end{equation}
\begin{equation} \label{Ce}
C_{jk } ^{e } := C_{jk } ^{e,\mathbb{M} }:=- \Pi _{\rm eff } ^1 \log \left( \tfrac{1 }{4 }
\big{(} (x_j - x_k )^2 + (y_j - y_k )^2 \big{)} \right) \Pi _{\rm
eff } ^1;
\end{equation}
Observe that (\ref{Cn}) and (\ref{Ce}) are related to (\ref{CjB})
and (\ref{CjkB}) by conjugation by $U_{xy}^B$.  See also remark
\ref{interpr:v_C} for a physical interpretation of these three
terms.

\begin{lemma} \label{AsympPot1} Let
$R_{00 } (-\beta ^2 ) = (- \tfrac{1 }{2 } \Delta _z + \beta ^2
)^{-1 } $, $\beta
> 0 $ be the free resolvent in dimension 1. There exists a
positive constant $C_{(\ref{asympVj}) } := C_{(\ref{asympVj}) }
(\mathbb{M } ) > 0 $ only depending on $\mathbb{M } $, such that
for all $B , \beta > 0 $,
\begin{equation} \label{asympVj}
\|R_{00}(-\beta^2 )^{1/2}
\left(\sqrt{B}V_j^1(\sqrt{B}z)-\left(q^B(z)+C_j^{n} \; \delta(z)
\right) \right) R_{00 } (-\beta ^2 ) ^{1/2 } \| \leq
\frac{C_{(\ref{asympVj}) } }{\sqrt{\beta } B^{1/4 } } ,
\end{equation}
and
\begin{equation} \label{asympVjk}
\| R_{00 } (-\beta ^2 ) ^{1/2 }\left( \sqrt{B } V_{jk } ^1
(\sqrt{B } z ) - \left( q^B (z) + C_{jk } ^{e } \; \delta(z)
\right) \right) R_{00 } (-\beta ^2 ) ^{1/2 } \| \leq \frac{
C_{(\ref{asympVj}) } }{\sqrt{\beta } B^{1/4 } } ,
\end{equation}
the norm being the operator norm on $L^2 (\mathbb{R } ,
F_{\mathbb{M } } ^1 ) $.
\end{lemma}

\noindent To simplify future estimates, we have taken the same
constant in both inequalities.
\medskip

\noindent {\it Proof.}  Recall the formulas (\ref{def:V1j})
and (\ref{def:V1jk}) for $V^1 _j (z) $ and $V^1 _{jk } (z) $.
We need the asymptotics of their Fourier-transforms at $0 $. By
\cite[9.6.21]{AS}, the
Fourier transform of $(1 + z^2 )^{-1/2 } $ equals
$$
\mathcal{F } \left ( (1 + z^2 )^{-1/2 } \right) (\zeta ) = 2 K_0
(|\zeta | ) ,
$$
where $K_0 $ is the Macdonald function. Since the projector $\Pi
_{\rm eff } ^1 $ effectively only acts in the $x $ and $y
$-variables, it follows that
\begin{eqnarray} \nonumber
\FF V_j ^1 (\zeta ) &=& \Pi _{\rm eff } ^1 \mathcal{F } _{z \to
\zeta } \left( (x_j ^2 + y_j ^2 ) ^{-1/2 } \left( 1 + ( (x_j ^2 +
y_j ^2
)^{-1 / 2 } z )^2 \right)^{-1/2 } \right) \Pi _{\rm eff } ^1 \\
&=& 2 \Pi _{\rm eff } ^1 K_0 \left( \sqrt{x_j ^2 + y_j ^2 } \cdot
\zeta \right) \Pi _{\rm eff } ^1 , \nonumber
\end{eqnarray}
with a similar formula for $\FF V_{jk }^1 . $

Now it is known that
$$
K_0 (|\zeta | ) = - \log |\zeta | + \log 2 - \gamma + O(| |\zeta
^2
 \log |\zeta | \ | ) , \ \ |\zeta | \to 0 ,
$$
and that $K_0 (|\zeta | ) $ is bounded on $|\zeta | \geq 1 $ (even
exponentially decreasing there): see e.g. \cite[9.6.13]{AS}. It then easily follows
that, as $\zeta \to 0 $ and as operators on $\mbox{Ran } \Pi
_{\rm eff } ^1 $,
\begin{eqnarray} \label{AsymptoticVchap}
&\ & \FF V_j ^1  (\zeta ) \simeq - 2 \log |\zeta | - 2 \gamma +
C_j
^{n } , \\
&\ & \FF V_{jk } ^1  (\zeta ) \simeq -2 \log |\zeta | - 2 \gamma +
C_{jk }^{e } , \nonumber
\end{eqnarray}
with an error of $O(| \zeta ^2 \log |\zeta | \ | ) . $ An appeal
to lemma \ref{Asymp-Lemma}, with $\lambda = \sqrt{B } $ and with
$C_0 = 2 $ and $C_1 = - 2 \gamma + C_j ^n $ respectively $C_1 = -2
\gamma + C_{jk } ^e $, then finishes the proof. \hfill QED
\medskip

\begin{remark} \rm{An explicit computation of the matrices of $C_j ^{n
} $ and $C_{jk } ^{e } $ with respect to the natural basis $X_m ^1
, m \in \Sigma (\mathbb{M } ) $ shows that $C_j ^{n } $ and $C_{jk
} ^{e } $ do depend on their indices $j $ and $j, k $,
respectively. }
\end{remark}

We will likewise need lemma \ref{AsympPot1a}  for $\mathfrak{v} =
V _j ^1 $. We can without loss of generality assume that $j = 1 $,
by permutational symmetry of $\Sigma (\mathbb{M } ) $. As we have
seen above, $\FF V^1 _1 (\zeta ) = 2\Pi ^1 _{\rm eff } K_0
(|\zeta| \sqrt{x_1 ^2 + y_1 ^2 } ) \Pi ^1 _{\rm eff } $. It can
easily be verified that $ \| \; (|\log |\zeta | \; | + 1 )^{-1 }
K_0 (\zeta ) \; \|_{\infty } = 1 $, so that, for example,
\begin{equation} \label{C_V1:1}
C_{V^1_{1 } } \leq C_{(\ref{C_V1:1}) } :=  2 +
2 \| \; \Pi _{\rm eff } ^1 \left | \log \sqrt{ x_1 ^2 + y_1
^2 } \right | \Pi _{\rm eff } ^1 \; \| .
\end{equation}
The operator norm on the right and side can be evaluated
explicitly, and behaves asymptotically for large positive
$\mathbb{M } $ as $ 2 \log(\mathbb{M } ) . $
\medskip

We next extend lemma \ref{AsympPot1} to multi-particle potentials.
Let us define the multi-particle potential $v_C $ by
\begin{eqnarray} \label{V-eff}
v_C (z) = v_C ^B (z) &=& - Z \sum _j \left( q^B (z_j ) + C_j ^{n }
\delta (z_j ) \right) \\
&+& \sum _{j < k } \left( q^B (z_j - z_k ) + C_{jk } ^{e } \delta
(z_j - z_k ) \right) . \nonumber
\end{eqnarray}

\begin{lemma} \label{AsympPot2} Let $R_{00 } (- \alpha ^2 ) =
(- \tfrac{1 }{2 } \Delta _z + \alpha ^2 )^{-1 } $, $\alpha > 0 $,
be the resolvent of the free Hamiltonian in $\mathbb{R }^N $. Then
\begin{equation} \label{V-eff1}
\|R_{00 } ( - \alpha ^2 ) ^{1/2 } \left( \sqrt{B } \mathcal{V }
^1 _{\rm eff } (\sqrt{B } z ) - v_C (z) \right) R_{00 } ( - \alpha
^2 ) ^{1/2 }
\|\leq {C_{(\ref{ConstV-eff}) } \over \sqrt{\alpha } B^{1/4 }},
\end{equation}
where
\begin{equation} \label{ConstV-eff}
C_{(\ref{ConstV-eff}) } := C_{(\ref{ConstV-eff}) } (N, Z,
\mathbb{M } ) := C_{(\ref{asympVj}) }  \; N^{1/4 } \left( Z +
\frac{1 }{2 } (N - 1 ) \right).
\end{equation}
\end{lemma}

\noindent {\it Proof.} We split both potentials into their
`electron-nucleus' and `electron-electron' parts:
$$
\mathcal{V }^1 _{\rm eff } = \mathcal{V }^1 _{{\rm eff }, n } +
\mathcal{V }^1 _{{\rm eff } , e } ,
$$
and similarly for $v_{C } $: $v_C = v_{C, n } + v_{C, e } =
v_{C, n } ^B + v_{C, e } ^B $. Writing $\mathcal{V } _{s,
\nu } ^B (z) $ for $\sqrt{B } \mathcal{V }_{s, \nu } ^1 (\sqrt{B }
z ) $ (with a mild abuse of notation), where $\nu = n $ or $e $,
we bound the left hand side of (\ref{V-eff1}) by
\begin{equation} \label{V-eff2}
\|R_{00}(-\alpha ^2 ) ^{1/2 } \left(\mathcal{V } _{{\rm eff } , n
} ^B\!-\!v_{C , n } \right) R_{00 } (-\alpha ^2 ) ^{1/2 } \| +
\|R_{00 } (-\alpha ^2 ) ^{1/2 } \left( \mathcal{V } _{{\rm eff } ,
e } ^B\!-\! v_{C , e } \right) R_{00 } (-\alpha ^2 ) ^{1/2 } \| ,
\end{equation}
and estimate the two terms separately.
%We will repeatedly use the
%elementary observation that for a bounded self-adjoint operator, $
%\| A \| \leq C $ iff  $A \leq C \ \mbox{and} \ -A \leq C $, the
%inequalities on the right being in operator sense; this is a
%trivial consequence of the spectral Theorem.
Let $R_{00 , j } (-\beta ^2 ) $ be the 1-dimensional resolvent in
the variable $z_j $, with a $\beta $ which will be picked below.
We will simply write $R_{00 } $ for $R_{00 } ( - \alpha ^2 ) $ and
$R_{00 , j } $ for $R_{00 , j } ( - \beta ^2 ) $. If we put
$$
\Delta V _j := V_j ^B (z_j ) - q^B (z_j ) - C_j ^n \delta (z_j ) ,
$$
and
$$
\Delta \mathcal{V }_n := Z \sum _j \Delta V_j = \mathcal{V }^B
_{{\rm eff } , n }  - v_{C, n } \; ,
$$
then, by (\ref{asympVj}),
\begin{eqnarray*}
R_{00 }^{1/2} \Delta \mathcal{V }_n R_{00 }^{1/2 } %&=& \sum _j R^{1/2 }
%\Delta V_j R^{1/2} \\
&=& Z \sum _j (R_{00 }^{1/2 } R_{00 , j } ^{-1 / 2 } ) \; ( R_{00
, j } ^{1/2 } \Delta
V_j R_{00 , j } ^{1/2 } )\; (R_{00 , j} ^{-1 / 2 } R_{00 }^{1/2 } ) \\
&\leq & \frac{C_{(\ref{asympVj}) } Z }{\sqrt{\beta } B^{1/4 } }
\sum _j R_{00 }^{1/2 } R_{00 , j }
^{-1 } R_{00 }^{1/2 } = \frac{C_{(\ref{asympVj}) } Z
}{\sqrt{\beta } B^{1/4 } } R_{00 } (- \alpha )^2
(-\frac{1 }{2 } \Delta _z + N \beta ^2 ) \\
&\leq & \frac{C_{(\ref{asympVj}) } Z }{\sqrt{\beta } B^{1/4 } }
\max _{\xi \in \mathbb{R }^N } \frac{|\xi |^2 / 2 + N\beta ^2
}{|\xi |^2 / 2 +
\alpha ^2 } =  \frac{C_{(\ref{asympVj}) }  Z \max (1, N \beta ^2
/ \alpha ^2 ) }{\sqrt{\beta
} B^{1/4 } } =  \frac{C_{(\ref{asympVj}) }  Z N^{1 / 4 }
}{\sqrt{\alpha } B^{1/4 } }
\end{eqnarray*}
if we pick $\beta = \alpha / \sqrt{N } $; this choice actually
minimizes $\beta ^{-1/2 } \max (1, N \beta ^2 / \alpha ^2 ) $ as a
function of $\beta \geq 0 $, as is easily checked. Similar
estimates  show that $- R_{00 }^{1/2 } \Delta \mathcal{V }_n
R_{00 }^{1/2 } $ is bounded from above, in operator sense, by the
same number, and we therefore conclude that the first norm in
(\ref{V-eff2}) is bounded by $C_{(\ref{asympVj}) } Z N^{1/4 } /
\sqrt{\alpha } B^{1/4 } . $
\medskip

To estimate the second term of (\ref{V-eff2}), we will use the
following lemma, which is analogous to lemma \ref{FE-special} from
section 2. Let $\Delta _j = - d^2 / dz_j ^2 $, $\Delta _k = - d^2
/dz_k ^2 $.

\begin{lemma} \label{FE-special'} Let $\mathfrak{v} =
\mathfrak{v}(z) $ be an $L(F) $-valued distribution on $\mathbb{R
} $ ($F $ a finite-dimensional Hilbert space), such that $R_{00 }
(-\beta ^2 )^{1/2 } \mathfrak{v }  R_{00 } (-\beta ^2 )^{1/2 } $
is self-adjoint, for $\beta > 0 $. Then for all $\mu > 0 $,
\begin{eqnarray}
&\ & \| \ (- \frac{1 }{2 } \Delta _j - \frac{1 }{2 } \Delta _k +
\mu ^2 )^{-1/2 } \; \mathfrak{v } (z_j - z_k ) \; (-\frac{1 }{2 }
\Delta _j -
\frac{1 }{2 } \Delta _k + \mu ^2 )^{-1/2 } \ \| \\
&\leq & \frac{1 }{2 } \| \ (- \frac{1 }{2 } \Delta _s + \frac{\mu
^2 }{2 } )^{-1/2 } \; \mathfrak{v}(s) \; (- \frac{1 }{2 } \Delta
_s + \frac{\mu ^2 }{2 } )^{-1/2 } \ \| , \nonumber
\end{eqnarray}
where the norm on the left hand side is of course taken in $L^2
(\mathbb{R }^2 , F ) . $
\end{lemma}

\noindent {\it Proof.} We use a similar change of variables as in
the proof of lemma \ref{FE-special}: $s = (z_j - z_k ) / \sqrt{2 }
$, $t = (z_j + z_k ) / \sqrt{2 } $. Then, with $\simeq $ denoting
unitary equivalence,
\begin{eqnarray*}
&\ & \left(- \frac{1 }{2 } \Delta _j - \frac{1 }{2 } \Delta _k +
\mu ^2 \right)^{-1/2 } \; \mathfrak{v } (z_j - z_k ) \; \left(-
\frac{1 }{2 } \Delta
_j - \frac{1 }{2 } \Delta _k + \mu ^2 \right)^{-1/2 } \\
&\simeq & \left( - \frac{1 }{2 } \Delta _s - \frac{1 }{2 } \Delta
_t + \mu ^2 \right)^{-1/2 } \; \mathfrak{v } (\sqrt{2 } s ) \;
\left( - \frac{1 }{2 } \Delta _s - \frac{1 }{2 } \Delta _t +
\mu ^2 \right)^{-1/2 } \\
&\simeq & \frac{1 }{2 } \left( - \frac{1 }{2 } \Delta _s - \frac{1
}{2 } \Delta _t + \frac{\mu ^2 }{2 } \right)^{-1/2 } \;
\mathfrak{v } (s ) \; \left( - \frac{1 }{2 } \Delta _s - \frac{1
}{2 } \Delta _t + \frac{\mu ^2 }{2 } \right)^{-1/2 } .
\end{eqnarray*}
Observing that
$$
\| \; \left( - \frac{1 }{2 } \Delta _s + \frac{\mu ^2 }{2 }
\right)^{1/2 } \left( - \frac{1 }{2 } \Delta _s - \frac{1 }{2 }
\Delta _t + \frac{\mu ^2 }{2 } \right)^{-1/2 } \; \| \leq 1 .
$$
on $L^2 (\mathbb{R }^2 )  $, the lemma follows. \hfill QED

\medskip

\noindent Let us write
\begin{equation} \label{Rjk}
R_{00 , jk } = R_{00 , jk }(-\mu ^2 ) = \displaystyle{\left(
-\frac{1 }{2 } (\Delta _j + \Delta _k) + \mu ^2 \right)^{-1/2 } }
,
\end{equation}
the 2-dimensional free resolvent, where $\mu $ will be optimized
at the end of the proof. Recall that $R_{00 } = R_{00 } (-\alpha
^2 ) $, and put
$$
\Delta V_{jk} = V_{jk } ^B (z_j - z_k ) - q^B (z_j - z_k ) - C_{jk
} ^{e } \delta (z_j - z_k ) .
$$
Then, using lemmas \ref{FE-special'} and \ref{AsympPot1}, $R ^{1/2
} (\mathcal{V }_{{\rm eff }, e } ^B - v_{C, e } ^B ) R_{00 }^{1/2
} $ can be estimated from above as follows:
\begin{eqnarray*}
R_{00 } ^{1/2 } \; \sum _{j < k
} \Delta V_{jk } \; R_{00 } ^{1/2 } &=& \sum _{j < k } \left( R_{00 } ^{1/2
} R_{00, jk } ^{-1/2 }
\right) \; \left( R_{00 , jk } ^{1/2 } \; \Delta V_{jk} \; R_{00 ,
j k } ^{1/2 } \right)
\; \left( R_{00 , jk }^{-1/2 } R_{00 }^{1/2 } \right) \\
&\leq & \frac{C_{(\ref{asympVj}) } }{2^{3/4 } \sqrt{\mu } B^{1/4 }
} \sum
_{j < k } R_{00 }^{1/2 } R_{00 , jk } ^{-1 } R_{00 }^{1/2 } \\
&=& \frac{C_{(\ref{asympVj}) } }{2^{3/4 } \sqrt{\mu } B^{1/4 } }
R_{00 } ( - \alpha ^2 ) \left( \sum _{j < k } (- \frac{1 }{2 }
(\Delta _j + \Delta _k ) + \mu ^2 )
\right) \\
&=& \frac{C_{(\ref{asympVj}) } (N - 1 ) }{2^{3/4 } \sqrt{\mu }
B^{1/4 } } R_{00 } (-\alpha ^2 ) \left( - \frac{1 }{2 }
\Delta _z + \frac{N \mu ^2 }{2 } \right) \\
&\leq & \frac{C_{(\ref{asympVj}) } (N - 1 ) }{2^{3/4 } B^{1/4 } }
\cdot \frac{1 }{\sqrt{\mu } } \max \left(1, \frac{N \mu ^2 }{2
\alpha ^2
} \right) \leq \frac{C_{(\ref{asympVj}) } (N - 1 ) N^{1/4 } }{2
\sqrt{\alpha } B^{1/4 } } ,
\end{eqnarray*}
where we minimized the right hand side over $\mu $ by choosing
$\mu = \alpha \sqrt{2/N } $. The similar upper bound for$ - R
^{1/2 } (\mathcal{V }_e - v_{C, e }^B ) R_{00 }^{1/2 } $ gives the
desired estimate for the second norm in (\ref{V-eff2}), and
combining the two estimates, we have proved lemma \ref{AsympPot2}.
\hfill QED
\medskip

We now derive a similar estimate for $\mathcal{W }^B $ as $B \to
\infty $.

\begin{lemma} \label{W-Est} Let $ R_{00 } ( - \alpha ^2 ) $ be the
free resolvent in dimension $N $, and let $U = U_{xy} ^B $ be the
unitary transformation defined by (\ref{def-U}). Then, if $\xi \le
0
$,
\begin{equation} \label{est-W}
\vert \vert R_{00 } ( - \alpha ^2 ) ^{1/2 } U^* \mathcal{W }^B U
R_{00 } ( - \alpha ^2 )^{1/2 } \vert \vert \leq
\frac{C_{(\ref{const-W}) } }{\alpha \sqrt{B } } ,
\end{equation}
with
\begin{equation} \label{const-W}
C_{(\ref{const-W}) } := C_{(\ref{const-W}) } (N, Z ) := 2 \pi
^{3/2 } N^{3/2 } \left( Z^2 + \frac{(N - 1 )^2 }{4 } \right) .
\end{equation}
\end{lemma}

\noindent {\it Proof.} Recall that $\mathcal{W }^B =  -
\mathcal{V }_{ {\rm eff } , \perp  } R \mathcal{V }_{\perp , {\rm
eff } } $, where $R = R(\xi ) = (H_{\perp } - \xi )^{-1 } $ on
$\mbox{Ran } \Pi_\perp^B $. Hence
$$
U^* \mathcal{W }^B U=-B \mathcal{V } _{ {\rm eff
} , \perp } ^1 ( \sqrt{B } z ) U^* R U \mathcal{V } _{\perp  ,
{\rm eff } } ^1 (\sqrt{B } z ) ,
$$
where $\mathcal{V } _{ {\rm eff } , \perp  } ^1 = \Pi _{\rm eff }
^1 \mathcal{V } \Pi _{\perp } ^1 $, and similarly for $\mathcal{V
} _{\perp  , {\rm eff } } ^1 $. Since for $\xi \le 0 $, $0 \leq R
\leq 2/B $ (see section 3), we have that, letting $\mathcal{V }
(\cdot , z ) $ be the function $(x, y ) \to \mathcal{V } (x, y, z
) $,
\begin{eqnarray} \nonumber
0 \leq  -  U^* \mathcal{W }^B U &\leq & 2 \Pi _{\rm eff } ^1
\mathcal{V } (\cdot , \sqrt{B } z )
\Pi _{\perp } \mathcal{V } (\cdot , \sqrt{B } z ) \Pi _{\rm eff } ^1 \\
&\leq & 2 \Pi _{\rm eff } ^1 \mathcal{V } (\cdot , \sqrt{B } z )^2 \Pi _{\rm eff } ^1 \nonumber \\
&\leq & 4 \left( \Pi _{\rm eff } ^1 \mathcal{V } _n (\cdot ,
\sqrt{B } z )^2 \Pi _{\rm eff } ^1 + \Pi _{\rm eff } ^1 \mathcal{V
} _e (\cdot , \sqrt{B } z )^2 \Pi _{\rm eff } ^1 \right) .
\label{est-W0.5}
\end{eqnarray}
As in the previous lemma, we treat the two terms separately. By
Cauchy-Schwarz,
$$
\mathcal{V } _n (x, y, \sqrt{B } z )^2 \leq Z^2 N \sum _{j = 1 }
^N \frac{1 }{\rho _j ^2 + B z_j ^2 } ,
$$
where $\rho _j ^2 = x_j ^2 + y_j ^2 . $ As before, let $R_{00 , j
} ( - \beta ^2 ) $ be the resolvent of $h_{00 , j } = -(1/2 ) d^2
/ dz_j ^2 $ on $\mathbb{R } $. We first estimate the $L^2 $-norm
of each
\begin{equation} \label{est-W1}
R_{00 , j } ( - \beta ^2 ) ^{1/2 } \Pi _{\rm eff } ^1 (\rho _j ^2
+ B z_j ^2 )^{-1 } \Pi _{\rm eff } ^1 R_{00 , j } ( - \beta ^2 )
^{1/2 } ,
\end{equation}
by conjugating with the Fourier transform $\mathcal{F } $. Since
$\mathcal{F } \left( (1 + z^2 ) ^{-1 } \right) (\zeta ) = \pi
e^{-|\zeta | } $, (\ref{est-W1}) then becomes an integral operator
with kernel
\begin{equation} \label{est-W2}
\displaystyle{\frac{1 }{2 } B^{-1/2 } \left( \tfrac{\zeta _j ^2
}{2 }+ \beta ^2 \right) ^{-1/2 } \left( \Pi _{\rm eff } ^1 \; \rho
_j ^{-1 } e^{- \rho _j B^{-1/2 } |\zeta _j - \eta _j | } \; \Pi
_{\rm eff } ^1 \right) \left(\tfrac{\eta _j ^2 }{2 } + \beta ^2
\right) ^{-1/2 }  } .
\end{equation}
The norm of (\ref{est-W2}) can be estimated by its Hilbert-Schmidt
norm, whose square can be bounded by:
$$
\frac{1 }{4B } \| \Pi _{\rm eff } ^1 \frac{1 }{\rho _j } \Pi _{\rm
eff } ^1 \|^2 \left( \int _{\mathbb{R } } (\tfrac{\xi ^2 }{2 } +
\beta ^2 )^{-1 } d\xi \right)^2 = \frac{\pi ^2 C_{(\ref{est-W:C})
} ^2 }{2 \beta ^2 B } ,
$$
 where we have put
\begin{equation} \label{est-W:C}
C_{(\ref{est-W:C}) } := \| \; \Pi _{\rm eff } ^1 \rho _j ^{-1 }
\Pi _{\rm eff } ^1 \; \| .
\end{equation}
Note that $C_{(\ref{est-W:C}) } $ is independent of $j $, because
of the permutational symmetry of $\Sigma (\mathbb{M } ) $. It
follows that:
\begin{equation} \label{est-W2a}
R_{00, j } (-\beta ^2 ) ^{1/2 } \; \Pi _{\rm eff } ^1 (\rho _j ^2
+ B z_j ^2 )^{-1 } \Pi _{\rm eff } ^1 \; R_{00, j } (-\beta ^2 )
^{1/2 }\leq \frac{\pi C_{(\ref{est-W:C}) } }{\sqrt{2 } \beta
\sqrt{B } } ,
\end{equation}
and therefore
\begin{eqnarray} \nonumber
0 &\leq&
R_{00 } (-\alpha ^2 ) ^{1/2 } \; \Pi _{\rm eff } ^1 \;
\mathcal{V } _n (\cdot , \sqrt{B } ) ^2 \; \Pi _{\rm eff } ^1 \;
R_{00 } (-\alpha ^2 ) ^{1/2}\\
&\leq&  \frac{\pi C_{(\ref{est-W:C}) } Z^2 N
}{\sqrt{2 } \beta B^{1/2 } }
R_{00 } (-\alpha ^2 ) \sum _j \left( h_{00 , j } + \beta ^2 \right) =
\frac{\pi C_{(\ref{est-W:C}) } Z^2 N }{\sqrt{2} \beta B^{1/2 } } R_{00 }
(-\alpha ^2 ) (h_{00 } + N \beta ^2 ) \\
&\leq&  \frac{\pi
C_{(\ref{est-W:C}) } Z^2 N^{3/2 } }{\sqrt{2 }
\alpha B^{1/2 } }\nonumber,
\end{eqnarray}
if we choose $\beta = \alpha / \sqrt{N } . $ The same inequality
then holds for the norm, since the operator we estimate is
positive.
\medskip

We next treat the interaction term $R_{00 } (\alpha ^2 )^{1/2 }
\mathcal{V }_e (\cdot , \sqrt{B } z )^2 R_{00 } (\alpha ^2 ) ^{1/2
} $ in a similar way as in the proof of lemma \ref{AsympPot2}.
First, by Cauchy-Schwarz again,
$$
\mathcal{V }_e (\cdot , \sqrt{B } z ) ^2 \leq \frac{N (N - 1 ) }{2
} \sum _{j < k } \frac{1 }{\rho _{jk } ^2 + B (z_j - z_k ) ^2 } ,
$$
where we have put $\rho _{jk } ^2 = (x_j - x_k )^2 + (y_j - y_k )
^2 $. Since the rotation $(r_j , r_k ) \to $ \newline
$\left( 2^{-1/2 } (r_j - r_k ) , 2^{-1/2 } (r_j + r_k ) \right) $
 commutes with $\Pi _{\rm eff } ^1 $ (since it commutes with
$H_0 ^1 $ and with $\mathbb{L }_z $), we find, after a unitary
transformation, that
\begin{eqnarray*}
&\ & \| \, R_{00, jk} (-\mu ^2 )^{1/2 } \Pi _{\rm eff } ^1 \left(
\rho _{jk } ^2 + B (z_j -
z_k ) ^2 \right)^{-1 } \Pi _{\rm eff } ^1 R_{00, jk} (-\mu ^2 )^{1/2 } \, \| \\
&\leq & \| \, \frac{1 }{2 } R_{00, j } (- \frac{\mu ^2 }{2 } )
^{1/2 } \Pi _{\rm eff } ^1 \left( \rho _j ^2 + B z_j
^2 \right)^{-1 } \Pi _{\rm eff } ^1 R_{00, j } (-
\frac{\mu ^2 }{2 } ) ^{1/2 } \, \| \\
&\leq & \frac{\pi C_{(\ref{est-W:C}) } }{2 \mu \sqrt{B } } ;
\end{eqnarray*}
compare the proof of lemma \ref{FE-special'}. Hence, using similar
arguments as before,
\begin{eqnarray*}
&\ & R_{00 } (-\alpha ^2 ) ^{1/2 } \; \Pi _{\rm eff } ^1 \;
\mathcal{V } _e (\cdot , \sqrt{B } ) ^2 \; \Pi _{\rm eff } ^1 \;
R_{00 } (-\alpha ^2 ) ^{1/2
} \\
&\leq & \frac{\pi C_{(\ref{est-W:C}) } N (N - 1 ) }{4 \mu \sqrt{B
} } \; R_{00 } ( -  \alpha ^2
) \sum _{j < k } (h_{00, jk } + \mu ^2 ) \\
&\leq & \frac{\pi C_{(\ref{est-W:C}) } N (N - 1 )^2 }{4 \mu
\sqrt{B } } R_{00 }
(-\alpha ^2 ) (h_{00 } + \frac{N \mu ^2 }{2 } ) \\
&\leq & \frac{\pi C_{(\ref{est-W:C}) } N^{3/2 } (N - 1 )^2
}{4\sqrt{2 } \alpha \sqrt{B } } ,
\end{eqnarray*}
if we choose $\mu = \alpha \sqrt{2/N } $. Adding this estimate to
the one for $\mathcal{V }_n ^2 $, and remembering the factor 4
from (\ref{est-W0.5}), we have proved (\ref{est-W} ) with the
constant $C_{(\ref{const-W}) }  = 2^{3/2 } \pi C_{(\ref{est-W:C})
}  N^{3/2 } \left( Z^2 + { (N - 1 )^2\over4} \right) $. A priori,
$C_{(\ref{est-W:C}) }  $ might still depend on $\mathbb{M } $, but
it in fact does not, as we will finally show. We compute
$C_{(\ref{est-W:C}) } $ in the Landau basis
(\ref{LLBF}) of $F{_\mathbb{M } } ^1 $, with respect to
which $\Pi _{\rm eff } ^1 \; \rho _1 ^{-1 } \; \Pi _{\rm eff } ^1
$ diagonalizes:
\begin{eqnarray*}
C_{(\ref{est-W:C}) } &=& \max _{0 \leq m \leq \mathbb{M } }
\frac{1 }{2^m m! } \int _0
^{\infty } \rho _1 ^{2m } e^{-\rho _1 ^2 / 2 } d\rho _1 \\
&=& \displaystyle{\max _{0 \leq m \leq \mathbb{M } } \frac{1 }{m!
\sqrt{2 } } \int _0 ^{\infty } s^{m - \frac{1 }{2 } } e^{-s } ds }
\\
&=& \max _{0 \leq m \leq \mathbb{M } } \frac{\Gamma (m + \frac{1
}{2 } ) }{\sqrt{2 } \Gamma (m + 1 ) } .
\end{eqnarray*}
It is known that
$$
\Gamma \left(m + \frac{1 }{2 } \right) = \frac{1 \cdot 3 \cdot 5
\cdots \cdot (2m - 1 ) }{2^m } \Gamma \left(\frac{1 }{2 } \right)
,
$$
cf. e.g. \cite{AS}, formula (6.1.12), page 255. Using this, one
easily finds that
$$
C_{(\ref{est-W:C}) } = \frac{1 }{\sqrt{2 } } \Gamma \left( \frac{1
}{2 } \right) = \sqrt{\frac{\pi }{2 } } ,
$$
which completes the proof of the lemma. \hfill QED
\medskip

To prove Theorem~\ref{CorrSMT} we will need to control the Sobolev
norm of $V_1 ^B - v_{\delta } ^B $. This is done in the following
lemma, which we formulate in slightly greater generality than
needed, with an eye to future applications.

%%%%%%%%%%%%%%%%%%%%%%%%%%%%%%%%%
\begin{lemma} \label{vdelta-Est}%
%%%%%%%%%%%%%%%%%%%%%%%%%%%%%%%%%
 Let $V_1^{B}:=V_1^{B,\mathbb{M}}$ be defined as in (\ref{defVeffietjk}),
$c > 0 $ and let $\alpha _c = \alpha_c (B) $ be the unique positive
solution of
\begin {equation} \label{alpha-c}
\alpha _c = \frac{2 }{c } \log \left( \frac{\sqrt{B } }{\alpha _c
} \right) ,
\end{equation}
Then the constant
\begin{equation}\label{supVchapMoinsDelta}
C_{(\ref{supVchapMoinsDelta} ) } := \left( \pi ^2 + 9 \log ^2 (2)
+ \frac{64 \sqrt{2 } }{\pi } + \sup _{|\zeta | \leq 1}\left|\FF
V_1^1(\zeta) + 2 \log( |\zeta| )\right|^2 + \frac{8 \sqrt{2 }
}{\pi } \sup _{|\zeta | \geq 1 } |\FF V_1 ^1 (\zeta ) |^2 \right)
^{1/2 }
\end{equation}
is finite and depends only on $\mathbb{M}$. Moreover for all $c
> 0 $ and all $B>0$
\begin{equation}\label{VeffMoinsDelta}
\|R_{00}(-\alpha _c ^2)^{1\over2}(V_1^B - c \alpha _c (B)
\delta)R_{00}(-\alpha _c ^2)^{1\over2}\| \le
{C_{(\ref{supVchapMoinsDelta} ) }\over\alpha _c (B) }.
\end{equation}
\end{lemma}

\noindent Observe that if $c = 2 $, then $\alpha _c (B) $ is the
$\alpha (B) $ from Theorem \ref{FMT}: cf. (\ref{FMT:alpha}). Also
note that the constant  $C_{(\ref{supVchapMoinsDelta})}$ is independent of
$c > 0 . $
\medskip

\noindent {\it Proof.} The first statement follows at once using
the explicit knowledge of $\FF V_1^1$ obtained in the
proof of lemma \ref{AsympPot1}. To prove the estimate
(\ref{VeffMoinsDelta}) we introduce the auxiliary function $X$
 by
\begin{eqnarray*}
\FF V_1^B(\zeta)-c \alpha _c \FF \delta(\zeta)&=& \FF
V_1^1({\zeta\over\sqrt{B}})+ 2 \log{|\zeta|\over\sqrt{B}}-
2 \log{|\zeta|\over\alpha _c }\\
&=:& X({\zeta\over\sqrt{B}})- 2 \log{|\zeta|\over\alpha _c } ,
\end{eqnarray*}
where in the first line we used equation (\ref{alpha-c}).  The
Fourier transform of $V_1^B-c \alpha _c (B)\delta$ acts as
convolution by a function, and we estimate the norm of
$Y:=R_{00}(-\alpha _c ^2)^{1\over2}(V_1^B-c\alpha _c
\delta)R_{00}(-\alpha _c ^2)^{1\over2}$ by its Hilbert-Schmidt
norm, as in the proofs of lemmas \ref{Asymp-Lemma},
\ref{AsympPot1a}. It follows that
$$
\|Y\|^2\le{\sqrt{2}\over \pi \alpha _c }\int_\mathbb{R}{|\FF
V_1^B(\zeta)-c \alpha _c \FF \delta(\zeta)|^2\over 8\alpha _c
^2+\zeta^2}d\zeta\le{2\sqrt{2}\over \pi \alpha _c
}\int_\mathbb{R}{|X({\zeta\over\sqrt{B}})|^2+|2
\log{|\zeta|\over\alpha _c }|^2\over 8\alpha _c ^2+\zeta^2}d\zeta.
$$
First,
$$
{1\over\alpha _c }\int_\mathbb{R} {4 \over8\alpha _c
^2+\zeta^2}\left|\log{|\zeta|\over\alpha _c
}\right|^2d\zeta={\pi\over2\sqrt{2} \alpha _c ^2
}(\pi^2+9\log^2(2)).
$$
We next look at the contribution of $X $, which we split in two
parts:
\begin{eqnarray*}
\int_{\sqrt{B}}^\infty {|X({\zeta\over\sqrt{B}})|^2d\zeta\over
8\alpha _c ^2+\zeta^2}&=& {1\over\sqrt{B}}\int_1^\infty
{|X(\zeta)|^2d\zeta\over8 B^{-1}\alpha _c ^2+\zeta^2} \le
{2\over\sqrt{B}}\int_1^\infty{|\FF
V(\zeta)|^2 + 4 (\log|\zeta|)^2\over\zeta^2}d\zeta\\
&\le& {2\over\sqrt{B}}\left( \sup _{|\zeta | \geq 1 } |\FF
V(\zeta)|^2 \; + \; 8 \right) ,
\end{eqnarray*}
since
$$
\int_1^\infty{(\log|\zeta|)^2\over|\zeta|^2}d\zeta=2 ,
$$
and
\begin{eqnarray*}
\int_0^{\sqrt{B}} {|X({\zeta\over\sqrt{B}})|^2d\zeta\over 8\alpha
_c ^2+\zeta^2}&\le& \sup _{|\zeta| \leq 1 } |X(\zeta)|^2 \;
\int_0^{\sqrt{B}} {d\zeta\over8\alpha _c ^2+\zeta^2}\\
%&\le&{1\over2\sqrt{2}\alpha} \arctan(\sqrt{B})
%\sup\{|X(\zeta)|^2,|\zeta|<1|\}\\
&\le&{\pi\over4\sqrt{2}\alpha _c } \sup _{|\zeta| \leq 1 }
|X(\zeta)|^2 .
\end{eqnarray*}
The rest is now elementary. Notice in particular that $\sup_{B>0}
\alpha _c (B ) /\sqrt{B}=1 $. \hfill {\rm QED}
\medskip

The limit potential (\ref{V-eff}) suggests defining an effective
Hamiltonian $h_C = h_C ^B $ by:
\begin{equation} \label{EffHam1}
h_C := h_{00 } + v_C .
\end{equation}
As it stands, this is just a formal expression, and our first task
is to give a meaning to $h_C $ as a self-adjoint operator on $L^2
(\mathbb{R }^N ; F^1 _{\mathbb{M } } ) $. We do this by showing
that $v_C $ is form-bounded with respect $h_{00 } $, with zero
relative form-bound. Let $\langle \cdot , \cdot \rangle $ denote
the duality between distributions and test functions.

\begin{lemma} \label{relformbound} The quadratic forms
$u \to \langle \delta , |u|^2 \rangle = |u(0) |^2 $ and $u \to
\langle \mbox{\rm{Pf}} (1 / |z| ) , |u|^2 \rangle $ are
well-defined on $H^1 (\mathbb{R } ) $, and form-bounded with
respect to $h_{00 } $, with relative bound zero. More precisely,
we have for all $\varepsilon > 0 $ that
\begin{equation} \label{rel-delta}
\delta \leq \tfrac{1 }{2 } (-\varepsilon \Delta_z + \varepsilon
^{-1 } ) ,
\end{equation}
and
\begin{equation} \label{relPf}
{\rm Pf } ( |z |^{-1 } ) \leq C_{(\ref{refPf1}) } ( |\log
\varepsilon | + 1 ) (-\varepsilon \Delta_z + \varepsilon ^{-1 } )
,
\end{equation}
where
\begin{equation} \label{refPf1}
C_{(\ref{refPf1}) } := \sqrt{{\pi^2\over2}+2(\log2)^2} +\gamma .
\end{equation}
\end{lemma}

\noindent {\it Proof.} This is well-known for $\delta $. For
$\mbox{Pf }(|z|^{-1 } ) $ we first note that, since $\mbox{Pf
}(|z|^{-1 } ) = -2 \mathcal{F }^{-1 } (\log |\zeta | ) - 2 \gamma
\delta _0 $, it suffices to prove the form-boundedness of
$\mathcal{F }^{-1 } (\log |\zeta | ) $. The latter will follow
from:
\begin{equation} \label{relFlog}
 \|(-\varepsilon \Delta + \varepsilon ^{-1 } )^{-1 / 2 }
\mathcal{F }^{-1 } (\log |\zeta | ) (-\varepsilon \Delta +
\varepsilon ^{-1 } )^{-1 / 2 }  \|^2
\leq  { \frac{1 }{2 } (\log \varepsilon )^2 + \frac{\pi ^2
+ 4 (\log 2 )^2 }{8 } },
\end{equation}
for all $\varepsilon > 0 $. To prove (\ref{relFlog}), observe that
after conjugation by the Fourier transform $\mathcal{F } $, and
estimating the operator norm by the Hilbert-Schmidt norm, the
square of (\ref{relFlog}) can be bounded by:
\begin{equation} \label{HS}
\frac{1 }{(2\pi )^2 } \int _{\mathbb{R } } \int _{\mathbb{R } }
\frac{ (\log |\zeta - \eta | )^2 } {(\varepsilon \zeta ^2 +
\varepsilon ^{-1 } ) (\varepsilon \eta ^2 + \varepsilon ^{-1 } ) }
d\zeta d\eta .
\end{equation}
Changing variables and using (\ref{AL2}), we find that (\ref{HS})
can be bounded by
\begin{eqnarray*}&&
 \frac{1 }{2\pi } \int _{\mathbb{R } } \frac{ (\log |\zeta  |
)^2 }{\varepsilon \zeta ^2 + 4 \varepsilon ^{-1 } } d \zeta  =
\frac{1 }{2\pi } \int _{\mathbb{R } } \frac{ (\log |\zeta /
\varepsilon | )^2 }{\zeta ^2 + 4 } d \xi \\
&\leq&  \frac{1 }{\pi } (\log \varepsilon )^2 \int _{\mathbb{R } }
\frac{d \zeta }{\zeta ^2 + 4 } + \frac{1 }{\pi } \int _{\mathbb{R
} } \frac{(\log \zeta )^2 }{\zeta ^2 + 4 } d\zeta = \frac{1 }{2 } (\log
\varepsilon )^2 + \frac{\pi ^2 + 4 (\log 2)^2 }{8 }.
\end{eqnarray*}
Hence, using (\ref{rel-delta}),
\begin{eqnarray*}
&&\left \vert \langle \mbox{Pf }(|x|^{-1 } ) , |u |^2 \rangle
\right \vert =  2 \cdot \; \left \vert \; \langle \mathcal{F }^{-1 } (\log
|\xi | ) , |u|^2
\rangle \; + \; \gamma \langle \delta , |u|^2 \rangle \; \right \vert \\
&\leq & \left \{ \left(2 (\log \varepsilon )^2 + \frac{\pi ^2 }{2
} + 2 (\log 2 )^2 \right) ^{1/2 } + \gamma \right \} \left \{ -
\varepsilon (\Delta u, u ) + \varepsilon ^{-1 } \|u
\|^2 \right \}\\
&=&\left(\sqrt{2}|\log\varepsilon|+C_{(\ref{refPf1})}\right) \left
\{ - \varepsilon (\Delta u, u ) + \varepsilon ^{-1 } \|u \|^2
\right \} ,
\end{eqnarray*}
which implies (\ref{relPf}) since $\sqrt{2}/C_{ (\ref{refPf1})}<1
. $ \hfill {\rm QED }
\medskip

Because of the terms $C_j ^n \delta (z_j ) $ and $C_{jk } ^e
\delta (z_j - z_k )$ we have to extend the first part of lemma
\ref{relformbound} to the vector-valued case, but this is
immediate:  we just note that if $C $ a linear operator on
$F^1 _{\mathbb{M } } $ (or any finite-dimensional vector space,
for that matter), then the right interpretation of $C \delta $ as
quadratic form on $H^1 (\mathbb{R } , F^1 _{\mathbb{M } } ) $ is
given by $\left \langle \delta (z) , \; (C u (z) , u (z) ) \right
\rangle $, where $(\cdot , \cdot ) $ is the inner product on $F^1
_{\mathbb{M } } $.

Finally, we lift lemma \ref{relformbound} to $\mathbb{R }^N $.
Recall that if $L : \mathbb{R }^N \to \mathbb{R } $ is a linear
map, then the pull-back $L^* \Lambda $ of a distribution $\Lambda
$ on $\mathbb{R } $ is well-defined, and can be computed by going
to linear coordinates $z' $ with respect to which $L(z') = z'_1 $.
It then immediately follows from lemma \ref{relformbound} that
$L^* \delta $ and $L^* q^B $ will be form-bounded with respect to
$h_{00 } $ on $\mathbb{R }^N $, with relative bound 0. Taking
$L(z) = L_j (z) = z_j $ and $L(z) = L_{jk } (z) = z_j - z_k $, we
see that the sesqui-linear form $t_C ^B (u) $ given by
\begin{eqnarray} \label{t-eff}
t_C  ^B (u) &=& \frac{1 }{2 } || \nabla u ||^2 - Z \sum _j \left
\langle L_j ^* q^B , \; |u|^2 \right \rangle + \left \langle (L_j
^*
\delta , \; (C_j ^n u, u ) \right \rangle  \\
&+ & \sum _{j < k } \left \langle L_{jk } ^* q^B , \; |u|^2 \right
\rangle + \left \langle L_{jk } ^* \delta , \; (C_{jk } ^e u, u )
\right \rangle , \nonumber
\end{eqnarray}
is well-defined on $H^1 (\mathbb{R }^N, F^1 _{\mathbb{M } } ) $,
and bounded from below by $-C ||u ||^2 $, for some constant $C $,
depending on $B, Z, N $ and $\mathbb{M } $. By the
Kato-Lax-Lions-Milgram-Nelson Theorem (cf. e.g. \cite[Theorem
X.17]{RS2}), $t_C ^B $ defines a unique self-adjoint operator,
which we will call $h_C = h_C ^{B , \mathbb{M } } $, and
informally write as (\ref{EffHam1}). In Appendix A we will give a
characterization of the operator domain of $h_C ^B $. Similar
arguments will define $h_{\delta } ^B $ as a self-adjoint
operator, cf. \cite{BD}.
\medskip

\section{{\bf Proof of Theorem \ref{FMT} } }

We will first compare the resolvents of $h_{\rm eff}
:=  h_{\rm eff} ^{B, \mathbb{M } } $ and of $h_{\rm eff } + \mathcal{W } $,
where $\mathcal{W } :=
\mathcal{W }^{B, \mathbb{M } } (\xi ) $ was defined at the
beginning of section 3. Put
$$
r_{\rm eff } := r_{\rm eff } (\xi ) :=
\left( h_{\rm eff} - \xi \right) ^{-1 } , \ \ R_{\rm
eff } ^{\mathcal{W } } := R_{\rm eff } ^{\mathcal{W } } (\xi ) :=
\left(h_{\rm eff} + \mathcal{W } - \xi \right) ^{-1
} ,
$$
and let
\begin{equation} \label{FMT':B}
B_{(\ref{FMT':B}) } := \frac{4 C_{(\ref{const-W}) } ^2 }{\alpha\!
\left(  C_{(\ref{const-W}) }  \right) ^2 } ,
\end{equation}
where $\alpha = \alpha (B) $ is the function defined by
(\ref{FMT:alpha}) , and
\begin{equation} \label{FMT':c-eff}
c_{\rm eff } := 2 \left( \frac{N }{2 \varepsilon_{\rm eff}^2 } + 1
\right) C_{(\ref{const-W}) } ,
\end{equation}
with $\varepsilon_{\rm eff}= \varepsilon_{\rm eff}(Z, \mathbb{M } ) $ the unique
positive solution of
\begin{equation}\label{FMT':eps}
Z C_{(\ref{AsympPot1a:c}) }  \; C_{V^1_1 } \; \varepsilon(|\log
\varepsilon| + 2 ) = {1\over 4} ,
\end{equation}
where $C_{V_1^1 } $ is defined by (\ref{HypAsympPot1a}) with
$\mathfrak{v} = V_1^1$ (see (\ref{C_V1:1}) for an upper bound).
Note that both constants only depend on $N $, $Z $ and $\mathbb{M
} $, and this in a controlled way.

We then have:
\medskip

\begin{theorem} \label{FMT'} If $B \geq
B_{(\ref{FMT':B}) } $, $\xi \leq 0 $ and
\begin{equation} \label{FMT':0.5}
c_{\rm eff } \frac{\alpha }{\sqrt{B } } \leq d _{\rm eff } (\xi )
\leq \frac{1 }{2 } \alpha ^2 ,
\end{equation}
with $\alpha = \alpha (B ) $ is as in Theorem \ref{FMT}, then $\xi
\in \rho (h_{\rm eff} + \mathcal{W } ) $, and $\|
R_{\rm eff } ^{\mathcal{W } } (\xi ) \| \leq 2 \| r_{\rm eff } (\xi ) \| $.
Furthermore,
\begin{equation} \label{FMT':1}
\| \, R_{\rm eff } ^{\mathcal{W } } (\xi )  -r_{\rm eff } (\xi )
\| \leq c_{\rm eff} \frac{\alpha }{d _{\rm eff } (\xi ) ^2 \sqrt{B
} } .
\end{equation}
\end{theorem}

\noindent {\it Proof of Theorem \ref{FMT'}.} It clearly
suffices to establish Theorem \ref{FMT'} after conjugation by
$U_{xy }^B $, defined by (\ref{def-U}). To simplify notations, we
will simply denote the conjugated operators by the same letters as
the original ones.  Using the symmetrized resolvent formula
we estimate
\begin{equation} \label{FMT':2}
\|\,R_{\rm eff }^{\mathcal{W } } (\xi )-r_{\rm eff } (\xi ) \, \|
\leq \frac{1 }{d _{\rm eff} (\xi ) } \left( \frac{\| K_{\rm eff }
(\xi ) \| }{1 - \|K_{\rm eff } (\xi ) \| } \right) ,
\end{equation}
where
\begin{equation} \label{FMT':3}
K_{\rm eff } (\xi ) := |r_{\rm eff } (\xi ) |^{1/2
}\mathcal{W }(\xi) r_{\rm eff } (\xi ) ^{1/2 } ,
\end{equation}
with the convention that $A^{1/2 } := \mbox{sgn }(A ) |A |^{1/2 }
$, if $A $ is a self-adjoint operator\footnote{note that since
$\xi $ is not necessarily below the infimum of the spectrum,
$r_{\rm eff } (\xi ) $ is not necessarily positive, and we use the
symmetrized resolvent formula in the following form: $R_{\rm eff }
^{\mathcal{W } } =r _{\rm eff } ^{1/2 } (1 + | r_{\rm eff } |^{1/2
} \mathcal{W }r_{\rm eff } ^{1/2 } )^{-1 }| r_{\rm eff } |^{1/2 }
$ }. Now let $\mu < \inf(\sigma (h _{\rm eff} ) $, to be specified
later. The following elementary lemma will allow us to replace
$\xi $ by $\mu $ in (\ref{FMT':3}).

%%%%%%%%%%%%%%%%%%%%%%%%%%%%%%%%%%
\begin{lemma} \label{FMT':lemmeI}%
%%%%%%%%%%%%%%%%%%%%%%%%%%%%%%%%%%
If $\mu < \inf \sigma (h_{\rm eff} ) $, then for all real $\xi$ in the resolvent
set $\rho(h_{\rm eff})$,
\begin{equation} \label{FMT':4}
\| \; r_{\rm eff } (\xi ) \; (h_{\rm eff} - \mu ) \; \| \leq \max
\left( \frac{|\mu | }{ d _{\rm eff } (\xi )  } , 1 \right) .
\end{equation}
\end{lemma}

\noindent {\it Proof.} We distinguish two cases: $\inf \sigma (h_{\rm eff} ) <
\xi < 0 $ and $\xi < \inf \sigma (h_{\rm eff} ) $. (Observe that
$\inf \sigma (h_{\rm eff} ) < 0 $, by the HVZ Theorem, since this
is already the case for $N = 1 $). In the first case, let $(\xi _-,
\xi _+ ) $ be the largest open interval in $\rho (h_{\rm eff} ) $ which
contains $\xi $. Since $[0, \infty ) $ is in the spectrum (it is already in the
essential spectrum), $\xi _+ \le 0 $.  It is easy to see
that the function $x \to |x -\mu | / |x -\xi | $ is increasing on $(-\infty ,
\xi _- )\cap\sigma(h_{\rm eff})$ and decreasing on $(\xi _+ , \infty
)\cap\sigma( h_{\rm eff})$. It follows that
$$
\|r_{\rm eff } (\xi ) (h_{\rm eff} - \mu ) \| = \sup _{x \in
\sigma (h_{\rm eff} ) } \frac{|x - \mu | }{|x - \xi | } = \max
\left( \frac{|\xi _- - \mu | }{|\xi _- - \xi | } , \frac{|\xi _+ -
\mu | }{|\xi _+ - \xi | } \right) \leq \frac{|\mu | }{ d _{\rm eff
} (\xi )  } ,
$$
as was to be shown. One shows in a similarly way that
(\ref{FMT':4}) equals $(\inf \sigma( h_{\rm eff} ) -\mu) /(\inf
\sigma (h_{\rm eff} )  -\xi) \leq|\mu |/ d _{\rm eff } (\xi )
 $, if $\mu < \xi < \inf \sigma ( h_{\rm eff} ) $, and is
equal to $1 $ if $\xi < \mu $. \hfill QED
\medskip

Substituting $Id = r_{\rm eff } (\mu )^{1/2 }
(h_{\rm eff } - \mu )^{1/2 } = (h_{\rm
eff } - \mu )^{1/2 } r_{\rm eff } (\mu )^{1/2 } $ at
the appropriate places in formula (\ref{FMT':3}), we see that
if $\xi \leq 0 $,
\begin{equation} \label{FMT':5}
\| K_{\rm eff } (\xi ) \| \leq \max \left( \frac{|\mu | }{ d
_{\rm eff } (\xi ) } , 1 \right) \| K_{\rm eff } (\mu ; \xi ) \| ,
\end{equation}
with
$$
K_{\rm eff } (\mu ; \xi ) := r_{\rm eff } (\mu
)^{1/2 } \mathcal{W }(\xi )r_{\rm eff } (\mu )^{1/2
} .
$$
Repeating the same argument for $K_{\rm eff } (\mu ;
\xi ) $ using $Id = ((h_{00 } + \alpha ^2 )^{1/2 } R_{00 }
(-\alpha ^2 ) ^{1/2 } $, we obtain from lemma \ref{W-Est} that
\begin{equation} \label{FMT':6}
\| K(\mu ; \xi ) \| \leq \frac{C_{(\ref{const-W}) } }{\alpha
\sqrt{B } } \|r_{\rm eff } (\mu )^{1/2 } (h_{00 } +
\alpha ^2 )^{1/2 } \|^2 .
\end{equation}
We will now estimate the norm on the right hand side, for suitably
chosen $\mu $.

\begin{lemma} \label{FMT':mu} Assume $B\ge e^2$. Define
\begin{equation} \label{FMT'mu1}
\mu _{\rm eff } = \mu _{\rm eff } (N, Z, \mathbb{M } ) := -
\frac{\alpha ^2 }{2 } \left( \frac{N }{2 \varepsilon_{\rm eff}^2 } + 1
\right) ,
\end{equation}
where $\alpha $ is as in Theorem \ref{FMT}, and where
$\varepsilon=\varepsilon_{\rm eff}$ is the unique positive
solution to the equation (\ref{FMT':eps}). Then $\mu _{\rm eff } <
\inf\sigma(h_{\rm eff})$, and
\begin{equation} \label{FMT':mu2}
\| r_{\rm eff } (\mu _{\rm eff } )^{1/2 } (h_{00 } +
\alpha ^2 )^{1/2 } \| \leq \sqrt{2 } .
\end{equation}
\end{lemma}

Assuming the lemma for the moment, we continue with the proof of
Theorem \ref{FMT'}: we have, by (\ref{FMT':5}), (\ref{FMT':6}) and
(\ref{FMT':mu2}), that if $ d  _{\rm eff } (\xi ) \geq c_{\# }
\alpha / \sqrt{B } $, then
\begin{eqnarray*}
\| \; K_{\rm eff } (\xi ) \; \| &\leq & 2 \max \left \{
\frac{\alpha ^2 }{2 } \left( \frac{N }{2 \varepsilon_{\rm eff}^2 }
+ 1 \right) \frac{1 }{ d  _{\rm eff } (\xi ) } , \; 1 \right \} \;
\frac{C_{(\ref{const-W}) } }{\alpha \sqrt{B } } \le \frac{1 }{2 }
,
\end{eqnarray*}
provided that $c_{\# } $ satisfies
$$
c_{\# } \geq 2 \left( \frac{N }{2\varepsilon_{\rm eff}^2 } + 1 \right)C_{(\ref{const-W}) } ,
\ \mbox{and } \ \frac{C_{(\ref{const-W}) } }{\alpha \sqrt{B } }
\leq \frac{1 }{4 } .
$$
Since $\alpha (B ) \sqrt{B } $ is increasing, and since $\alpha (B
) \sqrt{B } = x $ iff $ B = x^2 /(4 \alpha (x / 4 )^2) $, the last
inequality is implied by $B \geq B_{(\ref{FMT':B}) } $. Choosing
$c_{\# } = c_{\rm eff } $ defined by (\ref{FMT':c-eff}), we
conclude that if $\xi $ is such that $ d  _{\rm eff } (\xi ) \geq
c_{\rm eff } \alpha / \sqrt{B } $, then by (\ref{FMT':2} ) ,
(\ref{FMT':5}), (\ref{FMT':6}) and our choice of $\mu = \mu _{\rm
eff } $,
\begin{eqnarray*}
\| R_{\rm eff} ^{\mathcal{W } } - r_{\rm eff } (\xi ) \| &\leq &
\frac{4 }{ d  _{\rm eff } (\xi ) } \; \max \left \{ \frac{\alpha
^2 }{2 } \left( \frac{N }{2 \varepsilon_{\rm eff}} ^2 + 1 \right)
\frac{1 }{ d  _{\rm eff } (\xi ) } , \; 1 \right \}
\frac{C_{(\ref{const-W})
} }{\alpha \sqrt{B } } \\
&\leq & 2 C_{(\ref{const-W}) } \left( \frac{N }{2 \varepsilon_{\rm
eff}^2 } + 1 \right) \frac{\alpha }{ d  _{\rm eff } (\xi )^2
\sqrt{B } } ,
\end{eqnarray*}
provided that $ d  _{\rm eff } (\xi ) \leq \alpha ^2 / 2 $. This
proves Theorem \ref{FMT'}, modulo that of lemma \ref{FMT':mu}.
\hfill QED
\medskip

\noindent {\it Proof of lemma \ref{FMT':mu}.} We will use a
scaling argument. If we let $\simeq $ denote the unitary
equivalence induced by the change of variables $z \to z / \alpha
$, where $\alpha > 0 $ is for the moment a free
parameter, and if we write $\mathcal{V }_{\rm eff } ^B $ for
$\sqrt{B } \mathcal{V }_{\rm eff } ^1 (\sqrt{B } \cdot ) $, then
if $\mu < \inf \sigma (h_{\rm eff } ) $,
\begin{eqnarray} \nonumber
0 &\leq & (h_{00 } + \alpha ^2 )^{1/2 }r_{\rm eff }
(\mu ) (h_{00 } +
\alpha ^2 )^{1/2 } \\
&\simeq & (h_{00 } + 1 )^{1/2 } \left( -\frac{1 }{2 } \Delta _z +
\alpha ^{-2 } \mathcal{V } _{\rm eff } ^B (\tfrac{\cdot }{\alpha }
) - \alpha^{-2}\mu
\right)^{-1 } (h_{00 } + 1 )^{1/2 } \nonumber \\
&\leq & (h_{00 } + 1 )^{1/2 } \left( -\frac{1 }{2 } \Delta _z +
\alpha ^{-2 } \mathcal{V } _{{\rm eff }, n } ^B (\tfrac{\cdot
}{\alpha } ) - \alpha^{-2}\mu \right)^{-1 } (h_{00 } + 1 )^{1/2 }
, \label{FMT':7}
\end{eqnarray}
$\mathcal{V }_{{\rm eff }, n } ^1 $ being the
attractive part of  $\mathcal{V } _{\rm eff } $. We
now choose $\alpha = \log (\sqrt{B } / \alpha ) $, as in
(\ref{FMT:alpha}), and put $\lambda := \sqrt{B } / \alpha $.
Notice that $B\ge e^2$ implies that $\lambda\ge e$. Then, by lemma
\ref{AsympPot1a},
\begin{eqnarray*}
\frac{\sqrt{B } }{\alpha ^2 } \mathcal{V } ^1 _{ {\rm eff }, n } \left(
\frac{\sqrt{B } }{\alpha } z \right) &=& - Z
\sum _{j = 1 } ^N
\frac{\lambda }{\log \lambda } V_j ^1 (\lambda z_j ) \\
&\geq & -  2C_{(\ref{AsympPot1a:c})} ZC_{V^1_1 } \;
\varepsilon (|\log \varepsilon | + 2 ) \left( -\frac{1 }{2 }
\Delta _z + \frac{N }{2 \varepsilon ^2 } \right),\qquad
\varepsilon>0 \\
&=&  -\frac{1 }{2 } h_{00 } + \frac{N }{4 \varepsilon _{\rm
eff } } ,
\end{eqnarray*}
uniformly in $\lambda $, if we choose $
\varepsilon:=\varepsilon_{\rm eff}= \varepsilon_{\rm eff}(Z ,
\mathbb{M } ) $ such that (\ref{FMT':eps}) holds. We now take
$$
\mu = \mu _{\rm eff } = - \frac{\alpha ^2 }{2 } \left( \frac{N }{2
\varepsilon_{\rm eff}^2 } + 1 \right) .
$$
Then it follows that $\mu _{\rm eff } < \inf \sigma (h
 _{\rm eff} ) $, since $h_{\rm eff} -\mu_{\rm
eff } \geq \tfrac{1 }{2 } (h_{00 } + 1 ) $ $\geq {1 \over 2 } $.
Furthermore,  (\ref{FMT':7}) can be estimated by
$$
(h_{00 } + 1 )^{1/2 } \left( \frac{1 }{2 } h_{00 } - \left(
\frac{N }{4 \varepsilon_{\rm eff}^2 } + \frac{\mu _{ \rm eff
 } }{\alpha ^2 } \right) \right)^{-1 } (h_{00 } + 1 )^{1/2 }
= 2 ,
$$
which implies (\ref{FMT':mu2}). \hfill QED
\medskip

\noindent {\it Proof of Theorem \ref{FMT}}. It now suffices to
combine Theorem \ref{ME} and Theorem \ref{FMT'}, while carefully
keeping track of the constants. First of all, (\ref{FMT':0.5}),
implies that $\| R_{\rm eff } ^{\mathcal{W } } \| \leq 2 \|r_{\rm
eff } (\xi ) \| $, and therefore $d_{\rm eff}^{\mathcal{W } } (\xi
) \geq d _{\rm eff } (\xi ) / 2 $. Furthermore, $\xi < 0 $ if
$d_{\rm eff } (\xi ) > 0 $, and if
\begin{equation} \label{FMT:B-eff}
B \geq B_{\rm eff } := \max \{ B_{(\ref{constant-B})}, B_{(\ref{FMT':B}) } , e^2 \} ,
\end{equation}
then
\begin{eqnarray*} %\label{PfFMT:1}
& & \|  ( H- \xi )^{-1 }  - (h_{\rm eff } - \xi
)^{-1 }
\oplus (H_\perp-\xi)^{-1} \|\\
& \leq & \| (H - \xi )^{-1 } - R_{\rm eff } ^{\mathcal W } (\xi
)\oplus (H_\perp-\xi)^{-1} \| + \| R_{\rm eff } ^{\mathcal W }
(\xi )-r_{\rm eff } (\xi ) \| \nonumber\\
&\leq& \frac{C_{(\ref{constant-C}) } }{\sqrt{B } } \frac{1
}{d_{\rm eff }^{\mathcal{W} } (\xi ) }
 + c_{\rm eff } \frac{ \alpha }{d_{\rm eff } (\xi
)^2 \sqrt{B } } \le\left(2 C_{(\ref{constant-C})} + c_{\rm
eff}{\alpha\over d_{\rm eff}(\xi ) }\right) {1\over
d_{\rm eff}(\xi ) \sqrt{B}}\quad (\mbox{using $d_{\rm
eff} (\xi )/2 \leq d_{\rm eff}^\mathcal{W}(\xi)$  } )\nonumber
\\ &\leq & \left( C_{(\ref{constant-C}) }  +
\frac{c_{\rm eff } }{\alpha } \right) \frac{\alpha ^2 }{d_{\rm eff
} (\xi ) ^2 \sqrt{B } }\quad (\mbox{using $d_{\rm eff
} (\xi ) \leq \alpha ^2 / 2 $ again } ) \nonumber \\
&\leq & \left( C_{(\ref{constant-C}) } + \frac{c_{\rm eff }
}{\alpha (B_{\rm eff } ) } \right) \; \frac{\alpha ^2 }{d_{\rm eff
} (\xi ) ^2 \sqrt{B } } ,
\end{eqnarray*}
since $\alpha (B )^{-1 } $ is a decreasing function of $B$.
This proves Theorem \ref{FMT} with a constant $C_{\rm eff }
$ which is equal to
$$
C_{\rm eff } := C_{(\ref{constant-C}) } + \frac{c_{\rm eff }
}{\alpha (B_{\rm eff } ) }  . \qquad {\rm QED}$$

\section{{\bf Proof of Theorem \ref{SMT} } }
As a first step, we will compare the resolvents $r
_{\rm eff } (\xi ) $ of $h_{\rm eff } $ and $ r_C
(\xi ) := (h_C - \xi )^{-1 } $ of $h_C := h_C ^B $. Recall, that
$d_C (\xi ) := \mbox{\dist } (\xi , \sigma (h_C ) ) . $

\begin{theorem} \label{SMT'} Let $\alpha = \alpha (B) $ be defined by
(\ref{FMT:alpha}). There exist (computable) constants $B_C ' , C_C
' \geq 0 $, only depending on $Z, N $ and $\mathbb{M } $, such
that for all $B \geq B_C ' $ and all real  $\xi \leq 0$ satisfying
\begin{equation} \label{SMT':i}
d_C (\xi ) \geq C_C ' \; \alpha ^{3/2 } B^{-1/4 } ,
\end{equation}
then $\xi \in \rho (H_{\rm eff } ) $, with $\|r_{\rm eff } (\xi ) \| \leq 2 \|
r_C (\xi ) \| $. In addition, letting\footnote{with $\varepsilon _{\rm eff } $
defined in (\ref{FMT':eps})}
\begin{equation} \nonumber
C_C '' = \max\{ C'_C , 4 C_{(\ref{ConstV-eff}) } \left( {N\over 2
\varepsilon_{\rm eff }^2}  + 1 \right) \} \geq C'_C ,
\end{equation}
then if
\begin{equation} \label{SMT':iii}
d_C (\xi ) \geq C_C '' \; \alpha ^{3/2 } B^{-1/4 },
\end{equation}
we also have that $\|r_C (\xi ) \| \leq 2 \|r_{\rm
eff } (\xi ) \| $.
\medskip

\noindent Finally, if
\begin{equation} \label{SMT':i-bis}
C_C ' \; \alpha ^{3/2 } B^{-1/4 } \leq d_C (\xi ) \leq \frac{1 }{2
} \alpha ^2 ,
\end{equation}
then
\begin{equation} \label{SMT':ii}
\|r _{\rm eff } (\xi ) - r_C (\xi ) \| \leq C_C '
\frac{\alpha ^{3/2 } }{d_C (\xi ) ^2  B^{1/4 } } .
\end{equation}
\end{theorem}
\medskip

\noindent {\it Proof.} The proof is similar to the proof of
Theorem~\ref{FMT'}, with however some technical changes, due to
the fact that $v_C $ is not homogeneous of degree $-1$, and that
its electron-electron part is not positive anymore. As
before, we conjugate all operators by $U_{xy } ^B $, keeping the
same letters for the conjugated operators.

Arguing as in the proof of Theorem~\ref{FMT'}, one shows that
\begin{equation} \nonumber
\|r  _{\rm eff } (\xi ) - r_C (\xi ) \| \leq \frac{1 }{d_C (\xi )
} \; \frac{\| K_C (\xi ) \| }{1 - \| K_C (\xi ) \| } ,
\end{equation}
where $K_C := |r_C (\xi ) |^{1/2 } (h_{\rm eff} -
h_C ) r_C (\xi )^{1/2 } . $ Using lemma \ref{AsympPot2}, we find
that,  for any $\mu < \inf \sigma (h_C ) $,
\begin{equation} \label{SMT':2}
\| K_C (\xi ) \| \leq \frac{C_{(\ref{ConstV-eff}) } }{B^{1/4 }
\sqrt{\alpha } } \; \max \left \{ \frac{|\mu  | }{d_C (\xi ) } , 1
\right \} \;  \|r_C (\mu )^{1/2 } (h_{00 } + \alpha ^2 )^{1/2 }
\|^2 .
\end{equation}
We then use the following analogue of lemma \ref{FMT':mu}, of
which we only state a qualitative version.

\begin{lemma} \label{SMT':mu} There exists a constant $\nu_C = \nu_C
(Z, N, \mathbb{M } ) \geq 1/2 $ such that if  $B \geq e$, and if
$\mu _C := -\nu _C \alpha ^2 $, $\alpha $ defined by
(\ref{FMT:alpha}), then $\mu _C < \inf \sigma (h_C ) $, and
\begin{equation} \label{SMT':mu1}
\|r_C (\mu_C )^{1/2 } (h_{00 } + \alpha ^2 )^{1/2 } \|^2 \leq 2 .
\end{equation}
\end{lemma}

\noindent {\it Proof.}  As before,  we will use
scaling. However,  contrary to $\delta (z)  $, the
distribution $\mbox{Pf}(1 / |z | ) $ is {\it } not homogeneous of
degree $-1$ on $\mathbb{R } $. In fact, if $\rho _{\alpha } $ is
the dilation $\rho _{\alpha } (z) = \alpha^{-1 } z $ on $\mathbb{R
} $, $\alpha
> 0 $ arbitrary, then the pullback of $\mbox{Pf }(1 / | \cdot | $
by $\rho _{\alpha } $ equals
\begin{equation} \label{DilPf}
\rho _{\alpha } ^*  \mbox{Pf} \left( \frac{1 }{|\cdot | } \right)
= \alpha \mbox{ Pf } \left( \frac{1 }{|\cdot | } \right) - 2
\alpha \log \alpha \; \delta .
\end{equation}
Let us split the potential $v_C := v_C ^B $ of $h_C $ as
$$
v_C = \log B \ v_{\delta } + v_Q ,
$$
where $v_\delta$ is defined in (\ref{Hdelta2}) and
\begin{equation} \nonumber
v_Q = -Z \sum _j \left( \mbox{Pf } \frac{1 }{|z_j | } + C_j ^n
\delta (z_j ) \right) + \sum _{j < k } \left( \mbox{Pf } \frac{1
}{|z_j - z_k | } + C_{jk } ^e \delta (z_j - z_k ) \right) ,
\end{equation}
the (pseudo-) Coulombic part. If $\simeq $ denotes unitary
equivalence with respect to the dilation $\rho _{\alpha } $ (on
$\mathbb{R }^N $), then in view of our choice of $\alpha $
\begin{eqnarray*}
h_C-\mu &\simeq& \alpha ^2 \left( h_{00 } + \frac{\log B - 2 \log
\alpha }{\alpha } v_{\delta } + \frac{1 }{\alpha } v_Q
-\alpha^{-2}\mu\right)= \alpha^2(h_{00}+2v_\delta+
\frac{1 }{\alpha } v_Q-\alpha^{-2}\mu) \\
&\ge&\alpha^2\left({1\over2}h_{00}-b-\alpha^{-2}\mu\right)
\end{eqnarray*}
form some $b > 0 $  depending only on $Z, N $ and $\mathbb{M
} $,  since by lemma~\ref{relformbound} we know that
$2v_\delta+ \frac{1 }{\alpha } v_Q$ is $h_{00}$ form bounded with
relative bound 0. Recall that $B\ge e$ implies $\alpha\ge1$.
Choosing $\mu=\mu_C:=-\alpha^2({1\over2}+b) =: -\alpha^2\nu_c $
will insure that $(h_{00 } + \alpha ^2 ) r_C (\mu ) (h_{00 } +
\alpha ^2 )^{1/2 }\le2 $ which is what we want to prove. A more
careful argument, which we will skip, will yield an explicit $b .
$ \hfill QED
\medskip

We continue with the proof of Theorem~\ref{SMT'}. By
(\ref{SMT':mu1}) and (\ref{SMT':2}) with $\mu = \mu _C $, we find
that
$$
\| K_C (\xi ) \| \leq  \frac{2 C_{(\ref{ConstV-eff}) } }{B^{1/4 }
\sqrt{\alpha } } \; \max \left \{ \frac{\nu _C \alpha ^2 }{d_C
(\xi ) } , 1 \right \} \leq {1\over2} ,
$$
if both
$$
d_C (\xi ) \geq 4 C_{(\ref{ConstV-eff}) } \nu_C \; \frac{\alpha
^{3/2 } }{B^{1/4 } } =: C_C ' \; \frac{\alpha ^{3/2 } }{B^{1/4 } }
,
$$
and $\alpha ^{1/2 } B^{1/4 } \geq 4 C_{(\ref{ConstV-eff}) } $
which, since $B \mapsto \alpha (B )^2 B $ is increasing, is
equivalent to $B\ge 4^3 C_{(\ref{ConstV-eff})
}^4\alpha(4C_{(\ref{ConstV-eff}) } ^{2} )^{-2 } . $
Since we also need $B\ge e$ we  put
$$
 B_C ' := \max\left\{\frac{4^3 C_{(\ref{ConstV-eff}) }^4 }{4 \alpha\!\!
\left( C_{(\ref{ConstV-eff}) } ^2 \right) ^2 },\,e\right\} .
$$
This fixes our constants $C_C ' $ and $B_C ' $, and also implies
that $\|r_{\rm eff } (\xi ) \| \leq 2 \| r_C (\xi )
\| $, by the resolvent formula.
\medskip

To show that (\ref{SMT':iii}) implies that $\| r_C (\xi ) \| \leq
2 \|r_{\rm eff } (\xi ) \| $, we repeat the
argument with $r_C $ and $ r _{\rm eff } $
interchanged: by the resolvent formula,
$$
r_C (\xi ) = r_{\rm eff } (\xi ) ^{1/2 } \left( 1 +
\widetilde{K }_C (\xi ) \right) ^{-1 } |r_{\rm eff
} (\xi ) | ^{1/2 } ,
$$
with
$$
\widetilde{K }_C (\xi ) := |r_{\rm eff } (\xi ) |
^{1/2 } (h_C -h_{\rm eff } ) r_{\rm
eff } (\xi )^{1/2 } ,
$$
so that, using lemmas \ref{AsympPot2}, \ref{FMT':lemmeI} and
\ref{FMT':mu}, we arrive at
$$
\| \widetilde{K }_C (\xi ) \| \leq \frac{2C_{(\ref{ConstV-eff}) }
}{B^{1/4 } \sqrt{\alpha } } \; \max \left \{ \frac{|\mu _{\rm eff
} | }{ d  _{\rm eff } (\xi ) } , 1 \right \} \leq
\frac{2C_{(\ref{ConstV-eff}) } }{B^{1/4 } \sqrt{\alpha } } \; \max
\left \{ \frac{2 |\mu _{\rm eff } | }{d_C (\xi ) } , 1 \right \} ,
$$
if $d_C (\xi ) \geq C'_C \alpha ^{3/2 } B^{-1/4 } $, by the first
part. We therefore conclude that $\| \widetilde{K }_C (\xi ) \|
\leq 1/2 $, and hence $\| r_C (\xi ) \| \leq 2 \| r_{\rm
eff } (\xi ) \| $,  if both $\sqrt{\alpha } B^{1/4 } \geq 4
C_{(\ref{ConstV-eff}) } $, which will be satisfied if $B \geq B_C
' $ defined above, and if
$$
\frac{4 |\mu _{\rm eff } | C_{(\ref{ConstV-eff}) } }{\sqrt{\alpha
} B^{1/4 } d_C (\xi ) } \le \frac{1 }{2 } .
$$
The latter inequality is equivalent to
$$
d_C (\xi ) \geq 4 \left(  \frac{N }{2 \varepsilon _{\rm eff } ^2 }
+ 1 \right) \; C_{(\ref{ConstV-eff}) } \; \frac{\alpha ^{3/2 }
}{B^{1/4 } } ,
$$
which yields condition (\ref{SMT':iii}) .
\medskip

Finally, if $\xi $ satisfies (\ref{SMT':i-bis}), then
\begin{eqnarray*}
\|r_{\rm eff }(\xi) - r_C(\xi) \| &\leq & \frac{4
C_{(\ref{ConstV-eff}) } }{\alpha ^{1/2 } B^{1/4 } d_C (\xi ) }
\max \left \{ \frac{\nu _C
\alpha ^2 }{d_C (\xi ) } , 1 \right \} \\
&\leq & 4 \nu _C C_{(\ref{ConstV-eff}) } \frac{\alpha ^{3/2 }
}{d_C (\xi )^2 B^{1/4 } } \\
& = & C_C ' \frac{\alpha ^{3/2 } }{d_C (\xi )^2 B^{1/4
} } ,
\end{eqnarray*}
where we used that $d_C (\xi ) \leq \alpha ^2 / 2 \leq \nu _C
\alpha ^2 $. This finishes the proof of Theorem~\ref{SMT'}. \hfill
QED
\medskip

\noindent {\it Proof of Theorem~\ref{SMT}.} We define
\begin{equation} \label{PfSMT:1}
B_C := \max \{ B_{\rm eff } , B_C ' \} ,
\end{equation}
and
\begin{equation} \label{PfSMT:2}
c_C := \max \{ C_c '' , 2 c_{\rm eff } \alpha (  B_C  )^{-1/2 }
B_C ^{-1/4 } \} .
\end{equation}
Suppose that $B \geq B_C $, and that
$$
 c_C  \alpha ^{3/2 } B^{-1/4 } \leq d_C (\xi ) \leq
\frac{1 }{4 } \alpha ^2 .
$$
 By Theorem~\ref{SMT'}, $d_{\rm eff } (\xi ) \leq 2 d_C
(\xi ) \leq \alpha ^2 / 2 $. By the same theorem, and by
(\ref{PfSMT:2}),
$$
d_{\rm eff } (\xi ) \geq \frac{1 }{2 } d_C (\xi ) \geq c_{\rm eff
} \alpha ( B_C )^{-1/2 } B_C ^{-1/4 } \alpha^{3\over2}B^{-1/4}
 \geq c_{\rm eff } \alpha (B ) B^{-1/2 } ,
$$
since $\alpha (B ) / \sqrt{B } $ is a decreasing function of $B >
0 $. The conditions of Theorem ~\ref{FMT} are therefore met. Since
the condition (\ref{SMT':i-bis}) of Theorem~\ref{SMT'} is clearly
also satisfied, we conclude that the difference of the resolvents
(\ref{SMT:DiffRes}) can be estimated by
$$
C_{\rm eff } \frac{\alpha ^2 }{d_{\rm eff } (\xi )^2 \sqrt{B } } +
C_C ' \frac{\alpha ^{3/2 } }{d_C (\xi ) ^2 B^{1/4 } } \leq C_c
\frac{\alpha ^{3/2 } }{d_C (\xi )^2 B^{1/4 } } ,
$$
for $B \geq B_C $, with
\begin{equation} \label{PfSMT:3}
C_C := 4 C_{\rm eff } \frac{\sqrt{\alpha ( B_C  ) } }{B_C^{1/4 } }
+ C_C ' ,
\end{equation}
where we used that $\alpha (B ) / \sqrt{B } $ is decreasing.
\hfill QED

%%%%%%%%%%%%%%%%%%%%%%%%%%%%%%%%%%%%%%%%%%%%%%%%%%%
\section{{\bf Proof of Theorem \ref{CorrSMT} } }%
%%%%%%%%%%%%%%%%%%%%%%%%%%%%%%%%%%%%%%%%%%%%%%%%%%
This will be done by closely following the strategy of section 6.
First, we compare the resolvents $r_{\rm eff } (\xi
) $ of $h_{\rm eff } $ and $ r_\delta (\xi ) :=
(h_\delta - \xi )^{-1 } $ of $h_\delta := h_\delta ^{B, \mathbb{M
} } $. Recall, that $d_\delta (\xi ) := \mbox{\dist } (\xi ,
\sigma (h_\delta ) ) . $
\begin{theorem} \label{SMT''} Let $\alpha = \alpha (B) $ be defined by
(\ref{FMT:alpha}). There exist (computable) constants $B_\delta' ,
C_\delta ' \geq 0 $, only depending on $Z, N $ and $\mathbb{M } $,
such that for all $B \geq B_\delta ' $ and all real  $\xi $
satisfying
\begin{equation} \label{SMT'':i}
d_\delta (\xi ) \geq C_\delta ' \; \alpha ,
\end{equation}
we have that $\xi \in \rho (h_{\rm eff } ) $, with
$\| r_{\rm eff } (\xi ) \| \leq 2 \| r_\delta (\xi )
\| $. In addition, letting
\begin{equation} \label{ConstVeff-Vdelta}
C_\delta'' = \max\{ C'_\delta , 4 C_{(\ref{ConstVeff-Vdelta}) }
\left( {N\over 2 \varepsilon_{\rm eff }^2}  + 1 \right) \}
\quad{\rm with}\quad
C_{(\ref{ConstVeff-Vdelta})}:=\left(NZ+{N(N-1)\over2}\right)C_{(\ref{supVchapMoinsDelta})}
,
\end{equation}
and $\varepsilon _{\rm eff } $ given by (\ref{FMT':eps}), then if
\begin{equation} \label{SMT'':i-bis}
d_\delta (\xi ) \geq C_\delta '' \; \alpha ,
\end{equation}
we also have that $\|r_\delta (\xi ) \| \leq 2 \|r_{\rm eff } (\xi ) \| $.
Finally, if
\begin{equation} \label{SMT'':iii}
C_\delta ' \; \alpha \leq d_\delta (\xi ) \leq \frac{1 }{2 }
\alpha ^2 ,
\end{equation}
then
\begin{equation} \label{SMT'':ii}
\|r_{\rm eff } (\xi ) - r_\delta (\xi ) \| \leq
C_\delta ' \frac{\alpha}{d_\delta (\xi ) ^2} .
\end{equation}
\end{theorem}
\medskip
\noindent {\it Proof.}

As in the proof of Theorem~\ref{FMT'}, \ref{SMT'} one shows that
\begin{equation} \nonumber
\| r_{\rm eff } (\xi ) - r_\delta (\xi ) \| \leq
\frac{1 }{d_\delta (\xi ) } \; \frac{\| K_\delta (\xi ) \| }{1 -
\| K_\delta (\xi ) \| } ,
\end{equation}
where $K_\delta := |r_\delta (\xi ) |^{1/2 } (h_{\rm
eff} - h_\delta ) r_\delta (\xi )^{1/2 }$. Using lemma
\ref{vdelta-Est} with $c = 2 $, lemma \ref{FE-special'},
the triangle inequality, and similar comparison arguments as in
the proofs of Theorems \ref{FMT'} and \ref{SMT'}, we easily
find that, for any $\mu < \inf \sigma (h_\delta ) $,
\begin{equation} \label{SMT'':2}
\| K_\delta (\xi ) \| \leq \frac{C_{(\ref{ConstVeff-Vdelta}) }
}{\alpha  } \; \max \left \{ \frac{|\mu  | }{d_\delta (\xi ) } , 1
\right \} \;  \|r_\delta (\mu )^{1/2 } (h_{00 } + \alpha ^2 )^{1/2
} \|^2 ,
\end{equation}
where $C_{(\ref{ConstVeff-Vdelta}) } $ was defined above. We then
use the following analogue of lemmas \ref{FMT':mu} and
\ref{SMT':mu}.

\begin{lemma} \label{SMT'':mu}  Let $\nu_\delta =1/2+4NZ^2$, $\mu _\delta := -\nu _\delta \alpha ^2
$ and let $\alpha $ be defined by (\ref{FMT:alpha}). Then $\mu
_\delta < \inf \sigma (h_\delta ) $, and
\begin{equation} \label{SMT'':mu1}
\|r_\delta (\mu_\delta )^{1/2 } (h_{00 } + \alpha ^2 )^{1/2 } \|^2
\leq 2 .
\end{equation}
\end{lemma}

\noindent {\it Proof.} We will use , as before, the scaling
$z\mapsto z/\alpha$. We get, with the help of lemma
\ref{relformbound},
\begin{eqnarray*}
h_\delta-\mu&\simeq& \alpha^2(h_{00}-2v_\delta-\alpha^{-2}\mu) \ge
\alpha^2(h_{00}-2Z\sum_{j=1}^N\delta(z_j)-\alpha^{-2}\mu)\\
&\ge&\alpha^2(
h_{00}(1-2Z\varepsilon)- N\varepsilon^{-1}Z-\alpha^{-2}\mu)\qquad(\varepsilon>0)\\
&=&{1\over2}\alpha^2(h_{00}+1) ,
\end{eqnarray*}
if we choose $\varepsilon=\varepsilon_\delta:=1/(4Z)$ and
$\mu=\mu_\delta:=-\alpha^2({1\over2}+4NZ^2)$. \hfill QED
\medskip

We continue with the proof of Theorem~\ref{SMT''}. By
 (\ref{SMT'':2}) and (\ref{SMT'':mu1}) with $\mu =
\mu_\delta $, we find that
$$
\| K_\delta (\xi ) \| \leq  \frac{2 C_{(\ref{ConstVeff-Vdelta}) }
}{\alpha } \; \max \left \{ \frac{\nu_\delta \alpha ^2 }{d_\delta
(\xi ) } , 1 \right \} \leq {1\over2} ,
$$
if both
$$
d_\delta (\xi ) \geq 4 C_{(\ref{ConstVeff-Vdelta}) } \nu_\delta \;
\alpha
 =: C_\delta ' \; \alpha,
$$
and $\alpha \geq 4 C_{(\ref{ConstVeff-Vdelta}) } $ which, since $B
\mapsto \alpha (B )$ is increasing, is equivalent to
$$
B \geq B_\delta' := 16 C_{(\ref{ConstVeff-Vdelta})}^2e^{ 8
C_{(\ref{ConstVeff-Vdelta}) } }.
$$
This fixes our constants $C_\delta' $ and $B_\delta' $, and also
implies that $\|r_{\rm eff } (\xi ) \| \leq 2 \|
r_\delta (\xi ) \| $, by the resolvent formula. To show that
(\ref{SMT'':i-bis}) implies that $\| r_\delta (\xi ) \| \leq 2 \|
r_{\rm eff } (\xi ) \| $, we repeat the argument
with $r_\delta $ and $r_{\rm eff } $ interchanged:
by the resolvent formula,
$$
r_\delta (\xi ) =r_{\rm eff } (\xi )^{1/2 } \left(
1 + \widetilde{K }_\delta (\xi ) \right) ^{-1 } |r_{\rm eff } (\xi ) | ^{1/2 } ,
$$
with
$$
\widetilde{K }_\delta (\xi ) := |r_{\rm eff } (\xi )
| ^{1/2 } (h_\delta - h_{\rm eff } )  r_{\rm eff } (\xi )^{1/2 } ,
$$
so that, using lemmas \ref{vdelta-Est}, \ref{FMT':lemmeI} and
\ref{FMT':mu}, we arrive at
$$
\| \widetilde{K }_\delta (\xi ) \| \leq  2\max \left \{ \frac{|\mu
_{\rm eff } | }{ d  _{\rm eff } (\xi ) } , 1 \right \}
{C_{(\ref{ConstVeff-Vdelta}) }\over\alpha}\leq  2\max \left \{
\frac{2 |\mu _{\rm eff } | }{d_\delta (\xi ) } , 1 \right
\}{C_{(\ref{ConstVeff-Vdelta}) }\over\alpha} ,
$$
if $d_\delta (\xi ) \geq C'_\delta \alpha$, by the first part. We
therefore conclude that $\| \widetilde{K }_\delta (\xi ) \| \leq
1/2 $ if both $\alpha\geq 4 C_{(\ref{ConstVeff-Vdelta}) } $, which
is satisfied since $B \geq B_\delta ' $ defined above, and if
$$
\frac{4 |\mu _{\rm eff } | C_{(\ref{ConstVeff-Vdelta}) } }{\alpha
d_\delta (\xi ) } \le \frac{1 }{2 }.
$$
The latter inequality is equivalent to
$$
d_\delta (\xi ) \geq 4 \left(  \frac{N }{2 \varepsilon _{\rm eff }
^2 } + 1 \right) \; C_{(\ref{ConstVeff-Vdelta}) } \; \alpha ,
$$
which yields condition (\ref{SMT'':i-bis}) .
\medskip
Finally, if $\xi $ satisfies (\ref{SMT'':iii}), then
\begin{eqnarray*}
\|r_{\rm eff }(\xi) - r_\delta(\xi) \| &\leq &
\frac{4 C_{(\ref{ConstVeff-Vdelta}) } }{\alpha d_\delta (\xi ) }
\max \left \{ \frac{\nu _\delta
\alpha ^2 }{d_\delta (\xi ) } , 1 \right \} \\
&= & 4 \nu_\delta C_{(\ref{ConstVeff-Vdelta})}
{\alpha\over d_\delta (\xi )^2  } =C_\delta ' \frac{\alpha
}{d_\delta (\xi )^2} ,
\end{eqnarray*}
where we used that $d_\delta (\xi ) \leq \alpha ^2 / 2 \leq
\nu_\delta \alpha ^2 $. This finishes the proof of Theorem~\ref{SMT''}. \hfill
QED

\medskip
\noindent {\it Proof of Theorem~\ref{CorrSMT}.} We want to realize
the assumptions of Theorem~\ref{FMT} and \ref{SMT''} with
conditions on $B$ and $d_\delta$. We define
$$
B_\delta := \max \{ B_{\rm eff } , B_\delta' \} ,
$$
and
\begin{equation} \label{cdelta}
c_\delta := \max \{ C_\delta '' , 2 c_{\rm eff } B_\delta ^{-1/2 }
\} .
\end{equation}
Clearly $B \geq B_\delta $ and $ c_\delta \alpha \leq d_\delta
(\xi ) \leq \alpha ^2/4  $ will do, for under these conditions,
using Theorems~\ref{FMT} and \ref{SMT''}, the left hand side of
(\ref{4.11}) can be estimated by
$$
{C_{\rm eff} \alpha^2\over d_{\rm
eff}^2(\xi)\sqrt{B}}+{C'_\delta\alpha\over d_\delta^2(\xi)} \le
\left({4C_{\rm eff} \alpha(B_\delta)\over \sqrt{B_\delta}}
+C'_\delta\right){\alpha\over d_\delta(\xi)^2}=:C_\delta
{\alpha\over d_\delta(\xi)^2}
$$
since we know that both $2d_{\rm eff}  (\xi )\ge d_\delta(\xi)$ x

and $2 d_{\delta } (\xi ) \ge  d_{\rm eff } (\xi ) $, by
Theorem~\ref{SMT''}, and since $\alpha (B ) / \sqrt{B } $ is a
decreasing function of $B > 0 $. \hfill QED

%%%%%%%%%%%%%%%%%%%%%%%%%%%%%%%%%%%%%%%%
\section{{\bf The fermionic case } }   %
%%%%%%%%%%%%%%%%%%%%%%%%%%%%%%%%%%%%%%%%

We first prove Theorem~\ref{FermionizedMThms}. This is simply done
by repeating the proofs of Theorems~\ref{FMT}, \ref{SMT} and
\ref{CorrSMT} for the `fermionized' operators, that is, for the
operators sandwiched between $P^{AS } $. We have to check that the
main ingredients of these proofs remain valid. First of all,
corollary \ref{ME:Fermi} is used to compare the resolvents of
$H^{B, \mathbb{M } } _{\rm f } $ and $H_{\rm eff, f} ^{B,
\mathbb{M } } + \mathcal{W }_f $. Next, lemma \ref{W-Est} remains valid for
$\mathcal{W }_f := P^{AS } \mathcal{W } P^{AS } $, with the
operator norm being the one on $P^{AS } \left( \mbox{Ran } \Pi
_{\rm eff } ^{B, \mathbb{M } } \right) $, since $P^{AS } $
commutes with $h_{00 } $ on $\mbox{Ran } \Pi _{\rm eff } ^{B,
\mathbb{M } } = L^2 (\mathbb{R }^N, F_{\mathbb{M } }^B ) $ (as we
will explicitly see below, $P^{AS } $ not only mixes the
coordinates of $\mathbb{R }^N $, but also the different components
with respect to the natural basis of $F_{\mathbb{M } }^B $;
however, $h_{00 } $ acts in a scalar way). We then repeat the
proof of Theorem~\ref{FMT} in section 5, replacing $d_{\rm eff } $
everywhere by $d_{\rm eff, f } $. Similar remarks apply to the
proofs of Theorems~\ref{SMT} and \ref{CorrSMT}. \hfill QED
\medskip

We next turn to Theorem~\ref{ThMT}. The parameter $B $ here plays
a non-essential r\^ole, and we will simply drop it, writing $X_m
$, $\chi _m $, $F_{\mathbb{M } } $ for $X^B _m $, $\chi ^B _m $,
$F_{\mathbb{M } }^B $, etc. We start by analyzing the subspace of
anti-symmetric wave functions in the range of $\Pi _{\rm eff } :=
\Pi _{\rm eff } ^B $. Recall that
$$
\Sigma (\mathbb{M } ) = \{ m = (m_1 , \cdots , m_N ) : m_j \geq 0
, m_1 + \cdots + m_N = \mathbb{M } \} .
$$
The permutation group $S_N $ acts on $\Sigma (\mathbb{M } ) $ by
$\sigma \cdot (m_1 , \cdots , m_N ) = (m_{\sigma (1) } , \cdots ,
m_{\sigma (N ) } ) $ and $\Sigma (\mathbb{M } ) $  can therefore
be written as a disjoint union of orbits of $S_N $ :
\begin{equation} \nonumber
\Sigma (\mathbb{M } ) = \bigcup _{\overline{m } \in \mathcal{M } }
S_N \cdot \overline{m } ,
\end{equation}
$\mathcal{M } \subset \Sigma (\mathbb{M } ) $ being a set of
representatives of $_{\mbox{\Large{$S_N $}}} \backslash \; \Sigma
(\mathbb{M } ) $.
%In
%particular, if $\overline{m } \neq \overline{m }' $ are both in
%$\mathcal{M } $, then $S_N \cdot \overline{m } \cap S_N \cdot
%\overline{m }' = \emptyset . $
If we let
\begin{equation} \label{F.2}
V_{\overline{m } } = \mbox{Span } \{ X_{\sigma \cdot \overline{m }
} : \sigma \in S_N \} ,
\end{equation}
then, recalling that $F_{\mathbb{M } } = \mbox{Span } \{ X_m : m
\in \Sigma (\mathbb{M } ) \} $, we have the orthogonal
decomposition
\begin{equation} \nonumber
F_{\mathbb{M } } = \bigoplus_{\overline{m } \in \mathcal{M } }
F_{\overline{m } } .
\end{equation}
From this it follows that
$$
\Pi_{\rm eff } ^{1, \mathbb{M } } \left( L^2 (\mathbb{R }^{3N })
\otimes \mathbb{C }^{2N } \right) = L^2 \left(\mathbb{R }^N
\right) \otimes F_{\mathbb{M } } = \bigoplus _{\overline{m } \in
\mathcal{M } } L^2 (\mathbb{R }^N ) \otimes F_{\overline{m } } .
\label{F.3}
$$
Since $P^{AS } $ leaves each $L^2 (\mathbb{R } ^N ) \otimes
F_{\overline{m } } $ invariant, it suffices to analyze the
subspace of anti-symmetric wave functions in each of the latter.
We therefore fix an $\overline{m } \in \mathcal{M } $ and let
\begin{equation} \label{F.4}
G_{\overline{m } } = \{ \sigma \in S_N : \sigma \cdot \overline{m
} = \overline{m } \} ,
\end{equation}
the stabilizer of $\overline{m } $. Choose representatives $\sigma
_1 , \cdots , \sigma _K $, $K = K(\overline{m } )$, for the right
equivalence classes of $G_{\overline{m } } $ in $S_N $ : $ S_N /
G_{\overline{m } } = \{ \sigma _1 G_{\overline{m } } , \cdots ,
\sigma _K G_{\overline{m } } \} $ with  $\sigma _i G_{\overline{m
} } \cap \sigma _j G_{\overline{m } } = \emptyset $ if $i \neq j
$. Then $X_{\sigma _1 \cdot \overline{m } } , \cdots , X_{\sigma_K
\cdot \overline{m } } $ constitutes an orthonormal basis for
$F_{\overline{m } } $, and each element $\psi = \psi (x, y, z ) $
of $L^2 (\mathbb{R }^N ) \otimes F_{\overline{m } } $ can be
uniquely written as:
\begin{equation} \label{F.5a}
\psi =  \sum _{j = 1 } ^K a_j  X_{\sigma _j \cdot \overline{m } }
= \sum _{j = 1 } ^K a_j (z) X_{\sigma _j \cdot \overline{m } } (x,
y ) ,
\end{equation}
for suitable $a_j = a_j (z) \in L^2 (\mathbb{R }^N ) $. For any
such $\psi $ and $r = (x, y, z ) \in \mathbb{R }^{3N } =
(\mathbb{R } ^N )^3 $,
\begin{eqnarray*}
\psi (\tau \cdot r )&=& \sum _j a_j(\tau \cdot z ) X_{\sigma _j
\cdot \overline{m } }
(\tau \cdot x, \tau \cdot y ) \\
&=& \sum _j a_j (\tau \cdot z ) X_{(\tau ^{-1 } \sigma _ j ) \cdot
\overline{m } } (x, y ),
\end{eqnarray*}
where $\tau \cdot r = (r_{\tau(1) } , \cdots , r_{\tau (N ) } ) $,
and similarly for  $\tau \cdot x $, $\tau \cdot y $ and $\tau
\cdot z $, and where we used that
$$
X_m (\tau \cdot x , \tau \cdot y ) = \prod _j \chi _{m_j }
(x_{\tau (j) } , y_{\tau (j) } ) = X_{\tau ^{-1 } \cdot m } (x , y
) .
$$
It follows that $\psi \in L^2 (\mathbb{R }^N ) \otimes
F_{\overline{m } } $ is anti-symmetric iff, for any $\tau \in S_N
$,
\begin{equation} \nonumber
\sum _j a_j (\tau \cdot z ) X_{(\tau ^{-1 } \sigma _j ) \cdot
\overline{m } } = (-1 )^{\tau } \sum _j a_j (z) X_{\sigma _j \cdot
\overline{m } } .
\end{equation}
This is equivalent to the statement that $\psi $ is antisymmetric
iff
\begin{equation} \label{F.6}
\sum _j a_j (\tau \cdot z ) X_{(\tau \sigma _j ) \cdot \overline{m
} } = (-1 )^{\tau } \sum _j a_j (z) X_{\sigma _j \cdot \overline{m
} } ,
\end{equation}
since the two statements are equivalent when $\tau $ is a
transposition, and these generate $S_N $. The version (\ref{F.6}),
with no $\tau ^{-1 } $, will be more convenient to work with. We
next observe that the map $\sigma _j \to \tau \sigma _j $ gives
rise to a a permutation $\rho (\tau ) $ of $S_K $:

%%%%%%%%%%%%%%%%%%%%%%%%%%%%%%%
\begin{lemma} \label{F.I}%
%%%%%%%%%%%%%%%%%%%%%%%%%%%%%%%
\rm{For any $\overline{m} \in \mathcal{M } $
the map
$\rho = \rho _{\overline{m } } : S_N \to S_K $, $K = K(\overline{m
} ) $, such that
$$
\rho (\tau ) (i) = j \Leftrightarrow \tau \sigma _i \in \sigma _j
G_{\overline{m } }
$$
is a well defined homomorphism.
}
\end{lemma}

\noindent In other words, $\rho(\tau ) $ is characterized by:
$$
\tau \sigma _i \in \sigma _{\rho (\tau ) (i) } G_{\overline{m } }
.
$$
One easily verifies that $\rho (\tau ) $ is indeed a permutation
of $\{ 1 , \cdots , K \} $, and that $\rho $ is an homomorphism of
$S_N $ into $S_K . $
\medskip

With this notation, the left hand side of (\ref{F.6}) reads :
$$
\sum _j a_j (\tau \cdot z ) X_{\sigma _{\rho (\tau )(j) } \cdot
\overline{m } } ,
$$
and on replacing $j $ by $\rho (\tau )^{-1 } (j) $ and using the
fact that the $X_{\sigma _j \cdot \overline{m } } $ form a basis
of $F_{\overline{m } } $, we find that $\psi $ is anti-symmetric
iff, for all $j $ and all $\tau \in S_N $,
\begin{equation} \label{F.7a}
a_j (z) = (-1 )^{\tau } a_{\rho (\tau ^{-1 } )(j) } (\tau \cdot z
) .
\end{equation}
If we successively replace $\tau $ by $\tau ^{-1 } $ and $z $ by
$\tau \cdot z $, this becomes
\begin{equation} \label{F.7}
a_j (\tau \cdot z ) = (-1 )^{\tau } a_{\rho (\tau )(j) } (z) ,
\end{equation}
which implies that all $a_j $ are uniquely determined by any one of
them, $a_1 $, say, which we let, by definition, correspond to
$\sigma _1 = e $, the unity element of $S_N . $ More explicitly,
since $\rho (\sigma _j ) (1) = j $ (for $\sigma _j \sigma _1 =
\sigma _j \in \sigma _j G_{\overline{m } } = G_{\rho (\sigma _j
)(1) \cdot \overline{m } } $, by definition of $\rho $), equation
(\ref{F.7a}) with $\tau = \sigma _j $ implies the important
relation
\begin{equation} \label{F.8}
a_j (z) = (-1 )^{\sigma _j } a_1 (\sigma _j\cdot z ) .
\end{equation}
\noindent In particular, $P^{AS} (L^2 (\mathbb{R }^N ) \otimes
F_{\overline{m} } ) $ can be identified with a subspace of $L^2
(\mathbb{R }^N ) $, by sending $\psi $ to $a_1 $ (see (\ref{F9})).
We now analyze the symmetry properties of $a_1 $ imposed by the
anti-symmetry of $\psi . $

\begin{lemma} \label{F.II} Let $H $ be the subgroup of $S_N $ generated
by the set $\{ \tau \sigma _{\rho (\tau )(1) } ^{-1 } : \tau \in
S_N \} $. Then, for all $\sigma \in H $, $a_1 (\sigma \cdot z ) =
(-1 )^{\sigma } a_{1 }(z) $, and these are the only
symmetry-conditions which the anti-symmetry of $\psi $ imposes on
$a_1$.
\end{lemma}

\noindent {\it Proof.} Let $\tau \in S_N $ be arbitrary. Then by
(\ref{F.7}), $a_1 (\tau \cdot z ) = (-1 )^{\tau } a_{\rho (\tau )
(1) } (z) $ which, by (\ref{F.8}), equals $(-1 )^{\tau } (-1
)^{\sigma _{\rho (\tau )(1) } } a_1 (\sigma _{\rho (\tau ) (1) }
\cdot z ) $. Therefore
$$
a_1 (\tau \sigma _{\rho (\tau )(1) } ^{-1 } \cdot z ) = (-1 )
^{\tau \sigma _{\rho (\tau )(1) } ^{-1 } } a_1 (z) ,
$$
whence the lemma. \hfill QED

\begin{lemma} \label{F.III} The group $H $ of lemma \ref{F.II} is
generated by the union of all stabilizers $G_{\sigma _j \cdot
\overline{m } } $ of $\sigma _j \cdot \overline{m } $, $1 \leq j
\leq K . $
\end{lemma}

\noindent {\it Proof.} Let $\tau \in S_N $. Then $\tau \in \sigma
_j G_{\overline{m } } $, for some $j $. Since $\tau \sigma _1 =
\tau e \in \sigma _j G_{\overline{m } } $, we have that $\rho
(\tau )(1) = j $, and therefore $\tau \sigma _{\rho (\tau )(1) }
^{-1 } \in \sigma _j G_{\overline{m } } \sigma _j ^{-1 } =
G_{\sigma _j \cdot \overline{m } } . $

Conversely, if $\sigma \in G_{\sigma _j \cdot \overline{m } } $,
then $\sigma = \sigma _j \sigma ' \sigma _j ^{-1 } $, for some
$\sigma ' \in G_{\overline{m } } $. Put $\tau = \sigma _j \sigma '
$. Then $\rho (\tau )(1) = j $, since $\sigma _j \sigma ' \sigma
_1 \in \sigma _j G_{\overline{m } } $, and therefore $\sigma =
\tau \sigma _j ^{-1 } = \tau \sigma _{\rho (\tau )(1) } ^{-1 } $
is a generator of $H $. We conclude that the set of generators of
$H $ equals $\cup _j G_{\sigma _j \cdot \overline{m } } $, which
proves the lemma. \hfill QED

\begin{lemma} \label{F.IV} Let $H $ be the subgroup from lemma
\ref{F.II}. If $G_{\overline{m } } = \{e \} $, then $H = \{ e \}
$, while if $ G_{\overline{m } } \neq \{e \} $, then $H = S_N . $
\end{lemma}

\noindent {\it Proof.} It is obvious, from lemma \ref{F.III}, that
if $G_{\overline{m } } = \{ e \} $, then $H = \{ e \} $, and $a_1
$ does not have to satisfy any symmetry-conditions with respect to
the action of $S_N $, by lemma \ref{F.II}.

Now suppose that $G_{\overline{m } } $ is non-trivial. Then there
exist two indices $i $ and $j $ such that $\overline{m } _i =
\overline{m } _j $. We can suppose, without loss of generality,
that $i = 1 $ and $j = 2 $. In that case, the transposition $(12)
$ is in $G_{\overline{m } } $, and therefore $(\sigma (1) , \sigma
(2) ) = \sigma (12) \sigma ^{-1 } \in H $, for all $\sigma \in S_N
$, by lemma \ref{F.III} again. But then all transpositions will be
in $H $, which clearly implies that $H = S_N . $ \hfill QED
\medskip

Define a linear mapping
\begin{equation} \label{F9}
U_{\overline{m } } : P^{AS } \left( L^2 (\mathbb{R }^N ) \otimes
F_{\overline{m } } \right) \to L^2 (\mathbb{R }^N ) ,
\end{equation}
by
$$
U_{\overline{m } } (\psi )(z) = \sqrt{K } (\psi (\cdot , z ) ,
X_{\overline{m } } )_{L^2 (\mathbb{R }^{2N } } ,
$$
where we recall that $K = K(\overline{m } ) = \# (S_N /
G_{\overline{m } } ) .$ If $\psi $ is given by  (\ref{F.5a}), then
$ U_{\overline{m } } (\psi )(z) = \sqrt{K } a_1 (z) $. By
(\ref{F.7}), $\| U_{\overline{m } } \psi \|^2 = \|\psi \|^2 $, so
that $U_{\overline{m } } $ is unitary and therefore injective. If
$G_{\overline{m } } $ is non-trivial, then $H = S_N $, and the
image of $U_{\overline{m } } $  is contained in the space $L_{AS}
^2 (\mathbb{R }^N ) $ of anti-symmetric wave-functions on
$\mathbb{R }^N $, by lemmas \ref{F.II} and \ref{F.IV}. Since the
only symmetries of $a_1 $ are those imposed by $H $, it follows
that $U_{\overline{m } } $ is surjective onto $L^2 _{AS }
(\mathbb{R }^N ) $. Similarly, if $G_{\overline{m } } = \{ e \} $,
then $H = \{ e \} $, and the image of $U_{\overline{m } } $ is
$L^2 (\mathbb{R }^N ) $: indeed, if $a_1 = a_1 (z) $ is arbitrary,
then
$$
\psi _{a_1 } := \sum _{\sigma \in S_N } a_1 (\sigma \cdot z )
X_{\sigma \cdot \overline{m } } (y, z )
$$
is an anti-symmetric element of $L^2 (\mathbb{R }^N ) \otimes
F_{\mathbb{M } } $ such that $U_{\overline{m } } (\psi _{a_1 } ) =
a_1 . $
\medskip

\noindent {\it Proof of Theorem~\ref{ThMT}.} Recall that
$$
\mathcal{M }_1 = \{ \overline{m } \in \mathcal{M } :
G_{\overline{m } } = \{ e \} \ \} , \ \ \mathcal{M } _2 = \{
\overline{m } \in \mathcal{M } : G_{\overline{m } } \neq \{ e \} \
\} ,
$$
and define
\begin{equation} \label{F.10}
 U_{\mathbb{M } } =: U_{\mathbb{M } } ^B : P^{AS } \left( L^2 (\mathbb{R }^N
\otimes F_{\mathbb{M } } ) \right) \ \to \ \sum ^{\oplus }
_{\overline{m } \in \mathcal{M }_1 } L^2 (\mathbb{R }^N ) \oplus
\sum ^{\oplus }_{\overline{m } \in \mathcal{M }_2 } L^2 _{AS }
(\mathbb{R }^N ) ,
\end{equation}
by $U_{\mathbb{M } } := \oplus _{\overline{m } \in \mathcal{M } }
U_{\overline{m } } $. Then we have shown that $U_{\mathbb{M } } $
is a surjective isometry. The intertwining formula of $h_{\delta ,
{\rm f } } ^{\mathbb{M } } $ with $U_{\mathbb{M } } $ being
obvious, this proves Theorem~\ref{ThMT}. \hfill QED
\medskip

Contrary to $h_{\delta , {\rm f } } ^{\mathbb{M } } $, the
operator $U_{\mathbb{M } } h_{C, {\rm f } } ^{\mathbb{M } }
U_{\mathbb{M } } ^* $ will in general not act diagonally anymore
on the range of $U_{\mathbb{M } } $, but will contain terms which
couple anti-symmetric and boltzonic components in (\ref{F.10}),
that is, components in $L^2 _{AS } (\mathbb{R }^N ) $ and $L^2
(\mathbb{R } ^N ) $. The potentially problematic terms in
$h^{\mathbb{M } } _{C, {\rm f } } $ are $U_{\mathbb{M } } C^{n,
\mathbb{M } } _{{\rm av }:1 } U_{\mathbb{M } } ^* $  and \
$U_{\mathbb{M } } C^{e, \mathbb{M } } _{{\rm av }:2 } U_{\mathbb{M
} } ^* $ (still dropping the $B $ from our notations). The first
one is easily seen to act diagonally on the right hand side of
(\ref{F.10}): recall that
$$
C^{n, \mathbb{M } } _{{\rm av }:1 } = - \Pi _{\rm eff } ^1 \left(
\frac{1 }{N } \sum _{j = 1 } ^N \log ( \frac{1 }{4 } \rho _j ^2 )
\right) \Pi _{\rm eff } ^1 ,
$$
and identify this with a Hermitian operator on $F_{\mathbb{M } } =
\mbox{Span } \{ X_m : m \in \Sigma (\mathbb{M } \} $ (it acts as a
multiplication operator on $L^2 (\mathbb{R } ^N , F_{\mathbb{M } }
) $). The matrix of $C^{n, \mathbb{M } } _{{\rm av }:1 } $ in the
basis $X_m $, $m \in \Sigma (\mathbb{M } ) $, is easily seen to be
diagonal. Moreover,
\begin{eqnarray*}
\langle X_{\sigma \cdot m } \vert C^{n, \mathbb{M } } _{{\rm av
}:1 } \vert X_{\sigma \cdot m } \rangle &=& - \int _{\mathbb{R }
^{2N } } |X_m (\sigma ^{-1 } \cdot (x, y ) |^2 \log  \frac{1 }{4 }
\left( \Pi
_{j = 1 } ^N \rho _j ^2 \right) ^{1/N } \; dx dy \\
&=& \langle X_m \vert C^{n, \mathbb{M } } _{{\rm av }:1 } \vert
X_m \rangle .
\end{eqnarray*}
Hence, taking $B = 1 $, for simplicity, $C^{n, \mathbb{M } }
_{{\rm av }:1 } $ will act on $F_{\overline{m } } $ as scalar
multiplication by
\begin{eqnarray*}
\langle X_{\overline{m } } C^{n, \mathbb{M } } _{{\rm av }:1 }
\vert X_{\overline{m } } \rangle &=&  -\frac{1 }{N } \sum _{j = 1 }
^N \int _{\mathbb{R }^{2N } } \left( \log \tfrac{\rho _j ^2 }{4 }
\right) \; \Pi _{\nu = 1 } ^N |\chi _{\overline{m }_{\nu } } ^1
|^2 (\rho_{\nu } ) \; dx_{\nu } dy_{\nu
} \\
&=&   -\frac{1 }{N } \sum _{j = 1 } ^N \int _{\mathbb{R }^2 }
\left(
\log \tfrac{\rho ^2 }{4 } \right) \; |\chi _{\overline{m }_j } ^1
|^2 (\rho ) dx
dy \\
&=&    -\frac{1 }{N } \sum _{j = 1 } ^N \left( 2^{\overline{m }_j }
\overline{m }_j ! \right)^{-1 } \int _0 ^{\infty } \rho ^{2
\overline{m }_j + 1 } \log \tfrac{\rho ^2 }{4 } \; e^{-\rho ^2 / 2
} d\rho \\
&=&   \log 2-\frac{1 }{N } \left( \sum _{j = 1 } ^N \psi (\overline{m
}_j + 1) \right) ,
\end{eqnarray*}
where $\psi (z) = \Gamma '(z) / \Gamma (z ) $, the logarithmic
derivative of the $\Gamma $-function. Hence $U_{\mathbb{M } }
C^{n, \mathbb{M } } _{{\rm av }:1 } U_{\mathbb{M } } ^* $ simply
acts diagonally on the right hand side of (\ref{F.10}), and more
precisely, by scalar multiplication in each component.
\medskip

The operator $U_{\mathbb{M } } C^{e, \mathbb{M } } _{{\rm av }:2 }
U_{\mathbb{M } } ^* $ is more complicated: it is not going to be
diagonal in the natural basis and it will in general mix the
different components $\mbox{Ran } U_{\mathbb{M } } $, even those
with index in $\mathcal{M }_1 $ and $\mathcal{M }_2 $. This is
already the case in the simplest case in which both $\mathcal{M
}_1 $ and $\mathcal{M }_2 $ are non-empty, namely that of two
electrons, $N = 2 $, and a total angular momentum of $\mathbb{M }
= 2 $. In that case $\Sigma (\mathbb{M } ) $ is the union of two
orbits under $S_2 $, namely $\{ (0, 2 ) , (2, 0 ) \} $ and $\{(1,
1 ) \} $ the first having as stabilizer the identity, and the
second having as stabilizer the full group $S_2 $. We can
therefore take $\mathcal{M }_1 = \{ (0, 2 ) \} $ and $\mathcal{M
}_1 = \{ (1, 1 ) \} $. We will now compute the matrix element
\begin{equation} \label{Example:Ce}
\langle X^1 _{(0, 2 ) } \vert C^{e, 2 } _{{\rm av }:2 } \vert X^1
_{(1, 1 ) } \rangle ,
\end{equation}
and simply observe that the result is non-zero. For this
computation it is convenient to use complex notation for the
lowest Landau functions (\ref{chi-m}): if we let $\zeta = x + iy
$, then (taking $B = 1 $ again)
$$
\chi ^1 _m = c_m \zeta^m e^{-|\zeta |^2 / 4 } ,
$$
where $c_m = (2\pi 2^m m! )^{-1/2 } $. We then find that
(\ref{Example:Ce}) equals
$$
- c_0 c_2 c_1 ^2 \int _{\mathbb{C } } \int _{\mathbb{C } } \zeta _1
\zeta _2 \overline{\zeta }_2 ^2 \; \log \left( \frac{|\zeta _1 -
\zeta _2 | ^2 }{4 } \right) e^{-(|\zeta _1 |^2 + \zeta _2 |^2 ) /
2 } \; d\zeta _1  d\zeta _2 .
$$
Making the (by now familiar) change of variables $u = (\zeta _1 +
\zeta _2 ) / \sqrt{2 } , v = (\zeta _1 - \zeta _2 ) / \sqrt{2 } $,
this integral becomes
$$
-{c_0 c_2 c_1 ^2\over4} \int _{\mathbb{C } } \int _{\mathbb{C } }
(u^2 - v^2 ) (\overline{u } - \overline{v } )^2  \; \log \left( |v
|^2 / 8 )\right) \; e^{-(|u |^2 + |v | ^2 ) / 2 } \; du dv .
$$
Since $ (u^2 - v^2 ) (\overline{u }^2 - 2 \overline{u }
\overline{v } + \overline{v }^2 ) = |u |^4 - |v |^4 + 2i \; {\rm
Im } \; u^2 \overline{v }^2 + 2 \overline{u } \overline{v } (v^2 -
u^2 ) $, and since, in general,
$$
\int _{\mathbb{C } } \int _{\mathbb{C } } u^{\alpha } \overline{u
}^{\beta } v^{\nu } \overline{v }^{\kappa } g(|u|^2 , |v|^2 ) du
dv = 0 ,
$$
unless $\alpha = \beta $ and $\nu = \kappa $, we obtain that
(\ref{Example:Ce}) equals
\begin{eqnarray*}
&\ & - {c_0 c_2 c_1 ^2\over4} \int _{\mathbb{C } } \int
_{\mathbb{C } } (|u|^4 - |v|^4 ) \log
(|v|^2 / 8  ) e^{-(|u|^2 + |v|^2 ) / 4 } du dv \\
&=&  - {c_0 c_2 c_1 ^2\over4} \left \{  \int
_{\mathbb{C } } |u|^4 e^{-|u|^2 / 2 } du \int _{\mathbb{C } } \log
(|v|^2 / 8 ) e^{-|v|^2 / 2 } dv
\right. \\
&\ & - \left. \int _{\mathbb{C } } e^{-|u|^2 / 2 } du  \int
_{\mathbb{C } } |v |^4 \log (|v|^2 /  8   ) e^{-|v|^2 / 2 } dv
 \right \} = {3\over16\sqrt{2}} ,
\end{eqnarray*}
 as can be shown using the $\Gamma $-function and its
derivative.
%Observing that, for arbitrary $k $,
%$$
%\int _{\mathbb{C } } |u|^{2k } e^{-|u|^2 / 2 } du = 2 \pi \cdot
%2^k \cdot \Gamma (k + 1 ) ,
%$$
%and
%$$
%\int _{\mathbb{C } } |v|^{2k } \log (|v|^2 / 4 ) e^{-|v|^2 / 2 }
%dv = 2 \pi \cdot 2^k \cdot \left( \Gamma ' (k + 1 ) - \log 2 \;
%\Gamma (k + 1 ) \right) ,
%$$
%we finally conclude that (\ref{Example:Ce}) is a non-zero constant
%times
%\begin{eqnarray*}
%\Gamma (3) \Gamma '(1) - \Gamma (1) \Gamma '(3) &=& 6 \Gamma '(1)
%- \Gamma '(3) \\
%&=& -3 -4 \gamma \simeq -5.30888 .
%\end{eqnarray*}
So (\ref{Example:Ce}) is non-zero, and $U_{\mathbb{M } } h_{C,
{\rm f } } ^{\mathbb{M } } U_{\mathbb{M } } ^* $ mixes the two
sectors.

%%%%%%%%%%%%%%%%%%%%%%%%%%%%%%%%%%
\section{\bf Concluding remarks }%
%%%%%%%%%%%%%%%%%%%%%%%%%%%%%%%%%%
We finally want to give an idea what these effective Hamiltonians
might be good for. Our original motivation for introducing them
was for studying the structure of the bottom of spectrum of $H^{B,
\mathbb{M } } $, in connection with the maximum ionization problem
for atoms in strong magnetic fields; see below. To illustrate how
this works, we first consider the comparison with $H_{\delta }
^{B, \mathbb{M } } = h_{\delta } ^{B , \mathbb{M } } \oplus
H_{\perp } ^{B, \mathbb{M } } $; cf. Theorem~\ref{CorrSMT}.
Since, by construction, $B_{\delta }
> B_{(\ref{constant-B}) } $, we know from Theorem \ref{ME} that
$\sigma \left( (H_{\perp } ^{B, \mathbb{M } } \right) \subset (0,
\infty ) $ if $B > B_{\delta } $. Let $E_{\delta } := \inf
h_{\delta } ^B$, then clearly
$E_{\delta } < 0 $.

Since $h_\delta^B$ is unitarily equivalent to
$\alpha^2(-{1\over2}\Delta_z + 2 v_\delta)$ we conclude that
whenever $E_\delta$ is an isolated eigenvalue, its position as
well as its isolation distance is proportional to $\alpha^2$. That
$E_\delta$ is an eigenvalue is true for $Z$ large enough when $N$
is fixed; to determine how big $Z $ has to be exactly, relative to
$N $, for this to happen is an open problem (see below). Let us
assume henceforth that $E_{\delta } $ is an eigenvalue, which then
necessarily is simple. Consequently $E_\delta$ is an eigenvalue of
$H_\delta^{B,\mathbb{M}}$ with multiplicity $\dim F_{\mathbb{M}}^B
$, see (\ref{hdeltaBM}). Choose two points $\xi_\pm:=E_\delta\pm C
\alpha^2$ in $\rho(
H_\delta^{B,\mathbb{M}})\cap\rho(H^{B,\mathbb{M}})$, which satisfy
(\ref{IsolationConditionHHdelta}). This is possible when
$B>B_\delta$, see Theorem~\ref{CorrSMT}. Let $\Gamma$ be the
circle in the complex plane centered at $E_\delta$ with radius
$C\alpha^2$ and define $P$ and $P_\delta$ as the eigenprojections
associated to $H^{B,\mathbb{M}}$ and $h_\delta^{B,\mathbb{M}}$,
respectively, onto their spectrum inside $\Gamma$. To estimate
$P-P_\delta$ we need a bound on
$R(\xi)-R_\delta(\xi):=(H^{B,\mathbb{M}}-\xi)^{-1}-
(H_\delta^{B,\mathbb{M}}-\xi)^{-1}$ for all $\xi\in\Gamma . $ We
know already that $\|R(\xi_\pm) - R_\delta(\xi) \|\le
C_\delta\alpha/d_\delta(\xi_\pm)^2=C_\delta C^{-2}\alpha^{-3}$ by
theorem~\ref{CorrSMT}. To propagate this estimate on all of
$\Gamma$ we use the convenient formula (see \cite[IV.(3.10)]{K})
\begin{eqnarray*}
&\|R(\xi)-R_\delta(\xi)\| \le
{\left\|H_{\delta}^{B,\mathbb{M}}-\xi_\pm\over
H_{\delta}^{B,\mathbb{M}}-\xi\right\|^2 \|R(\xi_\pm)-
R_\delta(\xi_\pm) \|\over 1-|\xi-\xi_\pm| \left\|
H_{\delta}^{B,\mathbb{M}}-\xi_\pm\over
H_{\delta}^{B,\mathbb{M}}-\xi\right\|\|R(\xi_\pm)- R_\delta(\xi)
\|}\le {C_\delta C^{-2}\alpha^{-3}\over1-\sqrt{2}C_\delta
C^{-1}\alpha^{-1}} .
\end{eqnarray*}
Then integrating over the contour $\Gamma $ finally gives
$\|P-P_\delta\|=\OO(\alpha^{-1})$ as $B$ tends to infinity. This
shows that for $B $ large enough these two projections have the
same dimension and since they are continuous with respect to $B $
we finally get that for all $B>B_\delta$, $\dim P=\dim
P_\delta=\dim F_{\mathbb{M}}^B$.

Our conclusion is therefore that, for sufficiently large $B $,
$H^{B, \mathbb{M } } $ will have a cluster of eigenvalues in the
interval $(E_{\delta } - c_{\delta } \alpha ^2 , E_{\delta } +
c_{\delta } \alpha ^2 ) $ with total multiplicity of $\dim
F_{\mathbb{M}}^B $, and apart from this no eigenvalues at a
distance $C \alpha ^2 $ from $E_{\delta } $, (the allowed $B $'s
will depend on $C $), so that the cluster is separated from the
rest of the spectrum of by a distance proportional to $\alpha^2 $.
In the particular case when $\dim F_{\mathbb{M}}^B=1$, i.e.
$\mathbb{M}=0$ or $N=1$, we get an estimate on the difference of
the eigenvectors $\Phi-\Phi_\delta=\OO(\alpha^{-1})$ as
$B\to\infty$ ( in the $L^2$ norm). Here $\Phi$ denotes the
eigenvector of $H^{B,\mathbb{M}}$ and
$\Phi_\delta(x,y,z):=\varphi_\delta^B(z) X_{m=0}^B(x,y) $ if
$\mathbb{M } = 0 $, where $\varphi_\delta^B$ is the ground state
of $h_\delta^B$. In the case $N=1$ one has $\varphi_\delta^B(z)=
\sqrt{ 2 \alpha Z } \; e^{-2 \alpha Z |z| } $ .
\medskip

Consider now the comparison with $h_C^{B,\mathbb{M}}$. The
difficulty here is to find the necessary a priori information on
the structure of $\sigma(h_C^{B,\mathbb{M}})$. In the case of
$N=1$, using the invariance of $\sigma(h_C^{B,\mathbb{M}})$ under
the reflexion $z\mapsto -z$, and the characterisation of the
domain of $\sigma(h_C^{B,\mathbb{M}})$ in Appendix A, one sees
that the odd spectrum\footnote{that is, the spectrum of $h_C $
restricted to the odd wave functions} is $B$-independent and
coincides with the spectrum of the hydrogen in the $s$ sector of
symmetry. Since the even spectrum intertwines with the odd
spectrum and since it is monotonically decreasing with respect to
$B$, cf. (\ref{qBintro}), it is easy to realize that
$\sigma(h_C^{B,\mathbb{M}})\cap\mathbb{R}_-$, apart from the
ground state energy, is made up of clusters of two eigenvalues,
the clusters being separated by a distance of order 1 as $B \to
\infty $. Thus by using Theorem~\ref{SMT} and following a similar
strategy as above one can conclude that for $N = 1 $, and
arbitrarily small $\varepsilon $,
$\sigma(H^{B,\mathbb{M}})\cap(-\infty,-\varepsilon]$ has the same
cluster structure for $B>B_\varepsilon$ sufficiently large.
Moreover this spectrum deviates from the one of the Coulomb model
by at most $c_C \alpha(B)^{3\over2}B^{-{1\over4}}$ as $B\to\infty
. $
\medskip

The model operator $h _{\rm eff}^{B,\mathbb{M}}$ is for the moment
of mainly theoretical interest since it does not seem to be
solvable even in the one electron case. Notice however that one
could solve $h_{\rm eff}^{B,\mathbb{M}}$ numerically, at least for
few electrons and small $\mathbb{M } $, and subsequently use
Theorem~\ref{FMT} to approximate the true spectrum of
$H^{B,\mathbb{M}}$ for large $B $. Given the non-trivial dimension
reduction achieved by theorem \ref{FMT} (from wave-functions of
$3N $ variables to ones of $N $ variables, albeit vector-valued)
such a procedure would, from a numerical point of view, seem
preferable to attacking $H^{B, \mathbb{M } } $ directly.
\medskip

Whether the simpler models, i.e. the delta and the Coulomb model,
are solvable in the $N$-electron case, $N\ge2$, is a challenging
question in view of applications. We are thinking in particular of
the problem of {\sl determining the maximum number $N_c $ of
electrons which a clamped nucleus with charge $Z$ can bind when an
intense homogeneous magnetic field is applied}. \cite{LSY} has
shown that $\liminf N_c / Z \geq 2 $ as $Z, B/Z^3 \to \infty . $
Very little precise is known for fixed $Z $ and high $B $. It is
conjectured that there should be a $B $-independent absolute (that
is, non-asymptotic) upper bound of the form $N_c \leq a Z + b $,
similar to Lieb's bound $N_c \leq 2Z + 1 $ valid when $B = 0 $,
but this is as yet unproved. Some weaker results are known, of
which the best to date is the one of \cite{Sei}; see also
\cite{BR} for work on heuristic models related to our $h_{\rm eff
}^{B, \mathbb{M } } $. It is natural to first try to solve the
maximal binding question for $h_{\delta } ^B $, or any of our
other effective Hamiltonian, and use the approximation theorems of
this paper to draw conclusions for $H^{B, \mathbb{M } } $ itself.
Some modest progress is possible in this way. It is for example
known that the delta model with two electrons is at least
numerically solvable, see \cite{Ros}, and that this model
possesses a unique bound state at the bottom of its spectrum as
long as $Z>0.375 $. Therefore using Theorem~\ref{CorrSMT} we see
that for all $Z>0.375 $ there exists $B_Z\ge0$ such that for all
$B\ge B_Z$ one nucleus with such a charge can bind two electrons.
As a consequence, Lieb's bound of $N_c \leq 2Z + 1 $ is no longer
valid in strong magnetic fields. For general $Z $, no maximum
ionization bound for the $\delta $-model is known as yet.
\medskip

Let us now briefly turn to the effect of particle symmetry. It
follows from Theorems~\ref{FermionizedMThms} and ~\ref{ThMT} that
$H_{\rm f } ^{B,\mathbb{M}}$ can be approximated by a direct sum
of copies of $h_{\delta } ^B $ acting on anti-symmetric $L^2
(\mathbb{R } ^N ) $ plus a direct sum of copies of $h_{\delta } ^B
$ acting on $L^2 (\mathbb{R }^N ) $ {\it without any symmetry
condition}. The latter will occur iff $\mathcal{M }_1 \neq
\emptyset $, which is the case iff $\mathbb{M } \geq 0 + 1 +
\cdots + (N - 1 ) = \tfrac{1 }{2 } N (N - 1 ) $. For such
$\mathbb{M } $, the ground state energy will be approximately that
of boltzonic $h_{\delta } ^B $, which is also the ground state
energy of the bosonic $h_\delta^B $, and the same can be shown to
be the case for the ground state wave function (assuming there is
one), by standard permutation arguments. In fact,
Theorem~\ref{FermionizedMThms} plus Theorem~\ref{ThMT} predict the
existence of a cluster of $\# \mathcal{M }_1 $ eigenvalues at the bottom of the
spectrum of $H_{\rm f } ^{B,\mathbb{M } } $, at a distance of order $\alpha (B
)^2 $ from the origin as $B\to\infty . $

A further interesting corollary to Theorem \ref{ThMT} can be
obtained by considering $\mathbb{M } $ as a free parameter.
%: let us $h_{\delta , {\rm f } } ^B := h_{\delta }^B \vert _{L^2 _{AS } }
%(\mathbb{R }^N ) $. Then
If $h_{\delta } ^B $ possesses  a ground state, whose energy is
isolated in its spectrum,
%and strictly less than the $\inf \sigma (h_{\delta , {\rm f } } ^B ) $,
%\bleu n'est-ce pas toujours le cas?
then for sufficiently large $B $, $H^B_{\rm f} := P^{\rm AS }
H^{B} $ will assume its ground state for an $\mathbb{M } \geq
\tfrac{1 }{2 } N (N - 1 ) $. Stated otherwise, assuming there is a
mechanism for transfer of the angular momentum (e.g. emission and
absorption of photons), atoms in strong magnetic fields will have
an orbital angular momentum in the field direction of at least
$\tfrac{1 }{2 } N (N - 1 ) $. A natural conjecture is that we have
equality here. Notice that this conjecture was shown to be true in
the case of $N=1$ in \cite{AHS}, see also \cite{BaSe}.
\medskip

We mention one further application of these effective
Hamiltonians. After the location of the spectrum to leading order
one can now use regular perturbation theory to compute lower order
corrections. We have shown how this can be done in \cite{BeBDPe}.
This seems definitely more convenient than variational techniques
and more familiar than the Birman-Schwinger method used in
\cite{AHS} for the one electron case. Continuing for example the
above comparison of $H^{B,\mathbb{M}}$ with
$h_\delta^{B,\mathbb{M}}$, it is immediate to realize that
adding the first order perturbative correction will give an error
of order 1. In case $N=1 $ we get that the ground state
energy of $H^{B,\mathbb{M}}$ is equal to $\langle h_{\rm
eff}^{B,\mathbb{M}}\Phi_\delta,\Phi_\delta\rangle+\OO(1)$ with
$\Phi_\delta(x,y,z)=\sqrt{ 2\alpha Z } e^{- 2
\alpha Z \; |z| } \; \chi_\mathbb{M}^B(x,y)$. This should be
compared to Theorem~2.5 of \cite{AHS}. In fact, one can write the
ground state energy as a convergent power series, each term of
which being of order $\alpha^{-k}$ for $k$ running from $-2$ to
infinity. However, as pointed out in \cite{AHS}, in view of the
$\log(B)$ behaviour of $\alpha$ this series is of limited value.
The situation will be much better with the Coulomb model
$h_{C}^{B,\mathbb{M}}$ since the perturbation series will converge
much faster because of the $\alpha^{3\over2}B^{-{1\over4}}$
behaviour of the r.h.s. of (\ref{SMT:DiffRes}). We hope to tackle
this program soon.
\medskip

The above remarks concentrated on applications of our effective
Hamiltonians to the ground state energy of $H^{B, \mathbb{M } } $,
but their potential interest is not limited to that. As the
example of $h_C $ shows, other parts of the discrete spectrum of
$H^{B, \mathbb{M } } $ will also become amenable to analysis, if
this is the case for the effective operator. On a conceptual
level, Theorem \ref{FMT} gives a precise mathematical sense to,
and justification of, the physicist's attractive heuristic picture
of an atom in a strong homogeneous magnetic field as consisting of
electrons in their lowest Landau band states interacting through a
kind of "residual" electrostatic interaction. Finally let us note
that the technics developped in this article are expected to work
in other contexts. An interesting example is that of 2-dimensional
electronic systems
 on a cylinder which describe excitons in carbon nanotubes; cf.
\cite{CDP}.
\medskip

\appendix
\section{{\bf A characterization of the operator domain of $h_C $ } }

In this appendix we will characterize the operator domain of $h_C
= h_C ^B $. It will in fact be convenient to consider a slightly
more general situation. Let $\mathcal{L } = \{ L_{\nu } , 1 \leq
\nu \leq K \} $, be a finite collection of hyperplanes in
$\mathbb{R }^N $ (we might more generally consider non-singular
$C^1 $-hypersurfaces). Let $F $ be a finite dimensional complex
vector space,  with Hermitian inner product $( \cdot , \cdot ) $
and let $A_{\nu } , B_{\nu } $ be Hermitian operators on $F $. Let
$H^k (\mathbb{R}^N , F ) $ be the $k $-th Sobolev space on
$\mathbb{R}^N $, with values in $F $. Then the following
sesquilinear form is well-defined on $H^1 (\mathbb{R }^N , F ) $
and bounded from below:
\begin{equation} \label{tL1}
 t_{\mathcal{L } } (u, v ) = \frac{1 }{2 } (\nabla u ,
\nabla v ) + \sum _{\nu } \left \langle L _{\nu }^* \mbox{Pf }
(|\cdot |^{-1 } ) , (A_{\nu } u , v ) \right \rangle + \left
\langle L_{\nu } ^* \delta , (B_{\nu } u , v ) \right \rangle ,
\end{equation}
$\langle \cdot , \cdot \rangle $ denoting the duality between
distributions and test functions.  We let
$$ h_{\mathcal{L } } = -
\frac{1 }{2 } \Delta _z + \sum _{\nu } A_{\nu } \mbox{Pf } \left(
\frac{1 } {L(z) } \right) + B_{\nu }  \delta  (L_{\nu }
(z) ) .
$$
be the associated self-adjoint operator, whose existence is
guaranteed by the $\rm{KL^2 M N} $-Theorem. Let $\mathcal{R } $ be
the set of connected components of $\mathbb{R }^N \setminus \cup
_{\nu } \mbox{Ker } L_{\nu } $, so that $\mathbb{R }^N \setminus
\cup _{\nu } \mbox{Ker } L_{\nu } = \cup _{R \in \mathcal{R } } R
$ and $R \cap R' = \emptyset $, if $R, R' \in \mathcal{R } , R
\neq R' $. Note that, on any of these components $R $, $L_j ^*
\mbox{Pf} (1 / |\cdot | ) $ simply equals the function $ 1 / |L_j
(z) | . $ We will identify $L_j $ with an element of $\mathbb{R
}^N $, using the Euclidean inner product on $\mathbb{R } ^N $; the
latter will be denoted by a dot: $z \cdot w $, to distinguish it
from the Hermitian inner product $(v, w ) $ on $F $. If we let
$H^2 (R , F ) $ be the $F $-valued Sobolev space of order 2 on the
open subset $R \subset \mathbb{R }^N $, then the domain of
$h_{\mathcal{L } } $ can be characterized as follows:

\begin{theorem} \label{domain} Let $u \in L^2 (\mathbb{R }^N ,
F ) $. Then $u \in {\rm Dom } (h _{\mathcal{L } } ) $ iff the
following three conditions hold:
\medskip

\noindent (i) $ u \in H^1 (\mathbb{R }^N , F ) , $
\medskip

\noindent (ii) For each $R \in \mathcal{R } $,
$$
-\frac{1 }{2 } \Delta u + \sum _{\nu } \frac{1 }{|L_{\nu } (z) | }
A_{\nu } u \in H^2 (R , F ) ,
$$

\noindent (iii) For each $j $ and for each $x \in \{ L_j = 0 \}
\setminus \cup _{k \neq j } \{L_k = 0 \} $:
$$
L_j \cdot \nabla u \left(x + \varepsilon \frac{L_j }{\| L_j \| }
\right) - L_j \cdot \nabla u \left(x - \varepsilon \frac{L_j }{\|
L_j \| } \right) \simeq - 4 \log \varepsilon A_j u (x) + 2 B_j
u(x) + o(1) ,
$$
as $\varepsilon \to 0 . $
\end{theorem}

\noindent {\it Proof.} We will use the following characterization ( see
\cite{K}) of
${\rm Dom } (h _{\mathcal{L } } ) $:
\begin{equation} \label{Kato}
u\in{\rm Dom}(h_{\mathcal{L}})
\iff \left\{
\begin{array}{ll}
u \in {\rm Dom } (t _{\mathcal{L } } ) = H^1 (\mathbb{R } ^N , F ) , \\
\mbox{ and}
\\
| t _{\mathcal{L } } (u, v ) | \leq C_u \|v \|^2 , \ \forall v \in
{\rm Dom } (t _{\mathcal{L } } ) = H^1 (\mathbb{R } ^N , F ) ,
\end{array}\right.
\end{equation}
the norm on the right being the $L^2 $-norm. Here we may, and
will, suppose without loss of generality that $v \in C^1 _c
(\mathbb{R }^N , F ) $, the space of compactly supported $F
$-valued $C^1 $-functions.

To analyze (\ref{Kato}), we will first establish a convenient
integral expression for the pull-backs of $ \delta  $
and of $\mbox{Pf }(1/|\cdot | ) $ under a linear map $L :
\mathbb{R }^N \to \mathbb{R } $. Using the Euclidean inner
product, we can identify $L $ with an element of $\mathbb{R }^N $,
which we also denote by $L $. Using the definition of pull-back
(cf. H\"ormander \cite{H}, chapter 6), one easily shows that
 if $d\sigma _L $ denotes the Euclidean surface measure on
$\mbox{Ker} (L) $, and $\| L \| $ is the Euclidean norm of $L $,
then
\begin{equation} \label{L*delta}
\langle L^*  \delta  , \varphi \rangle = \int _{ \{L =
0 \} } \frac{\varphi }{\|L \| } d\sigma _L ,
\end{equation}
and
\begin{equation} \label{L*Pf}
\langle L^* \mbox{Pf }(1 / |\cdot | ) , \varphi \rangle = - \int
_{\mathbb{R ^N } } \frac{\mbox{sgn}(L (z) ) ) \log |L(z) | }{\| L
\|^2 } \ L \cdot \nabla u (z) dz .
\end{equation}
In fact, since taking pull-backs is coordinate invariant, it
suffices to verify these formulas in a orthogonal coordinate
system in which $L = (\| L \|, 0, \cdots , 0 ) $.
\medskip

We next observe that, for each $R \in \mathcal{R } $, there exists
a (unique) function $s_R : \{ 1 , \cdots , K \} \to \{ 0, 1 \} $,
such that
$$
R = \{ z \in \mathbb{R }^N : (-1 ) ^{s_R (j) } L_j (z) \geq 0 ,
 j = 1 , \; \cdots , K \} ,
$$
and if $\varepsilon > 0 $, we define $R_{\varepsilon } $ by:
$$
R_{\varepsilon } = \{ z \in \mathbb{R }^N : (-1 ) ^{s_R (j) } L_j
(z) \geq \varepsilon , \;  j = 1 , \cdots , K  \} .
$$
Observe that the boundary of $R_{\varepsilon } $, as well as that
of $R $, are polyhedrae, each of whose faces  are
contained in one the hypersurfaces $\{ (-1 ) ^{s_R (j) } L_j =
\varepsilon \} $ and $\{ L_j = 0 \} $, respectively. On such a
face, the outward-pointing normal $n_{R , \varepsilon } $ can be
identified with the vector $- (-1) ^{s_R (j) } L_j / \|L_j \| $
(translated to the relevant base point on the face, to be
precise). Note, that $n_{R , \varepsilon } $ is only defined a.e.
on the boundary (with respect to the surface measure). This will
not cause difficulties, though.
\medskip

 Suppose now that $u \in \mbox{Dom} (h_{\mathcal{L } } ) $
and in particular satisfies the conditions (\ref{Kato}).  We
then have,  for any $v \in C^1 _c (\mathbb{R }^N , F ) $,

\begin{eqnarray} \label{tL2}
t_{\mathcal{L } } (u, v ) &=& \lim _{\varepsilon \to 0 } \; \left
\{ \; \frac{1 }{2 } \sum _{R \in \mathcal{R } } \ \int
_{R_{\varepsilon} } ( \nabla u , \nabla v
) dz \right. \\
&- & \sum _{\nu } \sum _R \int _{R_{\varepsilon } }
\frac{\mbox{sgn } (L_{\nu } ) \log |L_{\nu } | }{\|L_{\nu } \|^2 }
\ (L_{\nu } \cdot \nabla ) (A_{\nu } u , v )
\ dz \nonumber \\
&+ & \left. \sum _{\nu } \int _{ \{ L_{\nu } = 0 \} }
\frac{(B_{\nu } u , u ) }{\|L_{\nu } \| } \ d\sigma _{L_{\nu } }
\right \} . \nonumber
\end{eqnarray}
We want to  apply Gauss' divergence Theorem to each of the
integrals over $R_{\varepsilon } $: this is allowed since if $u
\in H^1 (\mathbb{R }^N , F ) $ satisfies (\ref{Kato}), then by
choosing $v $ compactly supported in $R $, we see that
$$
- \frac{1 }{2 } \Delta _z u + \sum _{\nu } A_{\nu } \left( \frac{1
} {L(z) } \right) u \in L^2_{\rm loc } (R , F ) ,
$$
which obviously implies that $\Delta u \in L^2 _{\rm loc } (R , F
) $ (since we are staying away from the singularities on the
hyperplanes), and hence $u \in H^2 _{\rm loc } (R , F ) $. Now
\begin{equation} \label{tL3}
\int _{R_{\varepsilon} } ( \nabla u , \nabla v ) dz = \int
_{R_{\varepsilon} } (-\Delta u , v ) dz + \int _{\partial
R_{\varepsilon } } (n_{R, \varepsilon } \cdot \nabla u , v )
d\sigma _{R, \varepsilon } ,
\end{equation}
$d\sigma _{R, \varepsilon } $ being the euclidean surface measure
on the boundary. Furthermore,
\begin{eqnarray} \nonumber
&- \|L_{\nu } \|^{-2 } \int _{R_{\varepsilon } } \mbox{sgn }
(L_{\nu } ) \log |L_{\nu
} | \ (L_{\nu } \cdot \nabla ) \left [ (A_{\nu } u , v ) \right ] \ dz = \\
&\displaystyle{ \int _{R_{\varepsilon } } \frac{(A_{\nu } u, v )
}{|L_{\nu } (z) | } dz - \|L_{\nu } \|^{-2 } \int _{\partial
R_{\varepsilon } } \mbox{sgn}(L_{\nu } ) \; \log |L_{\nu } | \;
(A_{\nu } u , v ) \left( L_{\nu } \cdot n_{R , \varepsilon }
\right) \ d\sigma _{R, \varepsilon } . } \nonumber
\end{eqnarray}
Using this, the second term in (\ref{tL2}) can, after
re-arranging, be written as:
\begin{eqnarray} \nonumber
&\ & \sum _{R \in \mathcal{R } } \int _{R_{\varepsilon } } \left(
\sum _j \frac{ (A_j u , v ) }{|L_j (z) }]\right) \; dz
\\
&+ & \sum _j \sum _{\pm } \ \pm \| L_j \|^{-1 } \int _{ \tiny{
\begin{array}{ll} L_j = \pm \varepsilon, \\ |L_{\nu } |
> \varepsilon \ \mbox{if } \nu \neq j \end{array} } } \ \mbox{sgn} (L_j (z) ) \; \log |L_j (z)
| \; (A_j u , v )
%d\sigma _{ \{ L_j = \pm \varepsilon \} }
\nonumber \\
&+ & \sum _{j, k: j \neq k } \sum _{\pm } \ \pm \| L_j \|^{-1 }
\int _{\tiny{\begin{array}{ll} L_k = \pm \varepsilon , \\
|L_{\nu } |
> \varepsilon \ \mbox{if } \nu \neq k \end{array} } } \mbox{sgn} (L_j (z) )
\; \log |L_j (z) | \; (A_j u , v ) ,
%d\sigma_{ \{L_k = \pm \varepsilon \} }
\nonumber
\end{eqnarray}
the surface integrals being with respect to the natural surface
measures. Since $v $ is compactly supported, the third sum will
vanish, in the limit of $\varepsilon \to 0 $. This follows from
the local integrability of $u \cdot \log |L_j | $ on $\{ L_k = 0
\} $ (if $j \neq k $), which can be seen as follows. Since $u \in
H^1 $, its restriction $u |_{\{ L_k = 0 \} }$ is in
(vector-valued) $L^2 $. On the other hand, $\log |L_j | $,
restricted to $\{ L_k = 0 \} $ is in $L^2 _{\rm loc }
( \mathbb{R }  ^{N - 1 } ) $, and therefore $u \log
|L_j | $, restricted to $\{ L_k = 0 \} $ is locally integrable.
\medskip
Rearranging also the  sum over $\mathcal{R } $ of the
boundary terms in (\ref{tL3}) as a sum of integrals over the
various hypersurfaces $\{ L_j = \pm \varepsilon \} $, we conclude
that for $v \in C^1 _c (\mathbb{R }^N  , F  ) $,
\begin{eqnarray} \nonumber
t_{\mathcal{L } (u, v ) } &=& \lim _{\varepsilon \to 0 } \int
_{R_{\varepsilon } } \left( - \frac{1 }{2 } \Delta u + \sum _{\nu
} \frac{
A_{\nu } u }{|L_{\nu } (z) | } \ , v \right) dz \\
&+ & \sum _j \sum _{\pm } \ \mp \frac{1 }{2 } \int _{ L_j = \pm
\varepsilon, |L_{\nu } |> \varepsilon (\nu \neq j ) } \frac{ (L_j
\cdot \nabla u , v )
}{\|L_j \| } d \sigma _{\{ L_j = \pm \varepsilon \} } \nonumber \\
&+ & \sum _j 2 \log \varepsilon \sum _{\pm } \int _{ L_j = \pm
\varepsilon, |L_{\nu } | > \varepsilon (\nu \neq j ) } \frac{(A_j
u , v )}{\|L_j \| }
d \sigma _{\{ L_j = \pm \varepsilon \} } \nonumber \\
&+ & \int_{\{ L_j = 0 \} } \frac{(B_j u , v ) }{\|L_j \| } d\sigma
_{\{L_j = 0 \} } . \nonumber
\end{eqnarray}
Since $u \in H^1 $ implies that $\| u(\cdot + \varepsilon L_j /
\|L_j \| ) - u (\cdot ) \|_{L^2(\{ L_j = 0 \} ) } =
O(\sqrt{\varepsilon } ) $, and since $(u, v ) \in L^1 (\{ L_j = 0
\} ) $, we can replace the next to last term by
$$
2 \log \varepsilon \| L_j \|^{-1 } \int _{ \{ L_j = 0 \} } (A_j u
, v ) d\sigma _{\{L_j = 0 \} } + o(1), \ \ \varepsilon \to 0 .
$$
We can, for the same reason, replace $v $ in the second integral
by its restriction to $\{ L_j = 0 \} $.  Letting
$\varepsilon \to 0 $, it follows that if $u $ satisfies
(\ref{Kato}), then it satisfies (i), (ii) and (iii).

Conversely, suppose that $u $ satisfies conditions (i), (ii) and
(iii) of the theorem. Then running the argument backwards shows
that $|t_{\mathcal{L } } (u, v ) | \leq C_u \| v \| ^2 $, first
for all compactly supported $v $ and thence for all $v \in H^1
(\mathbb{R }^N , F ) $. Hence $u \in \rm{Dom} (h_{\mathcal{L } } )
$, by (\ref{Kato}).  \hfill QED

%}

\end{document}